\documentclass[12pt,letterpaper,hyper]{JHEP3}

\usepackage[utf8]{inputenc}
\usepackage{amsmath,epsfig}
\usepackage{amssymb,amsfonts}
\usepackage{graphicx}
\usepackage{dsfont}
\newcommand{\JJ}{{\cal J}}

\newcommand{\MM}{{\cal M}}
\newcommand{\NN}{{\cal N}}

\newcommand{\OO}{{\cal O}}

\newcommand{\KK}{{\cal K}}

\newcommand{\be}{\begin{equation}}
\newcommand{\ee}{\end{equation}}
\newcommand{\ben}{\begin{eqnarray}\displaystyle}
\newcommand{\een}{\end{eqnarray}}

\newcommand{\bea}[1]{\begin{eqnarray}\label{#1} }
\newcommand{\eea}{\end{eqnarray}}
\newcommand{\refb}[1]{(\ref{#1})}

\def\ZZZ{{\hbox{ Z\kern-1.6mm Z}}}
\def\RRR{{\hbox{ R\kern-2.4mm R}}}
\def\CCC{{\hbox{ C\kern-2.0mm C}}}
\def\zzz{{\hbox{z\kern-1mm z}}}

\newcommand{\lsim}{{<\atop \sim}}

\newcommand{\vt}{\vartheta}

\newcommand{\qeq}{{\hbox{=\kern-2.3mm ? \kern.5mm }}}
\renewcommand{\qeq}{=}

\newcommand{\eps}{\epsilon}

\newcommand{\wt}{\widetilde}
\newcommand{\wh}{\widehat}

\newcommand{\CN}{\mathcal{N}}
\newcommand{\cO}{\mathcal{O}}

\newcommand{\IZ}{\mathbb{Z}}

\newenvironment{myenumerate}{
\begin{enumerate}
   \setlength{\itemsep}{1pt}
   \setlength{\parskip}{0pt}
   \setlength{\parsep}{0pt}}{\end{enumerate}}
\newenvironment{myitemize}{
\begin{itemize}
   \setlength{\itemsep}{1pt}
   \setlength{\parskip}{0pt}
   \setlength{\parsep}{0pt}}{\end{itemize}}
\title{Supersymmetric Index from Black Hole Entropy
}

\preprint{}

\author{
Atish Dabholkar$^{1, 2}$, Jo\~ao Gomes$^{1}$,
Sameer Murthy$^{1}$ and Ashoke Sen$^{1, 3}$\\

\it $^1${Laboratoire de Physique Th\'eorique et Hautes Energies (LPTHE)\\
\it{Universit\'e Pierre et Marie Curie-Paris 6; 
CNRS UMR 7589}\\
\it{Tour 13-14, 5$^{\grave{e}me}$ \'etage, Boite 126, 4 Place Jussieu}, 
\it {75252 Paris Cedex 05, France}}\\

\it $^2$Department of Theoretical Physics\\
\it Tata Institute of Fundamental Research\\
\it Homi Bhabha Rd, Mumbai 400 005, India\\

\it $^3$Harish-Chandra Research Institute \\
\it Chhatnag Road, Jhusi, Allahabad 211019, India

}
\abstract{

For BPS black holes with at least four
unbroken supercharges, we describe how the macroscopic
entropy can be used to compute an appropriate index, which
can be then compared with the same index computed in the
microscopic description. We obtain \textit{exact} results 
incorporating  all higher order quantum corrections in the 
limit  when only one of the charges, representing momentum
along an internal direction, approaches infinity keeping 
all other charges fixed at arbitrary finite values. In this 
limit, we find that  the microscopic index is controlled by
certain anomaly coefficients whereas the macroscopic
index is controlled by
the coefficients of certain Chern-Simons terms
in the effective action. 
The equality between the macroscopic and the microscopic index
then follows as a consequence of anomaly inflow.
In contrast,  the absolute degeneracy does
not have any such simple expression in terms of the anomaly
coefficients or coefficients of Chern-Simons terms. 
We apply our analysis to several examples of spinning 
black holes in five dimensions and non-spinning black holes 
in four dimensions  to
compute the  index
exactly in the limit when only one of the charges 
becomes large, and find perfect agreement
with the result of exact 
microscopic counting.
Our analysis  resolves a puzzle involving M5-branes 
wrapped on a 5-cycle in $K3\times T^3$.}


\begin{document}

\section{Introduction and Summary of Results}

In a class of supersymmetric string theories with sixteen or
more unbroken supercharges we now have a near complete
understanding of the spectrum of BPS states  \cite{9607026,0412287,
0505094,0506249,0508174,0510147,
0602254,0603066,0605210,0607155,0609109,0612011,
0702141,0702150,0705.1433,0705.3874,0706.2363,
0708.1270,0802.0544,0802.1556,0803.2692,0806.2337,
0807.4451,0809.4258,0808.1746,0901.1758,0907.1410,
0911.0586,0911.1563,9608096,9903163,0506151,0506228,0803.1014,
0804.0651}.  This makes these theories  ideal
testing ground for a comparison between the statistical
entropy of an ensemble of states and  the thermodynamic
entropy of the corresponding BPS black hole.
In particular, given such an exact knowledge of the
microscopic degeneracy, one can aim for a possibly
exact comparison with an appropriately defined
macroscopic entropy that  includes  all subleading
corrections.  On the macroscopic side the subleading
classical corrections  arising from local higher
derivative terms in the effective action can be
incorporated using  the Wald formula  \cite{9307038}
whereas the subleading quantum corrections, both
perturbative and nonperturbative,  can be incorporated
using the framework of quantum entropy
function \cite{0809.3304, 0903.1477}.
These can then be compared with the subleading
corrections on the microscopic side after carrying out a
systematic asymptotic expansion of the exact formula.

In carrying out such a comparison one needs to be careful about an
important subtlety. On the macroscopic side, the black hole entropy
defined from the first law of thermodynamics calculates the logarithm of
the absolute degeneracy as required by the Boltzmann relation. On the
other hand, on the microscopic side, one normally computes a
supersymmetric index which receives contribution only from BPS states
and hence is protected from any change under continuous deformations of
the moduli of the theory. {\it A priori} the index and the degeneracy
are not the same, and one could question the rationale behind comparing
the degeneracy computed in the macroscopic side with the index computed
on the microscopic side.

One can proceed nevertheless following the dictum that whatever can get
paired up will generically get paired up, and hence in the
interacting theory the index  equals the degeneracy.
In many examples this strategy
has worked very well for the leading entropy. However, there is no
guarantee that it will work also for the subleading corrections. Indeed,
there are a number of puzzles in the context of
four-dimensional black holes where an appropriate
index
in the conformal field
theory describing a system of branes and the macroscopic
degeneracy computed from black hole entropy
apparently differ at a subleading order \cite{9906094,0712.3166}.
In the context of certain
five-dimensional black holes in M-theory on $K3\times T^2$
and $T^6$
even the leading
asymptotics of the microscopic index apparently
disagrees with the black hole
entropy since the microscopic index
vanishes \cite{9711067}.
One can remedy the situation in some cases by considering a
modified index as suggested
in  \cite{9611205,9708062,9708130,9903163,0702146}.
However, there are
examples such as the one-sixteenth
BPS black hole in $AdS_5$ where no
microscopic index appears to have the right asymptotic growth that
agrees with the black hole
entropy  \cite{0401042,0401129,0506029,0510251,
0807.0559}.
It is thus
desirable both conceptually and practically to develop clear physical
criteria for deciding when the black hole degeneracy is captured by a
microscopic index and which particular index is relevant under what
conditions.

An argument based on the symmetries of the near horizon geometry
of the black hole was suggested in  \cite{0903.1477}. The
basic idea is to use the black hole degeneracy as an input to
compute an index on the macroscopic side and then compare
this with the index computed on the microscopic side.
This relies on the existence of an $AdS_2$
factor in the near horizon geometry of extremal black holes. 
The natural boundary condition on the various fields 
in $AdS_2$ is such as to fix all the charges (including angular momentum) 
and let the dual
chemical potentials fluctuate. In particular a spherically symmetric
horizon, being invariant under rotation, will represent an ensemble
of states all of which carry zero angular momentum.
Thus if $J$ denotes the third component of the angular momentum, and
we
define an index 
with the  weight factor $(-1)^F := \exp{(2\pi i J)}$, then all
the states which account for the entropy associated with the horizon
will have $(-1)^F=1$   
and hence 
\begin{equation}\label{deg=ind}
  \textrm{Tr} (-1)^F = \textrm{Tr}\left( 1 \right)   \, .
\end{equation}
Furthermore, if
the 
black hole preserves at least four supersymmetries,  
then spherical symmetry is forced on us since
the closure of the symmetry algebra implies  that 
the supergroup of symmetries is $SU(1,1 |2)$. This 
contains an $SU(2)$ factor which can be 
identified with a subgroup of spatial rotations.
Thus for such black holes \refb{deg=ind} holds and
the index equals the degeneracy. 

This general argument
needs to be further supplemented by taking into account the possible
contribution from degrees of freedom living outside the horizon, 
-- the hair modes \cite{0901.0359,0907.0593}. These include in
particular the fermion zero modes associated with the broken supersymmetry
generators which account for the supermultiplet structure of a
BPS state. The end result of this analysis expresses an appropriate
index (helicity trace index) 
for the full black hole as the product of the degeneracy associated
with the horizon (or horizons in case of multi-centered black
holes) and the same helicity trace index 
for the hair degrees of freedom \cite{0903.1477}. Since the contribution
from the hair modes is usually small this explians why the black
hole entropy represents the logarithm of an index to
leading order. But this argument also tells us that at the
subleading order we must take into account the effect of the hair
modes while comparing the black hole entropy with the logarithm
of the microscopic index. Indeed, without the hair modes one
runs into internal inconsistencies when two different black hole
solutions have identical near horizon geometries \cite{0901.0359,0907.0593}.

The above line of argument thus gives us a precise route for
computing an index from the macroscopic viewpoint which can then be
compared with the microscopic results for the same index.
However, explicit computation of the index on the macroscopic side is often
quite challenging for two reasons. First, computing the
entropy associated with the horizon requires us to carry out a path
integral over the string fields in the near horizon geometry of the
black hole. Second this procedure requires us to explicitly identify
the hair modes by analyzing supersymmetric deformations of the 
(multi-) black
hole solution and then quantizing them. These difficulties have been
overcome in special cases in various approximations, often
leading to non-trivial agreement between the macroscopic and microscopic
results not only at the perturbative 
level \cite{0412287,0510147,0609109} 
but also at the
non-perturbative level \cite{0810.3472,0903.1477,0904.4253,
0908.0039,0911.1563,1002.3857}. Furthermore this formalism also
predicts correctly the sign of the index from the macroscopic side
which agrees with the results of the microscopic analysis in
a wide class of theories \cite{0903.1477,1008.4209}.

In this paper we  develop an alternative line of argument
for computing the index on the macroscopic side in a 
special limit when only one of the charges 
carried by the black hole, representing momentum along an
internal circle $S^1$ 
in some duality frame, becomes large.  Even though 
this does not allow us to access the most general charge 
configuration, it  provides a practical method for an exact 
computation for sufficiently general configurations for which 
all  charges except the momentum can take any finite value. 
Moreover, by  changing duality frames, one can  choose 
different charges to play the role of the momentum that is 
becoming large  and thus explore different regions 
of the charge lattice. 

In the limit described above, the near
horizon geometry of the black hole coincides with the near horizon
geometry of an extremal BTZ black hole times a compact internal
space $\KK$ \cite{9712251,9802198,0506176}. 
Furthermore, by taking 
the limit in which the
asymptotic radius of $S^1$ approaches infinity,
we
can ensure that the full black hole geometry has an intermediate
region where the space-time has the form of 
$AdS_3\times \KK$, and the
near horizon geometry is embedded in this geometry as an
extremal BTZ black hole \cite{0802.2257}. 
In this case, up to some additional contributions described
below, the degeneracy associated with this black hole can be
regarded as the degeneracy of states 
in the $CFT_2$ dual to the $AdS_3$, and in
the limit of large momentum along $S^1$ this is given by the Cardy
formula. 
Thus computation of the degeneracy reduces to the computation
of the central charge of the dual $CFT_2$, which, as will be
reviewed below, can be computed in terms of coefficients of the
Chern-Simons term in the action of the bulk 
theory \cite{0506176}.\footnote{Although we are using the
language of the holographically dual
$CFT_2$, the computation is based
on macroscopic analysis since the central charge is calculated from
the effective action rather than from a microscopic calculation.
This is also reinforced by the fact that for BTZ black holes
Wald's formula \cite{9307038} for the entropy takes the form of 
Cardy formula \cite{9909061,0506176,0601228}.}
Note that 
this degeneracy includes the
contribution from  the black hole horizon,
any hair modes which live outside the black hole horizon but
inside the
asymptotic $AdS_3\times \KK$ geometry, and also multi-centered
black hole configurations in $AdS_3$ (if they exist). 
This is not a problem since
these must be included in the counting of states anyway.
On the other hand this does not include the contribution from any
modes which might live at the boundary of $AdS_3\times \KK$
or between $AdS_3\times \KK$ and the asymptotic space-time.
By an abuse of notation we shall call these the exterior modes,
-- these will include for example the analog of the $U(1)$ gauge
fields for string theory in 
$AdS_5\times S^5$ \cite{9802150,9807205,0003129,0108152}.
Thus the contribution from these exterior 
modes need to be computed
explicitly and combined with the $CFT_2$ contribution to get the
full microscopic degeneracy.

Let us now turn to the computation of the index in the macroscopic
theory. 
For this we first need to know which index in
we should calculate.
In order that we can compare the macroscopic results with the
microscopic results it is important that we begin with an index
whose definition
does not require any prior knowledge of either the macroscopic
geometry or the microscopic description of the system, but only
on the charges and angular momenta of the state which can be
measured unambiguously by an asymptotic observer. We shall call such
an index a space-time index. 
In order that the index can be reliably computed on both sides we need
to pick an appropriate space-time index which receives contribution
from the BPS states under consideration but not from non-BPS states.
In four dimensions this involves computing appropriate
helicity supertraces \cite{9611205,9708062,9708130}
whereas in five dimensions one can use a
slightly different version described {\it e.g.} in
 \cite{9903163}. In either case this index involves computing a
trace of $P$ multiplied by some polynomial in the angular momenta
over states carrying a fixed set of charges, 
where $P$ -- the analog of $(-1)^F$ for the Witten index -- is a
$\ZZZ_2$ symmetry generator under which the unbroken supersymmetry
generators have odd parity.
The role of the
angular momentum factor is to soak up the fermion zero modes
arising from the $P$-odd broken supersymmetries.
In the macroscopic
description
the contribution to this index comes from two
separate sources: the bulk of $AdS_3$ and
the exterior modes. 
By carefully analyzing the traces over these modes,and taking
into account the fact that the fermion zero modes
arising from the $P$-odd broken supersymmetries are part of the
exterior modes,  one finds that
the full index involves a trace of $P$ in the $CFT_2$ dual to the
bulk of $AdS_3$ and the trace of $P$ together with the angular
momentum factors over the exterior modes. 

For black holes which preserve at least four supercharges, the $AdS_3$
background that appears in the intermediate region has at least $(0,4)$
supersymmetry in the associated supergravity theory. Thus the dual
$CFT_2$ is actually a (0,4) superconformal theory with an $SU(2)$
R-symmetry group. 
Furthermore this R-symmetry group can be identified with the
spatial rotation group or one of its subgroups. 
One finds that the operator $P$ restricted to this $CFT_2$
can be identified as
$Tr ((-1)^{2J_R})$ 
where $J_R$
denotes the generator of the $U(1)$ subgroup of $SU(2)_R$. Thus the 
relevant $CFT_2$ index that appears
in the expression for the space-time index is  $Tr ((-1)^{2J_R})$, with
the trace
taken over the Ramond sector states of the $CFT_2$ carrying different
values of $J_R$ 
but fixed values of $(L_0-\bar L_0)\equiv p$, and fixed values of all the
$U(1)$ charges associated with left-moving currents.
This index receives non-vanishing contribution only from the Ramond
sector
ground states of the right-moving excitations of the
$CFT_2$, \i.e.\  only from states with $\bar L_0=0$, $L_0=p$.
Now in the absence of the $(-1)^{2J_R}$ insertion in the trace
the large $p$ behaviour of this index is given by the Cardy formula
and is determined by the left-moving Virasoro central charge
$c_L$ as well as the levels of various left-moving U(1)
current algebras under which the state carries charges.
We shall argue in \S\ref{sdegind} that the insertion of $(-1)^{2J_R}$ does
not change this behaviour since the effect of $(-1)^{2J_R}$ under
a modular transformation is to introduce a twist on the right-movers
but does not affect the left-moving ground state. Thus the
contribution to the index from the CFT$_2$ is given by the
Cardy formula.
Combining this with the contribution from the exterior modes
we can then recover the full macroscopic index.

While this gives a procedure for computing the index, the
explicit computation still suffers from various technical complications.
First of all in this approach we need to identify the exterior
modes and compute their contribution to the index
explicity. Furthermore to compute the contribution to the
index from the bulk of $AdS_3$ we need the central charge and
the levels of the U(1) current algebra. While these can be
related to the coefficients of various Chern-Simons terms in the
intermediate geometry that contains the $AdS_3$ factor, we
still need to compute these coefficients {\it after taking into
account the effect of higher derivative and quantum corrections.}
There is however
a further simplification that allows us to calculate
the total index directly without having to compute separately the
exterior and the bulk contribution. We shall argue that when one
combines the contribution to the index from the bulk of $AdS_3$
and the exterior modes to compute the total index, the result is
determined in terms of coefficients of Chern-Simons terms
{\it computed in the asymptotic space-time in which the black
hole is embedded} instead of in the intermediate geometry
containing the $AdS_3$ factor. The former can be calculated
explicitly, yielding an exact expression for the total contribution
to the index in the $p\to\infty$ limit. Note that if instead of
computing the index we had been computing the degeneracy, then
no such simplification occurs, and we really need to compute
separately the contribution from the bulk and the exterior
modes and combine them to get the full result.

Armed with this result, 
we carry out explicit computation
of the macroscopic results for the space-time
index for four and five
dimensional black holes in type IIB string theory compactified on
$K3\times T^2$, $T^6$, $K3\times S^1$ and $T^5$
in different limits in which only one of the charges becomes 
large keeping the other charges fixed. 
We then compute the same space-time index on the microscopic side
and compare this with the macroscopic results.
For the microscopic
computation  we
use two different techniques: we can begin with the exact formula for the
index in string theories with 16 or 32 unbroken supersymmetries
and study its limit when one of the charges becomes 
large, or we can
represent the microscopic system as a configuration of M5-brane wrapped
on $P\times S^1$ where $P$ and $S^1$ are appropriate four and one cycles
of the compact space and then calculate its index in the limit of
large momentum along $S^1$ using a Cardy like formula. Note that
in the latter approach we need to use a generalization of the Cardy
formula that determines the growth of the index rather than the
degeneracy. 
In all  cases, we find that the  macroscopic prediction for the index
always agrees with the  microscopic index
in the large momentum limit even for finite values of the other
charges.

The results of our analysis
are summarized below in
tables \ref{Summary2} and
\ref{Summary}.
In these tables
$d_{macro}$ denotes the macroscopic result for the
appropriate space-time index and
$d_{micro}$ denotes the
result of
microscopic
computation of the same space-time index. Below
we give more detailed explanation of the various entries in these
tables.
\begin{table}[h]
  \vspace{5pt}
  \centering
  \begin{tabular}{c|c|c|c}
   $\MM$  & Limit & $ \log d_{macro} $ & $ \log d_{micro}$
\\ \hline
    $K3$ & ${ \textrm{Type-IIB}\atop\textrm{ Cardy}}$
 & $ 2\pi\sqrt{{ Q_1 Q_5\left(n- {J^2\over 4 Q_1 Q_5}\right)}}$
& $ 2\pi\sqrt{{Q_1 Q_5\left(n- {J^2\over 4 Q_1 Q_5}\right)}}$
\\ \hline
    $K3$  & ${ \textrm{Type IIA}\atop\textrm{ Cardy}}$
& $2\pi\sqrt{Q_5 (n + 3)\left( Q_1 - \frac{J^2}{Q_5(n-1)}\right)}$
& $2\pi\sqrt{Q_5 (n + 3)\left( Q_1 - \frac{J^2}{Q_5(n-1)}\right)}$
\\ \hline
    $T^4$ & ${ \textrm{Type-IIB}\atop\textrm{ Cardy}}$
 & $ 2\pi\sqrt{{ Q_1 Q_5\left(n- {J^2\over 4 Q_1 Q_5}\right)}}$
& $ 2\pi\sqrt{{ Q_1 Q_5\left(n- {J^2\over 4 Q_1 Q_5}\right)}}$
\\ \hline
  \end{tabular}
\caption{\small
 Results for five-dimensional black holes for Type-IIB
compactification on $\MM \times S^1$.}
\label{Summary2}
\end{table}
\begin{myitemize}
\item Five-dimensional black holes

Table \ref{Summary2}
shows the results for spinning
five-dimensional black holes in Type-IIB string
theory compactified on $\MM \times S^1$,
carrying
$Q_5$ units of D5-brane charge
wrapped on $\MM\times S^1$, $Q_1$ units of D1-brane charge wrapped on
$S^1$,  momentum $n$ along $S^1$ and angular momentum $J$.
The second column of this table 
contains information about the limits we consider and the
frame that we use for computing
$d_{macro}$ in these
limits. In particular while in the Type-IIB
Cardy limit $\left(n- {J^2\over 4 Q_1 Q_5}\right) 
\to \infty$,  we carry out
the macroscopic computation directly in the type IIB frame, in the
Type-IIA Cardy limit $\left( Q_1 - \frac{J^2}
{Q_5(n-1)}\right) \to \infty$,
we need to go to a dual type IIA frame where $Q_1$ appears as the
momentum.

\bigskip

  \item Four-dimensional black holes

  Table \ref{Summary} shows the results
for four-dimensional non-spinning black holes in  M-theory
compactified on $\MM\times T^2\times S^1$,
carrying $Q_1$ units of M5-brane charge wrapped on
$C_2\times T^2\times S^1$, $Q_5$ units of M5-brane charge wrapped on
$\wt C_2\times T^2\times S^1$, $K$ units of M5-brane charge wrapped
on $\MM\times S^1$ and $n$ units of momentum along $S^1$. Here
$C_2$ and $\wt C_2$ denote a pair of dual 2-cycles of 
$\MM$. The limit
we consider is $n\to\infty$ which corresponds to 
taking the $L_0$ eigenvalue large in the boundary CFT$_2$.
\begin{table}[h]
  \vspace{5pt}
  \centering
  \begin{tabular}{c|c|c}
   $\MM$ & $ \log d_{macro} $ & $ \log d_{micro}$ \\ \hline
   $K3$  & $2\pi\sqrt{(
Q_1 Q_5   K + 4  K) n }$ & $2\pi\sqrt{( Q_1 Q_5   K +
4  K ) n}$  \\
 \hline
     $T^4$  & $2\pi\sqrt{( Q_1 Q_5   K ) n}$ & $2\pi\sqrt{( Q_1 Q_5   K )n}$
\\ \hline
  \end{tabular}
  \caption{\small Results for four-dimensional black holes for  M-theory
compactified on $\MM\times T^2\times S^1$.}
\label{Summary}
\end{table}
\end{myitemize}
The results in both tables clearly
show that the macroscopic prediction $d_{macro}$ for
the space-time index always agrees with the microscopic
prediction $d_{micro}$ for the same index.

There are several novelties in our analysis
which are worth emphasizing: 
\begin{myenumerate}

\item The formul\ae\ quoted in the two tables are exact in the limits
mentioned, \i.e.\ they hold even when the charges other than the one
which is taken to infinity are finite. Thus, they go far beyond the
supergravity approximation and incorporate the effects
of $\alpha'$ and string loop corrections. On the macroscopic side this
is achieved by an exact computation of the coefficients of certain
Chern-Simons terms in the action whereas on the microscopic side
this is achieved by the use of an exact microscopic formula for the
index evaluated in the same limits as described above.

\item
In all cases, the limits that we consider can be regarded as a 
Cardy limit of a CFT$_2$ in an appropriate  duality frame.
If the underlying CFT$_2$ is
weakly coupled in this duality frame,  we can calculate
$d_{micro}$ with the
help of the Cardy like formula for the index and degeneracy.
This is the case for the Type-IIB Cardy limit in Table \ref{Summary2}.
However in 
some cases, the microscopic configuration may contain a
set of NS5-branes and as a result, a weakly 
coupled description of the CFT$_2$ may not be available. 
This is the case for the type IIA Cardy limit in Table 
\ref{Summary2}. 

\item Since $d_{micro}$ is an index which does not change under 
duality\footnote{In general, the index can also jump 
because of wall-crossings but in the  $\CN=4$ context 
these are exponentially subleading corrections not 
relevant to the present analysis.}, one might expect 
that $d_{micro}$ can always be computed in an appropriate 
duality frame where a weakly coupled CFT$_2$
description is available.
Indeed for all the examples 
in Table \ref{Summary2}, a weakly coupled CFT$_2$ 
description is
available in the Type-IIB frame, and this allows us to
compute $d_{micro}$. However, under this 
duality, the type IIA Cardy limit corresponds to an `anti-Cardy' 
limit ($L_0$ eigenvalue fixed and $c$ large) in the
Type-IIB frame. As a result, usual methods of asymptotic 
evaluations are not applicable. One can nevertheless 
compute the asymptotics in this limit from the exact 
formula using the methods of \cite{0807.0237,0807.1314} 
which cleverly exploit the additional symmetries of the 
exact counting function. 
\item Our result for four dimensional black holes resolves a puzzle
raised in  \cite{9906094,0712.3166}
involving black holes in M-theory compactified on
$K3\times T^3$. A naive application of the results of
 \cite{0506176} without accounting for the different treatment required
for the $CFT_2$ dual to the bulk of $AdS_3$ and the 
exterior
modes led to an apparent mismatch between black hole entropy
and the logarithm of the microscopic degeneracy. For example, 
if one evaluates the absolute degeneracy in the microscopic theory 
at weak coupling, then one obtains 
 $2\pi\sqrt{( Q_1 Q_5   K +
6  K ) n}$ for the logarithm of the absolute degeneracy which 
differs from the correct macroscopic answer at sub-leading order. 
In contrast, our
analysis  leads to a perfect agreement between the microscopic and the
macroscopic results as shown in Table \ref{Summary}. This example 
thus underscores the necessity and utility of defining a macroscopic 
supersymmetric index  from black hole entropy for correct comparisons  
with microscopic computations. 
\item Our analysis also gives explicit form of the entropy of
five dimensional spinning
black holes after taking into account the effect of higher derivative
corrections. Previous attempts to do this involved using a specific set
of higher derivative terms in the five dimensional effective
action \cite{0705.1847,0801.1863,0910.4907}.
In contrast our analysis relies on the ability to express the entropy in
terms of coefficients of certain Chern-Simons terms in the action,
and is exact in the limit considered. This also agrees with the
prediction from the microscopic side based on the exact formula
for the index.
\end{myenumerate}

For M5 branes wrapped
on $S^1$ times a four cycle of a generic Calabi-Yau manifold,
ref. \cite{0506176} presented an argument explaining why the
microscopic and the macroscopic entropy would always agree
in the Cardy limit. This argument 
was based
on the observation that in a (1+1) dimensional conformal
field theory with (0,4) world-sheet supersymmetry, the Virasoro
central charge $c_R$ carried by the right movers is related to the
level of the right-moving SU(2) R-symmetry current. This
in turn is related to the anomaly in this R-symmetry current. 
Using anomaly inflow and identifying
the SU(2) R-symmetry current as (a subgroup of) the spatial rotation
one can relate this to the coefficient of the $SU(2)$
Chern-Simons terms in the effective action. 
Furthermore
the difference $c_L-c_R$
between the left- and right-moving central charges
is related to the gravitational anomaly in the world-sheet theory
of the brane system which in turn is related to the coefficient of the
gravitational Chern-Simons term in the effective action of string
theory. 
Using these one can express the central charge $c_L$
of the left-moving
Virasoro algebra -- which controls the growth of the microscopic
degeneracy -- in terms of the gravitational and SU(2) 
Chern-Simons terms in the effective action. 
The latter in turn controls
the black hole entropy, leading to the equality between the macroscopic
and the microscopic entropy.

In our examples, the Calabi-Yau manifold is either $K3 \times T^{2}$ or $T^{6}$. 
Since the systems we analyze also have four unbroken
supersymmetries,
it is natural to ask if similar argument  can be used
to explain the agreement between the microscopic and the
macroscopic entropies in our systems. 
The main additional
complication that arises in our case is the failure of the 
identification of the R-symmetry current of the microscopic
theory with the spatial rotation group. We find that while for
the part of the microscopic system that controls most of the
entropy this identification is correct; it fails for a
small component.\footnote{A
similar mismatch was found in  \cite{9707093} between the modes
living on the Coulomb and the Higgs branch of the D1-D5
system. Here the disagreement
is between different components of the CFT at the same point
in the moduli space.} A simple example of this is provided by the
scalar modes representing transverse oscillation of the brane. These
are non-chiral modes on the brane world-volume and transform
in the $(2_L,2_R)$ representation of the rotation group 
$SU(2)_L\times SU(2)_R$ in five
dimensions and 3 representation of the rotation group 
$SU(2)$ in four dimensions.
For definiteness let us focus on the five dimensional case.
In order to identify the $SU(2)_R$ subgroup of the rotation group in
five dimensions
as the right-moving R-symmetry on the brane
world-volume this must act trivially on the left-movers. This
clearly fails for the left-moving part of the above scalars
which transform in the fundamental representation of
$SU(2)_R$.
As a result the total anomaly in the $SU(2)_R$
spatial rotation symmetry is not
related to the level of the $SU(2)$ R-symmetry current
in the world-sheet theory, 
and the growth of the degeneracy of the microscopic system is no
longer controlled by the anomaly coefficients which can be
directly related to the coefficients of the Chern-Simons terms in the
effective action. 
A similar
problem occurs in the macroscopic description. For the CFT
that is holographically dual to the bulk of the $AdS_3$ factor
appearing in the near horizon geometry, 
the R-symmetry can be identified
as the spatial $SU(2)_R$
rotational symmetry acting on the space transverse
to $AdS_3$. But this identification need not hold for the
exterior modes
which might live on the boundary of $AdS_3$ -- the analog of the
$U(1)$ super Yang-Mills theory for type IIB supergravity on
$AdS_5\times S^5$ -- or between $AdS_3$ and the asymptotic
infinity. 
In particular these modes include 
the transverse oscillation modes of the brane which
fail to satisfy the conditions needed for identifying the R-symmetry
with spatial $SU(2)_R$
rotation.
For this reason the coefficients of the Chern-Simons
terms in the effective action do not directly give us information
about the growth of the degeneracy obtained by combining the
black hole entropy with the contribution from these additional
exterior modes. 
Remarkably however we find that the results on both sides simplifiy
when we focus on an appropriate index rather than the 
absolute degeneracy. In the microscopic theory we find that
the growth of the
index is directly controlled by the gravitational and rotational 
anomaly coefficients exactly as they would have controlled the
growth of the degeneracy if the subtle difference between the
R-symmetry transformation and spatial rotation had been absent.
On the macroscopic side we find that total contribution to the
index from the black hole living in the bulk of $AdS_3$ and
the exterior modes  is controlled by the
coefficients of the Chern-Simons terms in the effective action in
the asymptotic space-time in which the black hole is embedded. 
Since the latter are related to the anomaly coefficients in the
microscopic theory this allows us to establish the equality between
the microscopic and the macroscopic index.

The rest of the
paper is organized as follows. In \S\ref{sdegind} we review the
argument relating the black hole entropy to an index, 
and give an alternative argument 
leading to similar results for special class of
black holes whose
near horizon geometry contains a locally $AdS_3$ factor.
In \S\ref{smacro} we compute the
macroscopic index of a class of 
spinning five dimensional
black holes and
non-spinning four
dimensional black holes in appropriate limit in which the
near horizon geometry develops an $AdS_3$ factor.
In \S\ref{shair} we complement the analysis of \S\ref{smacro}
by including the effect of the exterior contribution to the
macroscopic index. 
In \S\ref{smicro} we use
the known expressions for
the exact microscopic index of these systems
to extract its behaviour in the various Cardy limits
and find perfect agreement with the macroscopic
results of \S\ref{smacro} and \ref{shair}.
In \S\ref{smsw} we repeat the analysis of \S\ref{smicro} using
the M-theory description for the four dimensional black holes. While in this
description we cannot calculate the index exactly, we can compute it in the
Cardy limit and find precise agreement with the results of \S\ref{smicro}.
Both in \S\ref{smicro} and \S\ref{smsw} we also calculate the
microscopic degeneracy whenever there is an
underlying two dimensional weakly coupled
conformal field theory, and find that
in some cases they differ from the microscopic values of the space-time
index.
In \S\ref{sdis} we 
give a general proof of why the microscopic and the macroscopic
computation of the index must always agree. This argument is
a generalization of the argument of  \cite{0506176} by taking into
account existence of degrees of freedom for which the R-symmetry
generators of the world-sheet theory do not always match with the
spatial rotation generators -- a fact that was crucial in the argument of
 \cite{0506176}. This analysis also explains why the
degeneracy and index do not always grow at the same rate.
In appendix \ref{sd} we describe the computation of the coeffcients
of the Chern-Simons terms
which arise from dimensional reduction
of gauge invariant Lagrangian density in higher dimensions.
In appendix \ref{sa} we complement the analysis
of asymptotic growth of the exact microscopic index
in \S\ref{smicro} by demonstrating 
that some terms, which were ignored
in the analysis of \S\ref{smicro}, 
are indeed small compared to the leading terms.

\section{Computing the Index in the Macroscopic
Theory \label{sdegind}}

In this section we first introduce the relevant 
indices for counting BPS states in four and five dimensional
black holes and then 
review the argument of
 \cite{0903.1477,0908.3402} as to how the
degeneracy of a
supersymmetric black hole, computed by
exponentiating the entropy, can be used to compute a
macroscopic index that can be compared with a microscopic
index. We then give
an alternative version of this argument that applies to the special
case of black holes with locally $AdS_3$ factors in their near horizon
geometry.

We begin by defining the helicity trace index in four dimensions.
Due to Lorentz invariance the number of supercharges in a
four dimensional theory is always a multiple of 4; furthermore
the number of supersymmetries preserved by a state is also a
multiple of 4.
If we consider a black hole that breaks altogether $4k$ supercharges,
then the standard index for counting these states is the helicity
trace index $B_{2k}$ defined as \cite{9611205,9708062,9708130}
\be \label{edefb2k}
B_{2k} = {1\over (2k)!}\,
Tr\left[ (-1)^{F} (2h)^{2k}\right] =
{1\over (2k)!}\,
Tr\left[ e^{2\pi i h}\, (2h)^{2k}\right]\, ,
\ee
where $h$ is the third component of the angular momentum of
a state in the rest frame, and the trace is taken over all states
carrying a given set of charges. In order that a given state gives a
non-vanishing contribution to this index, the number of supersymmetries
broken by the state must be less than or equal to $4k$;
otherwise trace over the fermion zero modes associated with the
broken supersymmetries will make the trace vanish.
On the other hand if we have
states with
precisely $4k$ broken supersymmetries then $B_{2k}$ receives
contribution from these states, but not from any other state with
more than $4k$ broken supersymmetries.
Since quantization of each pair of fermion zero modes produces
a pair of states carrying $h=\pm{1\over 4}$,
the trace over the $4k$ fermion zero modes associated with the
broken supersymmetries is given by
\be \label{etrace}
(e^{i\pi/2}-e^{-i\pi/2})^{2k} (2k)!/2^{2k}=(-1)^k (2k)!\, .
\ee
The $(2k)!$ term arises from the binomial expansion of
$(2h)^{2k}$ after expressing   $h$ as the sum of contributions
from different pairs. This cancels the similar factor in the
denominator in \refb{edefb2k}, leaving behind a contribution
of $(-1)^k$.

It is easy to find a generalization of this in five dimensions.
The spatial rotation group in five dimensions is
$SU(2)_L\times SU(2)_R$. We shall denote by $J_L$ and $J_R$
their $U(1)$ generators.
Among the set of all the supersymmetry generators of the theory,
half belong to $(2_L,1_R)$ representation of 
$SU(2)_L\times SU(2)_R$ and the other half belong to the
$(1_L,2_R)$ representation of $SU(2)_L\times SU(2)_R$.
For a state preserving 4 supersymmetries,
the unbroken supersymmetry generators can be either in the
$(2_L,1_R)$ or in the $(1_L,2_R)$ representation; we shall
choose the convention in which they are in the $(1_L,2_R)$
representation. 
The rest of the supersymmetry
generators will be broken, giving rise to fermion
zero modes carrying the quantum numbers of the broken
generators.
Let $4k$ be 
the number of 
broken generators in the $(1_L,2_R)$ representation.
We now consider the index \cite{9903163}
\be \label{edefi2k}
C_{2k}\equiv {(-1)^k\over (2k)!}\,
Tr\left[ (-1)^{ 2 J_R} \, (2J_R)^{2k} \right]\, ,
\ee
where
the trace
is taken over all states carrying a fixed value of
$J_L$ and fixed set of charges
but all possible values of $J_R$.
Without the $(2J_R)^{2k}$ factor the trace over the 
$(1_L,2_R)$ fermion
zero modes carrying $(J_L,J_R)=(0,\pm{1\over 2})$ would make
the trace vanish. However the $(2J_R)^{2k}$ factor soaks up
the $2k$ pairs of fermion zero modes exactly as in the case of
four dimensional black holes and gives a non-vanishing result.
There are also $(2_L,1_R)$
fermion zero modes carrying $(J_L, J_R)=
(\pm{1\over 2},0)$, but they do not make the trace vanish since the
trace is taken over states carrying a fixed $J_L$.
It is also easy to see that the non-BPS states do not contribute to this
index. They would have additional
fermion zero modes in the $(1_L,2_R)$ representation and hence
trace over these fermion zero modes 
would make the index vanish.

As an example, we can consider the BMPV black hole \cite{9602065}
in type IIB string theory compactified on $K3\times S^1$. This
breaks 12 out of 16 supersymmetries. Eight of the broken supersymmetry
generators are in the $(2_L,1_R)$ representation,  four of the
broken generators are in the $(1_L,2_R)$ representation and
the four unbroken generators are in the $(1_L,2_R)$ representation.
Since there are four broken generators in the $(1_L,2_R)$
representation the argument given above shows that the
relevant index is $C_2$. Similarly if we consider BMPV black hole
in type IIB string theory on $T^4\times S^1$ then it breaks 28
of the 32 supersymmetries, with 16 broken generators in the
$(2_L,1_R)$ representation, 12 broken generators in the
$(1_L,2_R)$ representation and 4 unbroken generators in the
$(1_L, 2_R)$ representation. The index required for counting these
states is $C_6$.

Let us now compute the contribution to these indices from
BPS black holes with four supercharges. For definiteness
we begin with a four dimensional black hole
breaking $4k$ supersymmetries and compute
the index $B_{2k}$.
The net contribution to the index from
a black hole 
can be expressed as a sum of products of the contributions
from the horizon and the
hair \cite{0901.0359,0903.1477,0907.0593};
this could  involve contribution from
multiple horizons for multi-centered black holes.
Let us first focus on the contribution from single centered black holes.
Since the fermion zero modes associated with broken supersymmetries
live outside the horizon and hence are part of the hair degrees of
freedom of the
black hole \cite{0901.0359,0907.0593},\footnote{The
fermion zero mode associated with a
broken supersymmetry generator can be constructed as follows.
We make a
supersymmetry transformation of the original solution by an infinitesimal
parameter that approaches a constant
spinor corresponding to the broken generator
at infinity and vanishes
for $r<a$ for some constant $a$. By choosing $a$
such that the horizon lies
at $r<a$ we can ensure that
such deformations live outside the horizon
and hence are part of the hair degrees of freedom.}
we can express the contribution to
the index from the black hole as
\be \label{edecomp}
 B_{2k} = {1\over (2k)!}\, \left[Tr_{hor} (-1)^{2h_{hor}}\right] \,
\left[ Tr_{hair} (-1)^{2h_{hair}}
 (2h_{hair})^{2k}\right]\, ,
 \ee
 where $h_{hor}$ and $h_{hair}$ denote the helicities
 carried by the hair and the horizon. 
 For states carrying a fixed set of charges $\vec q$ this can be
 expressed as
 \be \label{eb2k}
 B_{2k}(\vec q) = \sum_{\vec q_{hor}} B_{0;hor}(\vec q_{hor})
 B_{2k;hair}(\vec q-\vec q_{hor})\, ,
 \ee
 where
 \be \label{eb0hor}
 B_{0;hor}(\vec q)=Tr_{hor;\vec q} (-1)^{2h_{hor}} \, ,
 \ee
 and
 \be \label{eb2khair}
 B_{2k;hair}(\vec q) = Tr_{hair;\vec q} (-1)^{2h_{hair}} 
 (2h_{hair})^{2k}\, .
 \ee
 Here $\vec q$ in the subscript of $Tr$ denotes that the trace is
 being taken over states carrying a fixed set of charges $\vec q$.
  We  now argue that
 if the black hole has 4 unbroken supersymmetries and
 if its near horizon geometry has an $AdS_2$ factor,
 then it must carry $h_{hor}=0$. The argument
 goes as follows.
 The closure of the
 SL(2,R) isometry of the near horizon geometry, and the
 unbroken supersymmetries requires that
 the near horizon geometry has the full $su(1,1|2)$ symmetry
 algebra.
 This includes $su(2)$ as a subalgebra, forcing the horizon
 to be spherically symmetric and hence carry zero angular
 momentum.\footnote{In asymptotically
Minkowski space-time
 or $AdS_d$ space-time with $d\ge 4$, where the asymptotic
 boundary conditions are set by the chemical potentials instead
 of the charges, the spherical symmetry
 of the background will correspond to evaluating the partition
 function at zero value of the chemical potential conjugate to the
 angular momentum. However the path integral over the 
 string fields in the near
 horizon $AdS_2$ geometry that is
 used to compute the horizon degeneracy
 must be carried out over configurations carrying
 fixed values of the total charges including
 angular momentum \cite{0809.3304,0809.4264}.
 Thus in
 this case spherical symmetry implies zero value of the angular
 momentum carried by the black hole. \label{f1}}
 This gives
 \be \label{etrbh}
B_{0;hor}(\vec q) = Tr_{hor;\vec q} (-1)^{2h_{hor}} =
Tr_{hor;\vec q} (1) =  d_{hor}(\vec q)
\, ,
 \ee
 where $d_{hor}(\vec q)$ is the degeneracy associated with the horizon
 degrees of freedom for charge $\vec q$. 
 In the classical limit it is given by the
 exponential of the Wald entropy, but more generally it can be
 computed from the path integral over the string fields in
 the near horizon geometry \cite{0809.3304}. Using \refb{eb2k}
  and \refb{etrbh} we get the contribution to
 $B_{2k}$ from the black hole
 \be \label{ebhcont}
 B_{2k}(\vec q)  = \sum_{\vec q_{hor}} d_{hor}(\vec q_{hor})
 B_{2k;hair}(\vec q -\vec q_{hair})\, .
 \ee
 $B_{2k;hair}(\vec q)$ can be computed once we have identified the
 hair degrees of freedom of the black hole. Thus \refb{ebhcont}
 can be used to make a prediction for the index $B_{2k}(\vec q)$
 from the macroscopic side.
 Note also 
 that since $d_{hor}(\vec q)$ is positive \refb{ebhcont} 
 makes a definite
 prediction for the sign of $B_{2k}$ provided we have
 sufficient knowledge of $B_{2k;hair}$. In particular in
 situations where the only hair modes are the fermion zero
 modes associated with broken supersymmetries, we have
 $\vec q_{hair}=0$, $B_{2k;hair}=(-1)^k$ and hence
 $(-1)^k B_{2k}=d_{hor}>0$. As  was shown in
  \cite{0903.1477,1008.4209}, 
  the macroscopic prediction for the sign of
 $B_6$ agrees with the result of explicit microscopic computation
 for all the $\NN=4$ supersymmetric string theories for which this
 index has been computed.
 The generalization of \refb{ebhcont} to multi-centered black holes
 is straightforward; since each center carries zero angular
 momentum due to supersymmetry, the contribution to
 $B_{2k}$ will be given by a formula analogous to \refb{ebhcont},
 with $d_{hor}$ replaced by the product of $d_{hor}$ from each
 center and we have to sum over all possible ways of
 distributing the total charge among the horizon and the hair.

This argument has a straightforward generalization to
five dimensions with $h$ replaced by $J_R$.
Incidentally, this reasoning also implies the well-known facts
that the horizon of a
supersymmetric black hole cannot carry any spin in
four dimensions, and that the
horizon of a supersymmetric black hole
can carry only the $SU(2)_L$ spin in five dimensions.
Also this argument does not generalize to the problematic
one-sixteenth BPS black holes in
$AdS_5$ since they have too little supersymmetry, and the completion of
the algebra containing the supersymmetry generators and the $SL(2,R)$
isometry of $AdS_2$
do not force us to have an $SU(2)$ symmetry in the near horizon
geometry. 

While this argument explains the relation between the index and
degeneracy, applying this argument to compute the contribution to
the index from the macroscopic side requires identifying 
explicitly the hair
modes of the black hole which is not always an easy 
task \cite{0901.0359,0907.0593}. Also this
would require computing $d_{hor}$ by evaluating the path integral
over string fields in the near horizon background 
geometry \cite{0809.3304} -- another
difficult problem. For these reasons we shall now give an alternative
approach to computing the index on the macroscopic side 
which is in the same
spirit but differs in details. If we consider a black hole for which one
of the charges can be identified as an internal momentum
along some circle $S^1$, and if we
consider a limit in which this momentum becomes large keeping
all the other charges fixed, then the near horizon geometry of such
a black hole is known to develop a locally $AdS_3$ factor by
combining the near horizon $AdS_2$ geometry with this
internal circle $S^1$ \cite{9712251,9802198}. 
Furthermore if we now adjust the asymptotic
moduli fields in such a way that we take the asymptotic value of the
radius of $S^1$ to infinity keeping all the other moduli fixed,
then the solution also develops a global $AdS_3$ factor in the
intermediate region, and the black hole solution can now be regarded
as the BTZ black hole living in this asymptotically 
$AdS_3$ space-time \cite{9803231,9809027,0802.2257}. 
The classical entropy of this black hole
has the form of a Cardy formula, with the central charge given by
some specific function of the parameters of the 
Lagrangian \cite{9909061,0506176,0601228}. 
Thus the classical black hole entropy can be
reinterpreted as the Cardy formula of the  CFT$_2$ that is
holographically dual to string theory in this geometry. Since the
Cardy formula in CFT$_2$ is expected to hold in the full quantum theory
this suggests that we can use Cardy formula as the quantum generalization
of the black hole entropy. The problem of computing the 
quantum corrected entropy of the
black hole then reduces to the problem of computing the 
quantum corrected central
charge. 
Since we do not have direct
knowledge of the CFT$_2$, this has to
be computed using the data 
in the bulk theory {\it after taking into account quantum
corrections to the bulk effective action}.
In this sense the entropy computed this way is still the
macroscopic entropy.

There are however several subtleties overlooked in the above discussion.
First of all the Cardy formula is supposed to count total degeneracy
of states in $CFT_2$ without caring about whether they are represented as
single or multicentered black holes inside $AdS_3$, or whether the
contribution comes from the horizon or the hair modes. So the above
definition of the black hole entropy includes all of these 
contributions.
This is not a serious problem since in order to compare the macroscopic
result with the microscopic result we need to sum over all the
contributions on the macroscopic side in any case. 
The microscopic degeneracy may also receive contribution from
configurations with multiple $AdS_3$ throat \cite{0802.2257}, but this can
be avoided by working in appropriate domains in the moduli space.
In any case in theories with 16 or more
supercharges the contribution from the multicentered black holes
is small and we shall ignore their contribution in our analysis.
The main
complication arises from the fact that the degeneracy of the CFT$_2$
dual to the theory living on the bulk of $AdS_3$ does not capture all
the degrees of freedom of the system. There may be additional
degrees of freedom living on the boundary of $AdS_3$ (analogous
to the $U(1)$ factor for $AdS_5$ \cite{9805112}), or in the
region between $AdS_3$ and the asymptotic infinity. This will in
particular include the Goldstino 
fermion zero modes associated with supersymmetries
which are broken by the
$AdS_3$ background. We shall collectively call all such modes
exterior modes.\footnote{The need for separating out the
exterior modes can be seen as follows. In the microscopic theory
where 
the dynamics is described by that of an oscillating string there
are a set of degrees of freedom associated with the center of mass
motion which are decoupled from the rest of the degrees of freedom.
This decoupling in the infrared limit follows from Goldstone's theorem
and is expected to be exact even in the full interacting theory. Thus
if the $CFT_2$ dual to $AdS_3$ had contained the full set of degrees
of freedom of the black hole then this CFT will be given by a sum
of two (or more) CFT's which do not interact with each other.
Thus we can define two stress tensors and hence there must be two
gravitons in the bulk theory, in contradiction to what we see. Furthermore
in the bulk theory the $SU(2)$ R-symmetry group of (0,4) supersymmetry
can be directly related to the spatial rotation group for four dimensional
black holes and the $SU(2)_R$ subgroup of the spatial rotation
group for five dimensional black holes. This identification fails
to hold for the CFT containing the center of mass modes, showing
again that these modes must live outside the bulk of $AdS_3$.}
Since in the limit we are considering -- taking the 
asymptotic radius
of $S^1$ to infinity keeping the momentum quantum
number fixed --
the physical momentum vanishes,  part of the black hole
solution lying
between the asymptotic space-time and the intermediate $AdS_3$ region
has full 1+1 dimensional Lorentz symmetry. Thus we would expect
the dynamics of the exterior
modes to be described by
some $(1+1)$ dimensional field theory. Their contribution
has to be combined with the Cardy formula to recover the total
degeneracy of states. 

So far we have talked about degeneracy, but our real interest is
in the index. Let us now see how the above discussion will change
when we try to compute the index instead of the degeneracy. 
Again for definiteness we shall first consider four dimensional
black holes and compute the index $B_{2k}$.
Denoting by $h_{bulk}$ and $h_{exterior}$ the contribution to
$h$ from the degrees of freedom living in the bulk and the exterior
of $AdS_3$, we can express
the trace appearing in \refb{edefb2k} as
\be \label{enewb2k}
B_{2k}=
{1\over (2k)!}\,
Tr\left[ e^{2\pi i (h_{bulk}+h_{exterior})}\, (2h_{bulk}
+2h_{exterior})^{2k}\right]\, .
\ee
For simplicity we shall assume that the
supersymmetries broken by
the black hole are also broken by the intermediate $AdS_3$ region,
\i.e.\ the black hole, when regarded as a solution in $AdS_3$, does
not break any further 
supersymmetry.\footnote{In some cases the unbroken 
supersymmetry generators get modified when we switch on
the charges on the black hole, e.g. when we switch on
M2-brane charges on an M5-brane\cite{9903163,0608059}. 
For the systems we shall
analyze this does not happen.
}
In this case all the fermion zero modes 
associated with broken
supersymmetry are part of the exterior degrees of freedom,
and in order to get a non-vanishing contribution to the trace
in \refb{enewb2k} we need to pick the 
factor of $(2h_{exterior})^{2k}$ from the binomial expansion
of $(2h_{bulk}
+2h_{exterior})^{2k}$. This gives
\be \label{enewk3s}
B_{2k}=
{1\over (2k)!}\,
Tr\left[ e^{2\pi i (h_{bulk}+h_{exterior})}\, 
(2h_{exterior})^{2k}\right]
=\sum_{\vec q} B_{bulk}(\vec q_{bulk}) 
B_{2k;exterior} (\vec q-\vec q_{bulk})\, ,
\ee
where $B_{bulk}= Tr_{bulk} e^{2\pi i h_{bulk}} $ in a fixed charge
sector. 

In the Cardy limit one of the charges, 
which we shall call $p$, becomes
large. We shall denote by $\vec Q$ the rest of the charges and
denote by $\wt{~~}$ the Fourier transform of various quantities 
$B_{2k}$,  $B_{2k;exterior}$ etc. with respect to
the charge $p$. For example
\be \label{edeft1}
\wt B_{2k}(\vec Q, \tau) = \sum_p \, B_{2k}(\vec Q, p) \, e^{2\pi i p\tau}
\, ,
\ee
etc. 
We shall now make the assumption that the 
exterior modes
do not carry any charge other than $p$, so that in the sum 
in \refb{enewk3s}
$\vec Q_{bulk}$ is always equal to $\vec Q$. 
Then \refb{enewk3s} takes the form:
\be \label{edeft2}
\wt B_{2k}(\vec Q,\tau)
=\wt B_{bulk}(\vec Q, \tau) 
\wt B_{2k;exterior} (\tau)\, .
\ee
Our goal is to compute the behaviour of $B_{2k}(\vec Q,p)$
for large $p$. This is controlled by the
behaviour of $\wt B_{2k}(\vec Q,\tau)$ for small $\tau$. To determine
this we need to find the small $\tau$ behaviour of 
$\wt B_{bulk}(\vec Q, \tau)$ and
$\wt B_{2k;exterior} (\tau)$. 
First we focus on $\wt B_{bulk}(\vec Q,\tau)$.
If instead of the index $B_{bulk}(\vec Q,p)$ we had been
interested in the degeneracy $d_{bulk}(\vec Q,p)\equiv Tr(1)$ 
of left-moving
excitations in the CFT$_2$, then for large $p$ it would grow as
$\exp[2\pi\sqrt{c_L^{bulk} p/6}]$ according to the Cardy formula,
where $c_L^{bulk}$ is the central charge of the left-moving 
Virasoro algebra of the CFT$_2$. 
This
implies
\be \label{edeft3}
\wt d_{bulk}(\vec Q, \tau) \sim \exp[\pi i c_L^{bulk} / 12\tau]\, ,
\ee
for small $\tau$. 
We shall now argue that for small $\tau$ the behaviour of
$\wt B_{bulk}(\vec Q,\tau)$ 
 is given by the same formula.
The argument goes as follows. With the help of a 
modular transformation in the two dimensional CFT, 
the behaviour of $d_{bulk}$
in the Cardy limit can be related to the ground state energy
of the left-moving sector, and this is what leads to 
\refb{edeft3}, with $-c_L^{bulk}/24$ interpreted as the
ground state energy of the left-moving sector. 
Now if instead of $\wt d_{bulk}$ we consider the
index $\wt B_{bulk}$, then following the same logic we can relate
its small $\tau$ behaviour 
to the ground state energy in the left-moving sector, but this
time with
a $(-1)^{2h_{bulk}}$ twisted boundary condition under
$\sigma\to\sigma+2\pi$, $\sigma$ being the world-sheet space
coordinate. 
Now quite generally when
the black hole (and the associated $AdS_3$) has four unbroken
supersymmetry generators, they combine with the
conformal
symmetry of the $AdS_3$ to generate a $(0,4)$ superconformal algebra.
This includes an $SU(2)$ R-symmetry current whose global part can
be identified as the spatial rotation symmetry.
Due to this identification, $h_{bulk}$ can be interpreted as the
zero mode of the $U(1)\subset SU(2)$  R-symmetry
current of the CFT$_2$.
Since the twist by the zero mode of the right-moving
$U(1)\subset SU(2)$ R-symmetry current of the CFT$_2$ is
not expected to
affect the ground state energy in the left-moving sector, this
energy will continue to be given by $-c_L^{bulk}/24$, and hence
the small $\tau$ behaviour 
of $\wt B_{bulk}$ is also given by the Cardy formula:
\be \label{ebtau}
\wt 
B_{bulk}(\vec Q, \tau) \sim \exp[\pi i c_L^{bulk} / 12\tau]\, .
\ee
We shall
see in \S\ref{shair} that  for small $\tau$
$\wt B_{2k;exterior} (\vec 0, \tau)$ is given by a 
formula similar
to \refb{ebtau}:\footnote{We should emphasize here 
that while the modularity of
$\wt B_{2k;bulk}(\vec Q,\tau)$ follows from the
fact that in the $CFT_2$ dual to the $AdS_3$ the action
of $h_{bulk}$ is chiral, 
the function $\wt B_{2k;exterior}(\vec Q,\tau)$  
is not \textit{a priori} a modular form
since the action of $h_{bulk}$ on the exterior
modes is not chiral.  Hence, 
to derive this asymptotics it is necessary to examine the 
behavior of  $\wt B_{2k;exterior} (\tau)$ explicitly as we 
describe in \S\ref{shair}.}
\be \label{edeft4}
\wt B_{2k;exterior}(\vec 0, \tau) \sim 
\exp[\pi i c_{L,eff}^{exterior} / 12\tau]\, ,
\ee
for some constant $c_{L,eff}^{exterior}$.
Substituting \refb{edeft3} and \refb{edeft4} into \refb{edeft2}
we get
\be \label{edeft5}
\wt B_{2k}(\vec Q,\tau) \sim \exp[\pi i c_{L,eff}^{macro} / 12\tau],
\qquad c_{L,eff}^{macro} \equiv c_L^{bulk} + c_{L,eff}^{exterior}\, ,
\ee
and hence, for large $p$,
\be \label{ebkgrow}
B_{2k}(\vec Q, p) \sim \exp[2\pi \sqrt{c_{L,eff}^{macro} p / 6}]\, .
\ee
This is our general expression for the index $B_{2k}$
for four dimensional black holes computed in the
macroscopic theory.
We shall describe the computation of $c_L^{bulk}$ and 
$c_{L,eff}^{exterior}$ in sections  
\ref{smacro} and \ref{shair} respectively.
We shall in fact
see that $c_{L,eff}^{macro}$ 
is simpler to calculate than the individual
contributions from the bulk and the exterior since the former
is directly 
related to the coefficients of certain Chern-Simons terms
in the effective action in the asymptotic space-time in which the
black hole is embedded.

Let us now consider five dimensional black holes.
The analysis goes through more or less in the same manner
with $h$ replaced by $J_R$
provided that all the 
$SU(2)_L$ singlet supersymmetry
generators which are broken by the black hole solution
are also broken by the $AdS_3$.
 The main difference arises from the
fact that the exterior modes of the five dimensional black
holes carry both $J_L$ and $J_R$ quantum numbers besides
the momentum along $S^1$. Since we are summing over
$J_R$ but keeping $J_L$ and the momentum along $S^1$
fixed in defining the index, the analog of \refb{enewk3s} now
takes the form
\be \label{ecardy5}
C_{2k}(\vec q) = \sum_{\vec q_{bulk}} C_{bulk}(\vec q_{bulk})
C_{2k,exterior}(\vec q - \vec q_{bulk}) \, ,
\ee
where the charge vector $\vec q$ now also includes the
$J_L$ quantum number, and $C_{bulk}$ denotes the trace
of $e^{2\pi i J_R}$.
We now separate out two charges from the
set $\vec q$, -- the momentum $p$ along $S^1$ and the $U(1)_L
\subset SU(2)_L$ charge $J_L=J/2$ -- and call the rest of the charges
$\vec Q$. Denoting by $\wt{~ ~}$ 
the Fourier transforms in the charges $p$ and $J$, by $\tau$ and $z$
the variables conjugate to $p$ and $J$, and assuming that the
exterior modes only carry $p$ and $J$ quantum numbers, we
can express \refb{ecardy5} as
\be \label{edeft21}
\wt C_{2k}(\vec Q,\tau,z)
=\wt C_{bulk}(\vec Q, \tau,z) 
\wt C_{2k;exterior} (\tau,z)\, .
\ee
In order to find the behaviour of $C_{2k}$ in the Cardy limit we need
to find the behaviour of $\wt C_{2k}$ for small $\tau$.
The behaviour of $\wt C_{bulk}(\vec Q, \tau,z)$ for small $\tau$
can be found as follows.
First we note that in CFT$_2$ dual to the bulk of
$AdS_3$ the $SU(2)_L$ and $SU(2)_R$
spatial rotations can be identified as the left- and right-moving
SU(2) R-symmetry currents. From this it follows that
if instead of $C_{bulk}$ we had considered the degeneracy
$d_{bulk}$ of the left-moving excitations then
for large $p$ and 
$J\lsim \sqrt p$, $d_{bulk}(\vec Q,p,J)$ grows as
$\exp\left[2\pi\sqrt{c_L^{bulk} \left(p - {J^2\over 4 k_L^{bulk}}
\right)}\right]$. Equivalently for small $\tau$ and
$z\lsim 1$ we have
\be \label{edeft7pre}
\wt d_{bulk}(\vec Q, \tau,z)\equiv \sum_{p,J}
d_{bulk}(\vec Q,p,J) e^{2\pi i p\tau+2\pi i Jz}
\sim \exp\left[{\pi i c_L^{bulk} \over 12\tau} -2\pi i {k_L^{bulk} z^2
\over \tau}
\right]\, .
\ee
\refb{edeft7pre} is a consequence of the modular symmetry of the
CFT$_2$, and the exponent $-{c_L^{bulk} \over 12 }
+ k_L^{bulk} z^2$ has the interpretation of the ground state
energy of the left-moving sector of the
CFT with the boundary condition twisted by
$e^{2\pi i Jz}$ under $\sigma\to\sigma + 2\pi$. 
Now following the same logic as in the case of $\wt B_{bulk}$ we
can argue that  for small
$\tau$, $\wt C_{bulk}$ will have the same behaviour as 
$\wt d_{bulk}$, since under modular transformation the extra
insertion
of $(-1)^{2J_R}$ in the trace will mapped to a twist by
$(-1)^{2J_R}$, and this, being a twist by the zero mode of a
right-moving current, should not affect the ground state energy
of the left-moving sector. 
Thus we get
\be \label{edeft7}
\wt C_{bulk}(\vec Q, \tau,z)
\sim \exp\left[{\pi i c_L^{bulk} \over 12\tau} -2\pi i {k_L^{bulk} z^2
\over \tau}
\right]\, .
\ee
Furthermore we shall find in \S\ref{shair}
that for small $\tau$ and $z\lsim 1$,
$\wt C_{2k;exterior} (\tau,z)$ is given by a similar formula
\be \label{edeft8}
\wt C_{2k;exterior} (\tau,z) \sim
\exp\left[{\pi i c_{L,eff}^{exterior} \over 12\tau} -2\pi i 
{k_{L,eff}^{exterior} z^2
\over \tau}
\right]\, .
\ee
Eq.\refb{edeft21} now gives
\ben \label{edeft9}
&& \wt C_{2k} (\vec Q,\tau,z) \sim
\exp\left[{\pi i c_{L,eff}^{macro} \over 12\tau} -2\pi i {k_{L,eff}^{macro} z^2
\over \tau}\right], \nonumber \\
&& c_{L,eff}^{macro} \equiv c_L^{bulk} 
+ c_{L,eff}^{exterior},
\qquad k_{L,eff}^{macro} \equiv k_L^{bulk} + k_{L,eff}^{exterior}
\, ,
\een
and hence
\be \label{edeft10}
C_{2k}(\vec Q,p,J)
\sim \exp\left[2\pi\sqrt{c_{L,eff}^{macro} \left(p - {J^2\over 4 k_{L,eff}^{macro}}
\right)}\right]\, .
\ee
Again we shall find that $c_{L,eff}^{macro}$ and $k_{L,eff}^{macro}$ are
given in terms of the coefficients of certain Chern-Simons terms
in the effective action in the asymptotic space-time, and hence are
easier to calculate than the individual contributions from bulk and
the exterior modes.

\section{Macroscopic Results for Four and
Five Dimensional Black Holes}
\label{smacro}

In this section we  examine the
macroscopic formul\ae\ for the entropy of a certain class of four
and five dimensional black holes in appropriate limits.
Much work has been devoted to the study of corrections
to black hole entropy due to a specific class of higher derivative
terms obtained by supersymmetrizing the curvature squared
terms, both in four and five
dimensions \cite{9812082,9906094,0007195,0510147,0607155,0609109,
0703087,0705.1847,0801.1863}.
However in this approach there is no {\it a priori}
justification of including only a specific subset of higher derivative
corrections to the effective action for computing the entropy.
Our approach will be based on the method advocated in
 \cite{0506176} where in certain limits the
higher derivative corrections to the black hole
entropy can be related to the coefficients of certain Chern-Simons terms
in the effective action. Since these coefficients are integers, possible
corrections to them are severely limited, and hence can often be
computed. This will allow us to compute the black hole entropy
in appropriate limits after including the effect of all possible higher
derivative corrections.

In all subsequent discussions we shall use units in which $\alpha'=1$,
normalize the ten dimensional Einstein-Hilbert + dilaton action
so that it takes the form 
\be \label{emetnor}
(2\pi)^{-7} \int d^{10}x \, \sqrt{-\det G} \, e^{-2\Phi}\,
\left[ R + 4(\nabla\Phi)^2\right]
\ee
and normalize the 
$p$-form field strength so that its kinetic term
has the form 
\be \label{epnorm}
-{1\over 2}\, {1\over p!}\,
(2\pi)^{-7} \int d^{10}x \, \sqrt{-\det G} \,
e^{\kappa\Phi} \, F_{M_1\cdots M_p} F^{M_1\cdots M_p}
\ee
for some appropriate constant $\kappa$.

\subsection{D1-D5-p system in type IIB on $K3\times S^1$} 
\label{szero}

We consider
a  system of $Q_5$ D5-branes wrapped on
$K3\times S^1$, $(Q_1+Q_5)$
D1-branes wrapped on $S^1$ and $n$ units of
momentum along $S^1$.
We
 choose the convention  in which positive $n$ denotes left-moving
momentum along $S^1$ and take $n$ to be positive.
Since a D5-brane wrapped on K3 carries $-1$ unit of D1-brane
charge, $Q_1$ represents the physical D1-brane charge carried
by this system.
Besides these charges we also make the system carry angular
momentum. 
In five dimensions the spatial rotation group is
$SU(2)_L\times SU(2)_R$. We shall 
consider D1-D5-p system of the type described above
carrying $U(1)_L\subset SU(2)_L$ charge $J_L=J/2$.
Supersymmetry then forces the corresponding black hole 
solution to be
invariant under $SU(2)_R$, \i.e.\ carry zero $SU(2)_R$
charge.
The entropy of a supersymmetric 
black hole carrying these charges, calculated using the two derivative 
action of the supergravity theory and the classical Bekenstein-Hawking
formula, is given by \cite{9602065}
\be \label{ezeroent}
2\pi\sqrt{Q_1 Q_5 n - {J^2\over 4}}\, .
\ee
Our goal will be to understand corrections to this formula in
two different limits:
\begin{enumerate}
\item Type IIB Cardy limit: $n\to\infty$ with $Q_1$, $Q_5$ fixed.
$|J|$ must be bounded by a term of order $\sqrt n$ so that
 $Q_1 Q_5 n - {J^2\over 4}$ scales as $n$.
\item Type IIA Cardy limit: $Q_1\to\infty$ with $Q_5$, $n$ fixed.
$|J|$ must be bounded by a term of order $\sqrt{Q_1}$ so that
 $Q_1 Q_5 n - {J^2\over 4}$ scales as $Q_1$.
\end{enumerate}
The type IIB
Cardy limit clearly
corresponds to taking the momentum along the circle 
$S^1$ to
infinity keeping other charges fixed in a type IIB frame. As 
we shall see,
the type IIA Cardy limit corresponds to taking the momentum along
the dual circle 
to infinity keeping the other charges fixed in a dual type IIA
frame.

\subsubsection{Type IIB Cardy limit}
We begin by writing 
down the near horizon
geometry of the black hole \cite{9602065,9601029}
in the normalization convention of
\cite{0611143,0901.0359}
for the action and the solution:
\ben \label{erotate}
dS^2&=&r_0 {d\rho^2\over \rho^2} + dy^2
+r_0(dx^4 + \cos\theta d\phi)^2
+{ J \lambda^2\over 8 r_0 R V} dy (dx^4 + \cos\theta d\phi)
- 2\sqrt{r_0} \rho dy d\tau
 \nonumber \\ &&  +r_0  \left(
 d\theta^2 + \sin^2\theta d\phi^2\right)
+ \wh g_{mn} du^m du^n \, , \qquad y\equiv y+2\pi R\nonumber \\
e^{\Phi} &=& \lambda\, , \nonumber \\
F^{(3)} &=& {r_0\over \lambda}\, \left[\eps_3 + *\eps_3
+{ J \lambda^2\over 16\, r_0^2 R  V}
\, dy\wedge \left({1\over \rho}\, d\rho\wedge
(d \, x^4 +\cos\theta\, d\phi)
+ \sin\theta\, d\theta\wedge d\phi\right)\right]\, , \nonumber \\
\een
where $dS^2$ denotes the string metric, $\Phi$ denotes the dilaton,
$F^{(3)}$ is the RR 3-form field strength, $g_{mn}$ is the
metric on K3 with  volume $(2\pi)^4\, 
V$, $u^m$'s are the coordinates
on $K3$, 
$(x^4,\theta,\phi)$ are the coordinates labelling a 
3-sphere $S^3$,
$\eps_3\equiv \sin\theta\, dx^4\wedge d\theta\wedge d\phi$ 
is the volume form
on this 3-sphere satisfying $\int_{S^3}
\eps_3 = 16\pi^2$ and $*\eps_3$ denotes the Hodge-dual
of $\eps_3$ in six dimensions.
The attractor equations determine
the near horizon parameters in terms of the charges via the
relations
\begin{equation} \label{ep5}
r_0 = {\lambda Q_5  \over 4
}, \quad   V={Q_1\over Q_5}, \quad R  = \sqrt {\lambda n\over Q_1}
\, .
\end{equation}
Note that $\lambda$, labeling the string coupling, is undetermined
on the horizon.
If $Q_1$, $Q_5$, $n$ are large but finite
then by adjusting $\lambda$ we can keep the
string coupling small, and the parameter $r_0$, that controls the
length scale of the near horizon geometry,
large. Thus in this case we have a systematic expansion in $\alpha'$
and the string coupling, with the leading term in the expansion
given by the Bekenstein-Hawking entropy.  We shall try to go beyond
this by taking only one of the charges to be large,
keeping the other charges finite.

By a coordinate change
\ben \label{ecchange}
&& x^4 = \tilde x^4 - {J\lambda^2 \over 16 r_0^2 R V} y, \quad
y = \tilde y \left(1 - {J^2 \lambda^4\over 256 r_0^3 R^2 V^2}
\right)^{-1/2}
= \tilde y \left(1 - {J^2 \over 4 Q_1 Q_5 n}\right)^{-1/2}\, ,
\nonumber \\
&& \tau =  \wt \tau \left(1 - {J^2 \over 4 Q_1 Q_5 n}\right)^{1/2}
\, ,
\een
we can bring the metric to the form
\ben \label{enewback}
dS^2&=&r_0 \left({d\rho^2\over \rho^2} -
\rho^2 d\tilde\tau^2\right) + (d\tilde y
-\sqrt{r_0}\, \rho d\tilde \tau)^2
+ \wh g_{mn} du^m du^n\nonumber \\
&& + r_0 \left((d\tilde x^4 + \cos\theta d\phi)^2
+ d\theta^2 + \sin^2\theta d\phi^2
\right) \, .
\een
Except for the global identification implicit in the periods
of the coordinates $(x^4,\theta,\phi, \tilde y)$ this metric 
has no dependence on $J$.  
In fact it
has locally an $AdS_3\times S^3$ factor, with
the coordinates $(\rho,\tau,y)$ labelling $AdS_3$ and 
$(\theta,\phi,x^4)$ labelling $S^3$ \cite{9805097}. The appearance of the
$AdS_3\times S^3$ factor allows us to apply the general
reasoning given in  \cite{0506176} which we shall now review.

We begin with the observation that
the classical Wald entropy given in
\refb{ezeroent} can be written 
in the form \cite{9204099,9909061,0506176,0601228}
\be \label{ep6}
S_{BH} = 2\pi \sqrt{{c^{bulk}_L \over 6} \left(n - {J^2 \over 
4 k_L^{bulk}}
\right)}\, ,
\ee
where 
\be\label{ezerock}
c^{bulk}_L = 6 Q_1 Q_5, \qquad k_L^{bulk} = Q_1 Q_5\, .
\ee
A physical explanation of this formula may be
given as follows. If we take the limit in which
the asymptotic radius $R_{as}$ 
of the circle $S^1$ goes to infinity {\it keeping fixed
all the quantized charges and adjusting the other moduli
so that the asymptotic geometry approaches a finite
six dimensional background}, then the black hole
solution develops an intermediate region which contains an 
$AdS_3\times S^3$ factor and the near horizon configuration given in
\refb{erotate} appears as the near horizon geometry of an
extremal BTZ black hole sitting inside the 
$AdS_3$ \cite{0802.2257}.\footnote{The asymptotic 
boundary
of this $AdS_3$ space is the (1+1) dimensional space
labelled by $y$ and $\tau$, and the symmetry of the
intermediate $AdS_3\times S^3$
includes the Lorentz transformation in this (1+1) dimensional
space as well as the full rotation group of $S^3$. This may
appear surprising since the black hole
carries $-n$ units of 
momentum along $S^1$ which breaks Lorentz symmetry
in the $y-\tau$ plane and angular momentum $J_L=J/2$ which
breaks the $SO(4)$ rotational symmetry of $S^3$ to its $SU(2)_R$
subgroup. The reason that this is not inconsistent is that if we take
$R_{as}$ to infinity keeping $n$ and $J$
fixed then the physical momentum $n/R_{as}$ and the angular
momentum per unit length $J/R_{as}$ both vanish. Since these
are the parameters which enter directly the black hole solution,
it is not surprising that from the point of view of an
asymptotic observer we recover the Lorentz invariance in the
$\tau-y$ plane as well as the $SO(4)$ rotational invariance in
this limit.}
Furthermore this black hole carries a $U(1)_L$ charge
$J/2$, with the $U(1)_L$ interpreted as the abelian subgroup of the
$SU(2)_L\subset 
SU(2)_L\times SU(2)_R$ gauge group 
arising out of dimensional reduction on $S^3$. By AdS/CFT correspondence
the states represented by this charged extremal BTZ black hole 
in this asymptotically $AdS_3$ geometry
can now be regarded as RR sector
states with $(\bar L_0=0, L_0=n)$
in the holographically dual $CFT_2$. Furthermore
in CFT$_2$ the $SU(2)_L\times SU(2)_R$
rotational symmetry of $S^3$ appears as the zero mode subalgebra of an
$SU(2)_L\times SU(2)_R$
current algebra, with $SU(2)_L$ being a left-moving current algebra
and $SU(2)_R$  a right-moving current algebra.
Thus  $J/2$ represents the charge carried by the global
part of the $U(1)_L\subset SU(2)_L$ current algebra. 
Eq.\refb{ep6} can now be interpreted as the Cardy 
formula for
the growth of  states in the two dimensional conformal
field theory, with
$c^{bulk}_L$ 
representing the central charge carried
by the left moving component of the stress tensor of the
CFT$_2$, and $k_L^{bulk}$ representing the level of the $SU(2)_L$
current algebra. 

In order to check that this interpretation is correct we must
independently compute $c_L^{bulk}$ and $k_L^{bulk}$ from first
principles and check that the result agrees with \refb{ezerock}
computed from black hole entropy. For this
it is also useful to introduce the quantities $c_R^{bulk}$
which  represents the central charge carried
by the right moving component of the stress tensor of this
CFT$_2$ and $k_R^{bulk}$ that gives the level of the
right-moving $SU(2)_R$ current algebra.
In the classical limit $c^{bulk}_L-
c^{bulk}_R$ is given by
the coefficient $c_{grav}^{bulk}$
of the Lorentz Chern-Simons term in the bulk theory,
and $k_L^{bulk}$ and $k_R^{bulk}$ are given by the
coefficients of the  Chern-Simons terms
involving $SU(2)_L$ and $SU(2)_R$ gauge fields in the
bulk theory.
Furthermore 
using the supersymmetry of the bulk theory one finds that
the boundary CFT$_2$ possesses (0,4) superconformal
symmetry.\footnote{In fact in 
this particular example the CFT$_2$ has (4,4)
superconformal supersymmetry and this allows us to relate
$c_L^{bulk}$ directly to the coefficient 
$k_L^{bulk}$ of the $SU(2)_L$
Chern-Simons terms in the bulk action via 
$c_L^{bulk}=6 k_L^{bulk}$.
However in order to maintain a uniform discussion of all the
cases we shall only make use of the (0,4) supersymmetry
of the CFT$_2$.}
Thus
the
$SU(2)_R$ current algebra can be identified as the R-symmetry
algebra of the (0,4) superconformal algebra, leading to the
relation
$c_R^{bulk}=6k_R^{bulk}$\cite{0506176}.
This gives:
\be \label{ecrkr}
c_L^{bulk}-c_R^{bulk} = c_{grav}^{bulk},
\qquad c_R^{bulk}=6 k_R^{bulk}\, .
\ee
This allows us to express $c_L^{bulk}$ as
\be \label{eexp}
c_L^{bulk} = c_{grav}^{bulk} + 6 k_R^{bulk}\, .
\ee
In the specific example under consideration,
there is no Lorentz
Chern-Simons term
in the supergravity approximation. 
Thus we have $c_{grav}^{bulk}=0$ and so
$c_R^{bulk}=6 k_R^{bulk}$. Eq.\refb{ezerock} would
then follow if we have $k_L^{bulk}=k_R^{bulk}
=Q_1Q_5$. The proof of this, given in
 \cite{0606230} has been reviewed in appendix \ref{sd} where we also
give a generalization of this result.

So far we have just reinterpreted the classical Bekenstein-Hawking
formula. But now we can turn the argument around to give a
definition of the black hole entropy in the full quantum theory
in the type IIB Cardy limit defined earlier. 
The main ingredient is the
observation that for states carrying large $L_0$ 
the Cardy formula is valid in the CFT$_2$
even in the quantum theory. 
Thus we can use \refb{ep6} to compute the full
quantum entropy associated with the bulk of $AdS_3$
in the large $n$ limit,
provided $c_L^{bulk}$ represents the left-moving
central charge in the full quantum theory, and $k_L^{bulk}$
is the level of the $SU(2)_L$ current algebra in the full
quantum theory.\footnote{If we assume that the effect of
quantum corrections can be encoded in a local 1PI
action in $AdS_3$, then \refb{ep6} can be derived directly
in the bulk theory, either via euclidean action 
formalism \cite{0506176} 
or
via Wald's formula \cite{0601228}.} Furthermore \refb{eexp} will also
continue to hold in the full quantum theory. Thus the problem
reduces to the
computation of $k_L^{bulk}$, $k_R^{bulk}$ and $c_{grav}^{bulk}$.
As argued in \S\ref{sdegind} these quantities also determine the
contribution to the index from the modes living in the bulk of
$AdS_3$. We still need to compute separately the contribution from
the exterior modes to which we shall come back later.

Let us now discuss the computation of these quantites
after taking into account higher derivative and quantum
corrections.
Since the coefficients of the
Chern-Simons terms are quantized,  $c^{bulk}_{grav}$,
$k_R^{bulk}$ and $k_L^{bulk}$ are
quantized. It then follows from \refb{eexp} that
$c_L^{bulk}$ is also quantized.
Thus these coefficients must be polynomial in
the charges $Q_1$, $Q_5$ and cannot, for example, carry any inverse
powers in the charges. This severely restricts the form of the corrections.
Furthermore, we can use a generalization of
the scaling argument of  \cite{0908.3402} to determine
in which order in perturbation theory a given correction could arise.
If we take an extremal
black hole carrying NS-NS sector electric charges
$\vec q^{(el)}_{NSNS}$, NS-NS sector magnetic charges
$\vec q^{(mag)}_{NSNS}$, and RR sector charges 
$\vec q_{RR}$, then the
argument of  \cite{0908.3402} implies that the $l$-loop contribution
to any of the coefficients $c^{bulk}_{grav}$,
$k_R^{bulk}$ and $k_L^{bulk}$ -- collectively denoted by 
by $c^{(l)}$ -- satisfies the scaling law:
\be \label{escaling}
c^{(l)}\left(\vec q^{(mag)}_{NSNS}, \lambda^2
\vec q^{(el)}_{NSNS}, \lambda \vec q_{RR}\right)
= \lambda^{2-2l} c^{(l)}\left(\vec q^{(mag)}_{NSNS},
\vec q^{(el)}_{NSNS}, \vec q_{RR}\right)\, .
\ee
In our example, $Q_1$, $Q_5$ are RR sector charges. 
Thus the scaling relation takes the form
\be \label{enewscale}
c^{(l)}(\lambda Q_1, \lambda Q_5) =
\lambda^{2-2l} c^{(l)}(Q_1, Q_5)\, .
\ee

Clearly the leading contribution to $k_L^{bulk}$ and $k_R^{bulk}$,
given by $Q_1Q_5$,
satisfies \refb{enewscale}
with $l=0$, showing that this arises at the tree level.
A correction to any of the coefficients
$c^{bulk}_{grav}$,
$k_R^{bulk}$ and $k_L^{bulk}$
 linear in $Q_1$ or $Q_5$ 
will be suppressed with respect to
the leading term by a power of $1/\lambda$ under the scaling
given in \refb{enewscale}. According to \refb{enewscale} this must arise
at $l=1/2$, \i.e.\ at the `half loop' order. Since close string
perturbation theory includes only contributions from integral number
of loops we see that we cannot get corrections to the central
charge which are linear in $Q_1$ or $Q_5$. Put another way, a correction
that is suppressed by a single power of RR charges must come from
terms in the action involving odd number of RR fields. Such terms are
not present in type IIB string theory.
By following the same line of argument we see that
a constant term in the central charge will produce an effect at the
one loop order. Thus we might ask whether one loop correction in type IIB
string theory could produce corrections to the Lorentz,
$SU(2)_L$ or $SU(2)_R$
Chern-Simons term in the theory living on $AdS_3$. 
We can consider two possibilities. The first possibility is that such
a term could arise from a one loop correction to the ten
(or six) dimensional effective action integrated over $K3\times S^3$
(or $S^3$). 
Since the term we are looking for is
independent of $Q_1$ and $Q_5$, it cannot involve the 3-form fluxes and
must be purely gravitational in nature. Now
in an even dimensional theory
it is impossible to write down a purely gravitational Chern-Simons
 term. Thus we do not get a constant contribution to $c_L^{bulk}$ by
integrating a higher dimensional Chern-Simons term on $S^3$.
The second possibility is that there can be one loop
contributions to the Lorentz and/or $SU(2)_R$ Chern-Simons terms
which arise in the theory after compactification on $K3\times S^3$
and cannot be seen in the ten or six dimensional type IIB string theory.
{\it A priori} we cannot rule out such a possibility; so let us
denote such one loop contributions to $c_{grav}^{bulk}$, $k_L^{bulk}$ 
and $k_R^{bulk}$ by $A$, $B$ and $C$ respectively. 
This gives
\be \label{eexactchor}
c^{bulk}_L = 6 Q_1 Q_5+A + 6 C, \qquad k_L^{bulk} = Q_1 Q_5
+ B \, .
\ee

$c_L^{bulk}$ and $k_L^{bulk}$ given in \refb{eexactchor}
control the contribution to the black hole 
degeneracy/index from the bulk of $AdS_3$. 
To
determine the full contribution to the macroscopic index 
using \refb{edeft9}, \refb{edeft10}
we must combine this with the contribution from the exterior
degrees of freedom mentioned in the previous section.
We shall show in \S\ref{shair} that 
the exterior contributions $c_{L,eff}^{exterior}$ and
$k_{L,eff}^{exterior}$ to the index precisely
cancel the constant shifts $(A + 6 C)$ and $B$ 
in eq.\refb{eexactchor}, leading to:
\be \label{eexact}
c_{L,eff}^{macro} = 6 Q_1 Q_5, \qquad k_{L,eff}^{macro} = Q_1 Q_5
\, .
\ee
Using \refb{edeft10} we now see that
the leading supergravity formula for the entropy is the complete
contribution to the index in the Cardy limit:
\be \label{dhor1}
\ln \, d_{macro}(n,Q_1, Q_5, J)
\simeq 2\pi\sqrt{Q_1 Q_5 n - {J^2\over 4}}\, .
\ee
Here $\simeq$ denotes equality up to corrections suppressed by
inverse powers of $n$. 
The macroscopic result \refb{dhor1}
agrees with the
microscopic result \refb{eiibgen} which will be derived in
\S\ref{smicro}. 

\subsubsection{Type IIA Cardy limit}
Let us turn to the type IIA Cardy limit: $Q_1\to\infty$ at fixed
$n, Q_5$ and $J\lsim \sqrt{Q_1}$ \cite{0807.0237}.
The strategy will be to examine the black hole in a different
duality frame in which $Q_1$ appears as a momentum along
a circle, and then apply the same line of reasoning to find an exact
formula for the black hole entropy in the limit $Q_1\to\infty$
at fixed $n, Q_5$. For this we first make an S-duality transformation
in the ten dimensional type IIB string theory to map this system
to an NS 5-brane, fundamental string, momentum system, and
then make a T-duality along the circle $S^1$ to map this into a system
in type IIA string theory on $K3\times \wt S^1$ with
$Q_5$ NS 5-branes wrapped along $K3\times \wt S^1$, $n$
fundamental strings wrapped along $\wt S^1$ and $Q_1$ units
of momentum along $\wt S^1$. By following the duality transformation
rules and making a change of coordinates
one finds that the near horizon geometry of the black hole
in the type IIA variables, denoted by $\sim$, takes the form
\ben \label{ep9iia}
dS^2&=&\wt r_0 {d\rho^2\over \rho^2} + dy^2
+\wt r_0(dx^4 + \cos\theta d\phi)^2
+{ J \wt\lambda^2\over 8 \wt r_0 \wt R \wt V} dy (dx^4 + \cos\theta d\phi)
- 2\sqrt{\wt r_0} \rho dy d\tau 
 \nonumber \\ &&  +\wt r_0  \left(
 d\theta^2 + \sin^2\theta d\phi^2\right)
+ \wh g_{mn} du^m du^n\, , \qquad y\equiv y+2\pi \wt R \nonumber \\
e^{\Phi} &=& \wt\lambda\, , \nonumber \\
\wt H^{(3)} &=& {\wt r_0}\, \left[\eps_3 + *\eps_3
+{ J \wt\lambda^2\over 16\, \wt r_0^2 \wt R  \wt V}
\, dy\wedge \left({1\over \rho}\, d\rho\wedge
(d \, x^4 +\cos\theta\, d\phi)
+ \sin\theta\, d\theta\wedge d\phi\right)\right]\, , \nonumber \\
\een
where
$\wt H^{(3)}$ is the NS-NS 3-form field strength.
The  near horizon parameters are now given
in terms of the charges and the parameter $\wt\lambda$ via the
relations
\begin{equation} \label{ep5iia}
\wt r_0\ = { Q_5  \over 4
}, \quad   \wt V=\wt\lambda^{2}\,
{n\over Q_5}, \quad \wt R  = \sqrt { Q_1\over n}
\, .
\end{equation}
With the help of the same coordinate transformation 
\refb{ecchange} we can remove the 
explicit $J$ dependence of the solution
except for in the periodic identification of the new coordinates.
The space-time spanned by the coordinates $(\rho,\tau,y,\theta,\phi,x^4)$
is now locally $AdS_3\times S^3$. 
If we take the limit in which the asymptotic radius $\wt R_{as}$ of
$\wt S^1$ goes to infinity keeping fixed the quantized charges
and the six dimensional background, then the solution develops an
$AdS_3\times S^3$ factor in the intermediate region, and
the
near horizon geometry described in \refb{ep9iia} can be
regarded as that of an
extremal charged BTZ black hole embedded in this asymptotically
$AdS_3\times S^3$ geometry.
In the holographically dual CFT$_2$ the BTZ black hole can now be
regarded as an RR sector 
state with $L_0=Q_1$, $\bar L_0=0$ and $U(1)_L\subset
SU(2)_L$ charge $J/2$. Thus 
the entropy of the black hole in the
limit of large $Q_1$ should be given by the Cardy formula
\be \label{ep6rothetear}
S_{BH}\simeq
2\pi \sqrt{\tilde c^{bulk}_L \left(Q_1 - {1\over 4} (\tilde
k^{bulk}_L)^{-1} J^2\right)/6}
\, ,
\ee
where now $\tilde c_L^{bulk}$, $\tilde c_R^{bulk}$, $\tilde k_L^{bulk}$,
and $\tilde k_R^{bulk}$  denote respectively the central charges of
the left and right-moving Virasoro algebras and the levels of
$SU(2)_L$ and $SU(2)_R$ current algebras in the CFT$_2$.
As before, $\tilde c_{grav}^{bulk}\equiv \tilde c_L^{bulk}
- \tilde c_R^{bulk}$ is related to the coefficient
of the Lorentz Chern-Simons term in the bulk and $\tilde k_L^{bulk}$
and $\tilde k_R^{bulk}$ are related to the coefficients of the
$SU(2)_L$ and $SU(2)_R$ Chern-Simons terms. Furhermore 
using the supersymmetries of the bulk theory one can
show that the
CFT$_2$ has $(0,4)$ supersymmetry. This leads to the relation
$\tilde c_R^{bulk} = 6 \, \tilde k_R^{bulk}$ and gives
\be \label{ezfab}
\tilde c_L^{bulk} = \tilde c_{grav}^{bulk}+ 6 \, \tilde k_R^{bulk}\, .
\ee
Comparison with \refb{ezeroent} shows that in the supergravity
approximation we have $\tilde c_L^{bulk}=6n Q_5$ and 
$\tilde k_L^{bulk}=n Q_5$. Since in this approximation there
is no Lorentz Chern-Simons term in the action, 
$\tilde c_{grav}^{bulk}$ vanishes and \refb{ezfab}
gives $\tilde k_R^{bulk} = n Q_5$. 
Direct computation of
$\tilde k_L^{bulk}$ and $\tilde k_R^{bulk}$ 
can be performed using the procedure reviewed in
appendix \ref{sd} and agrees with the values given above. 
Our goal now is to compute the corrections
to $\tilde c_{grav}^{bulk}$, $\tilde k_L^{bulk}$ and $\tilde k_R^{bulk}$
due to higher derivative and string loop corrections.

Since $\tilde c_{grav}^{bulk}$, $\tilde k_L^{bulk}$ and $\tilde k_R^{bulk}$
are all quantized, corrections to them
could involve terms linear in
$Q_5$ and/or $n$ and constant term. 
Now since $n$ represents an NSNS sector electric
charge and $Q_5$ an NSNS sector magnetic charge, the
scaling relation \refb{escaling} takes the form
\be \label{eiiascaling}
\tilde c^{(l)}(\lambda^2 n, Q_5) = \lambda^{2-2l}
\tilde c^{(l)}(n, Q_5)\, ,
\ee
where $\tilde c^{(l)}$ stands for $l$ loop contribution
to any of the quantities
$\tilde c_{grav}^{bulk}$, $\tilde k_L^{bulk}$ and $\tilde k_R^{bulk}$.
This shows that
a term linear in $n$, if present,
must arise at
string tree level.
Since this term would be linear in $n$, representing the NS-NS 3-form
flux $\wt H^{(3)}$ through $AdS_3$, it
will have to arise from a six dimensional Chern-Simons term
of the form $\int ~^D\wt H^{(3)}\wedge \Omega_{CS}$ where
$\Omega_{CS}$ is a Lorentz Chern-Simons 3-form in six dimensions, 
and $~^D$
denotes the dual field strength obtained by taking the Hodge dual
of the flux $\delta S/\delta \wt H^{(3)}$ \cite{0608182}.
But tree level
type
IIA string theory does not have such a term in the action since the
gauge invariant three form field strength in
type II string theories
do not involve a Lorentz Chern-Simons term.
This shows that there are no corrections linear in $n$.
According to the
scaling relation \refb{eiiascaling} the constant
term, if present, must arise at one loop. Since
it does not involve any charges, it will have
to either come from a purely gravitational term in ten
dimensions which upon
dimensional reduction on $K3\times S^3$ will produce a Lorentz
Chern-Simons term in $AdS_3$, or arise as a one loop effect
in the theory after compactification on $S^3$. Since 
there are no purely gravitational Chern-Simons terms
in ten or six dimensions, we can rule out the first
possibility. But as in the case of type IIB Cardy limit,
we cannot rule out
the second possibility. Let us denote such contributions to
$\tilde c_{grav}^{bulk}$, $\tilde k_L^{bulk}$ and $\tilde k_R^{bulk}$, if
present, by $\wt A$, $\wt B$ and $\wt C$ respectively. 

Finally a term linear in $Q_5$, if present, must arise
at one loop order, and come from a
term proportional to $\int \wt H^{(3)}\wedge \Omega_{CS}$ in
six dimensions.
Are there such one loop corrections to the Chern-Simons term?
The ten dimensional
type IIA string theory indeed contains a one loop 
Chern-Simons term
of the form 
\begin{equation} \label{ei8}
- \frac{1}{2\pi}\int \wt B\wedge I_8(X),
\end{equation}
where $\wt B$ is the NS-NS 2-form
field and 
$I_8(X)=\frac{1}{48}\left(p_2(X)-\frac{p_1^2(X)}{4}\right)$, 
$X$ being the ten dimensional space and $p_n$ denoting the
$n$th Pontryagin class \cite{9505053}.
Upon dimensional reduction on $K3$ this generates a
term proportional to
$\int \wt H^{(3)}\wedge \Omega_{CS}$.
Thus $\tilde c_{grav}^{bulk}$, $\tilde k_{L}^{bulk}$
and $\tilde k_L^{bulk}$ 
can all receive corrections linear in $Q_5$.
To compute the coefficients of these terms
we introduce the quantities $I_7^0$ and $p_1^0$ via the
relations $I_8=dI_7^0$ and $p_1=dp_1^0$.
Since $\wt H^{(3)}$ has nontrivial flux over $S^3$, 
the 2-form field $\wt B$ is not well defined. 
Thus instead of taking the coupling \refb{ei8}
we shall take 
\begin{equation} \label{eintpart}
 \frac{1}{2\pi} \int \wt H^{(3)}\wedge I_7^0
\end{equation}
by integration by parts. 
Now the spin 
connection in the Kaluza-Klein reduction is 
simply a direct sum of the connections on $AdS_3\times S^3\times K3$. 
Using the fact that the total pontryagin class of a direct 
sum satisfies $p(E\oplus F)=p(E)p(F)$, that
$\int_{K3} p_1 = 48$, and that 
$p_1 =  -d\omega_v(\Gamma)/8\pi^2$ where
\be \label{ecsx2pre}
\omega_v(\Gamma) = Tr_v\left( \Gamma\wedge d\Gamma +
{2\over 3}\Gamma\wedge \Gamma\wedge \Gamma\right)\, ,
\ee
the trace being taken over the vector representation,
we can express the
contribution from \refb{eintpart} as
\begin{equation} \label{ekm1}
-{1\over 32\pi^3}\int_{AdS_3\times S^3} 
H^{(3)}\wedge \omega_v(\Gamma)\, ,
\ee
where $\Gamma$ now stands for 
the spin connection on $AdS_3\times S^3$.
Using eqs.\refb{ecsx1}, \refb{eonelooppre} 
we see that the effect of \refb{ekm1} is to generate the
following corrections to 
$\tilde c_{grav}^{bulk}$, $\tilde k_R^{bulk}$ and 
$\tilde k_L^{bulk}$:
\be \label{eoneloop}
\Delta \tilde c_{grav}^{bulk} = 12 Q_5, \qquad \Delta 
\tilde k_R^{bulk} = Q_5,
\qquad \Delta \tilde k_L^{bulk} = - Q_5\, .
\ee
We can check the consistency of the overall 
sign and normalization by
setting $Q_5=1$; in this case the system is equivalent
to a fundamental heterotic string which has $c_{grav}=12$.
Combining \refb{eoneloop} 
with the leading supergravity results and the
constant shifts we arrive at the relations:
\ben \label{ecomb}
&& \tilde k_R^{bulk} = Q_5 (n+1) + \tilde C, \quad 
\tilde k_L^{bulk} = Q_5 (n - 1) + \tilde B, \quad
\tilde c_{grav}^{bulk} = 12 \, Q_5 +\tilde A, \nonumber \\
&& \tilde c_L^{bulk} = \tilde c_{grav}^{bulk} + 6 \tilde k_R^{bulk}
= 6 \, Q_5 (n+3) + \tilde A + 6\tilde C\, .
\een

We now need to use \refb{edeft9}, \refb{edeft10} 
to find the asymptotic formula
for the index. 
Again we shall see 
in \S\ref{shair} that the net effect of the exterior contribution
$\tilde c^{exterior}_{L,eff}$ and $\tilde k^{exterior}_{L,eff}$ 
is to cancel the terms proportional to $\wt A+6\wt C$ and $\wt B$
in $\tilde c_L^{bulk}$ and $\tilde k_L^{bulk}$.
Thus the growth of the
macroscopic index $d_{macro}$
in the type IIA Cardy limit $Q_1\to\infty$ for fixed $Q_5$, $n$
will be controlled by the constants
\be \label{ecombnew}
\tilde k_{L,eff}^{macro} = Q_5 (n - 1), \quad
\tilde c_{L,eff}^{macro} 
= 6 \, Q_5 (n+3) \, ,
\ee
and $\ln d_{macro}$
given by
\be \label{efinent}
\ln d_{macro}(n,Q_1,Q_5) \simeq 2\pi \sqrt{Q_5 (n+3) \left( Q_1
- {J^2\over 4Q_5(n-1)}\right)}\, ,
\ee
where $\simeq$ implies equality up to corrections suppressed
by powers of $Q_1$.
This agrees with the result found in
 \cite{0703087,0807.0237} for small $J$ and large $n$ 
computed using a
particular four
derivative correction to the five dimensional effective action.
Also the result for
$\tilde c_{L,eff}^{macro}$ agrees with the one computed in
 \cite{0912.0030,1001.1452} 
 (see also \cite{0710.3886,0809.4954}) assuming a specific structure of 
all the higher derivative
correction to the effective action.\footnote{Earlier results 
on this can be found in
 \cite{9812027}.}
Most importantly \refb{efinent}
agrees with the microscopic answer
\refb{ehetgen} 
which will be derived in
\S\ref{smicro}.

\subsection{Entropy of some four dimensional black holes}
\label{sfour}

We now consider a four dimensional theory obtained by
compactifying type IIB string theory on $K3\times S^1\times \wt S^1$.
In this theory we
take the non-spinning D1-D5-p system analyzed in \S\ref{szero}
and
place it in the background of $K$ Kaluza-Klein
(KK) monopoles associated
with the circle $\wt S^1$.
Since for $K=1$
this system has the same near horizon geometry as the five dimensional
D1-D5-p system analyzed in \S\ref{szero}, the macroscopic computation of the
index is identical to that in \S\ref{szero}
except for the difference in the contribution 
due to the exterior modes.
We shall however keep $K$ arbitrary and compute
the entropy in a different duality frame in which we regard them as
black holes in M-theory on $K3\times T^3$ carrying 
M5-brane charges and
internal momentum.
For this we first make a mirror symmetry transformation
in $K3$ to take the D1-D5 system to a D3-D3 system with
$Q_1$ D3-branes wrapped on $C_2\times S^1$  and
$Q_5$ D3-branes wrapped on a $\wt C_2\times S^1$
where $C_2$ and $\wt C_2$ are
a pair of dual 2-cycles of $K3$. We then
make a T-duality along the circle $\wt S^1$ to take the D3-branes
to D4-branes and the KK monopoles to NS 5-branes wrapped
on $K3\times S^1$. If we denote by $\wh S^1$ the T-dual
circle then we have
$Q_1$ D4-branes along $C_2\times S^1 \times \wh S^1$,
$Q_5$ D4-branes wrapped along $\wt C_2\times S^1 \times \wh S^1$,
and $K$ NS 5-branes along $K3\times S^1$, carrying $n$ units of
momentum along $S^1$. We can now regard the type IIA string
theory as M-theory compactified on a new circle $S^1_M$, so
that we have M-theory on $K3\times
S^1\times \wh S^1\times S^1_M$. The dyon configuration now
corresponds to $Q_1$ M5-branes along $C_2\times S^1 \times
\wh S^1\times S^1_M$,
$Q_5$ M5-branes wrapped along $\wt
C_2\times S^1 \times \wh S^1
\times S^1_M$,
and $K$ M5-branes wrapped along $K3\times S^1$,
carrying $n$ units of
momentum along $S^1$.

Our goal in this section
will be to analyze the black hole solution
corresponding to these charges and find the macroscopic
entropy of this system in the limit $n\to\infty$,
keeping the other charges
fixed. 
Since the analysis proceeds more or less in the same way as
for five dimensional black holes, our discussion will be brief.
As in the case of the D1-D5-p system one finds
that
near the horizon the $AdS_2\times S^2$ appearing in the
near horizon geometry of the black hole
combines with the circle $S^1$ to produce a locally $AdS_3\times S^2$
factor \cite{0506176}.
Furthermore if we take the limit in which the asymptotic radius
of $S^1$ approaches 
infinity, {\it keeping fixed all other quantized charges
and the five dimensional geometry in the M-theory frame} then the
M-theory background develops an intermediate $AdS_3\times S^2$
geometry, and the near horizon geometry of the black hole
appears as the near horizon geometry of an extremal BTZ black hole
embedded in this asymptotically $AdS_3\times S^2$ space. 
Thus applying the Cardy formula we see that
the entropy is given by the formula
\be \label{ement}
S_{BH} \simeq 2\pi \sqrt{c^{bulk}_L n/6}\, ,
\ee
where $c^{bulk}_L$ is the central charge of the left-moving Virasoro
algebra of the holographically dual CFT$_2$. In the supergravity
approximation $c^{bulk}_L= 6 Q_1 Q_5 K$, reproducing the 
Bekenstein-Hawking result
$2\pi\sqrt{Q_1Q_5 K n}$ for the entropy\cite{9507090,9512031}.

In the limit $n\to\infty$ with $Q_1$, $Q_5$, $K$ fixed,
the complete contribution to the entropy
(and the index) from the bulk modes on $AdS_3$ continues to be
given by \refb{ement} provided $c_L^{bulk}$ represents the exact
central charge of the left-moving Virasoro algebra
after
taking into account higher derivative and quantum corrections.
As usual $(c^{bulk}_L-
c^{bulk}_R)$ is given by the
coefficient $c_{grav}^{bulk}$
of the Lorentz Chern-Simons term in $AdS_3$. On the
other hand 
using the supersymmetries of the bulk geometry one can
show that the dual CFT$_2$ on the boundary has (0,4)
superconformal symmetry
acting on the right-movers.
As a result
$c^{bulk}_R$ can be
related to the level $k_R^{bulk}$ 
of the $SU(2)$ R-symmetry current in the
CFT$_2$ via the relation 
$c_R^{bulk}=6 k_R^{bulk}$.\footnote{Although there is now a
single $SU(2)$ we shall label its anomaly
coefficient by $k_R$.}
Since this $SU(2)$ current in the boundary theory is
holographically dual to the $SU(2)$ gauge fields in the bulk
arising from dimensional reduction on $S^2$, $k_R^{bulk}$ is
given by the coefficient of the $SU(2)$ Chern-Simons term in the
bulk. This allows us to determine $c_L^{bulk}$ in terms of the
coefficients of the Chern-Simons terms in $AdS_3$ via the relations
\be \label{eclb}
c_L^{bulk} = c_{grav}^{bulk} + 6 \, k_R^{bulk}\, .
\ee

The relevant Chern-Simons terms were evaluated
in  \cite{0506176} for
M-theory compactified on $M\times S^1$ where $M$ is a
general Calabi-Yau 3-fold. In this theory, consider a black hole
corresponding to M5-brane wrapped on $P\times S^1$ where $P$
is some general 4-cycle in $M$. Using the isomorphism
between 4-cycles and 2-forms we can associate
with $P$ a 2-form on $M$ which we shall also denote by $P$.
Then the result of  \cite{0506176} for $c^{bulk}_L$ and
$c^{bulk}_R$ are:
\be \label{emiccenpre}
c^{bulk}_R = \int_M \left(P\wedge P\wedge P + {1\over 2}
P\wedge c_2(M)\right)+\bar A_R, \qquad
c^{bulk}_L = \int_M (P\wedge P\wedge P + P\wedge c_2(M))
+\bar A_L
\, ,
\ee
where $c_2(M)$ is the second Chern class of $M$. Note that we have
allowed for constant shift $(\bar A_L,\bar A_R)$
in the central charges due to one loop effects arising after
compactification of $M$-theory on $K3\times T^2\times S^2\times 
AdS_3$. Computation in   \cite{0506176} was carried out by integrating the
quantum corrected ten dimensional Lagrangian density on
$K3\times S^3$, and ignored possible quantum corrections
which could arise after compactification on $K3\times S^3$.
Evaluating this
for the configuration we have, we get
\be \label{emiccen}
c^{bulk}_R =6 K(Q_1 Q_5 + 2)+\bar A_R, \qquad
c^{bulk}_L
=6 K (Q_1 Q_5 + 4)+\bar A_L\, .
\ee

Again we shall see in \S\ref{shair} that when we compute the
full index in the macroscopic theory  using 
\refb{edeft5}, \refb{ebkgrow}, 
the net effect of the exterior contribution
$c_{L,eff}^{exterior}$ is to cancel the
$\bar A_L$ term in $c_L^{bulk}$, giving rise to
\be \label{emiccentot}
c_{L,eff}^{macro}
=6 K (Q_1 Q_5 + 4)\, .
\ee
Eq.\refb{ebkgrow} now shows that the index computed in the
macroscopic theory grows as
\be \label{efbh}
\ln d_{macro}(n,Q_1, Q_5, K) \simeq 2\pi \sqrt{K(Q_1 Q_5 + 4)n}
\quad \hbox{for large $n$}\, .
\ee
This is in perfect agreement with the microscopic result
\refb{e4dgengen}
to be derived in \S\ref{smicro}.

\subsection{Black holes in
toroidally compactified type IIB string theory}
\label{storusmacro}

In this subsection we shall repeat the analysis of the previous
subsections for black holes in
toroidally compactified type IIB string theory. Since the analysis
proceeds in a more or less identical manner we shall mainly
state the results without going through the details of the analysis.

First we consider the D1-D5-p system wrapped on $T^4\times S^1$.
We shall use the same notation for the charges as in the case of
$K3\times S^1$ compactification, except that now $Q_1$ represents
the actual number of D1-branes since D5-branes wrapped on $T^4$
do not carry any D1-brane charge. In the  limit when $Q_1$, $Q_5$
are fixed and $n$ becomes large, we get the result:
\be \label{etx1}
\ln d_{macro}(n,Q_1, Q_5, J) \simeq \pi\sqrt{4 Q_1 Q_5 n- J^2}\, .
\ee
In the limit of fixed $n$, $Q_5$ and
$Q_1$ large, we have
\be \label{etx2}
\ln d_{macro}(n,Q_1, Q_5, J) \simeq \pi\sqrt{4 Q_1 Q_5 n -J^2}\, .
\ee
Derivation of \refb{etx1} is a straightforward generalization of the
similar analysis for type IIB on $K3\times S^1$ leading to
\refb{dhor1}.
The main difference between the analysis leading to
\refb{etx2} and that leading to \refb{efinent}
is that the dimensional reduction of the $\int \wt B\wedge I_8$ term
on $T^4$ does not produce any Chern-Simons term.
Thus all corrections to $\tilde c^{bulk}_L$ and $\tilde k_L^{bulk}$
from the supergravity
results, except for possible constant shifts from one
loop corrections, vanish.
The constant shift is cancelled by the contribution from the exterior
modes due to the
results of \S\ref{shair}. 
Using these results we arrive at \refb{etx2}.
This is in perfect agreement with the microscopic result
\refb{eap4} to be derived in \S\ref{smicro}.

If we now consider a four dimensional black hole obtained by
placing this system
in the background of $K$ KK monopoles,
and go to the duality frame in which the system is described by
momentum carrying M5-brane
wrapped on $T^7$, then we can analyze the
macroscopic entropy of the system following the same procedure
as in \S\ref{sfour}.
In this case the near horizon
geometry is locally $T^6\times AdS_3\times S^2$. The central charges
$c^{bulk}_L$ and
$c^{bulk}_R$ associated with this $AdS_3$ are given by
formul\ae\ similar to those given in \refb{emiccen} except that
now $\int P\wedge c_2(M)$ vanishes. 
Possible constant shift in $c^{bulk}_L$ due to one loop correction
is exactly cancelled by the hair contribution. This gives
\be \label{etx3}
\ln d_{macro}(n,Q_1, Q_5, K) \simeq 2\pi\sqrt{Q_1Q_5 K n}
\quad \hbox{for large $n$}\, .
\ee
This is in complete agreement with the macroscopic result
\refb{emicpred}.

\section{Analysis of the Exterior Contribution} \label{shair}

In this section we shall 
compute the coefficients $c_{L,eff}^{exterior}$ and
$k_{L,eff}^{exterior}$ appearing in \refb{edeft4} and
\refb{edeft8} and show that their effect is to cancel the charge
independent constant terms in the expressions for
$c_{L,eff}^{macro}$ and $k_{L,eff}^{macro}$ 
which arise from one loop quantum corrections and
{\it which cannot be
obtained as the dimensional reduction of the 1PI action in ten
dimensions on the intermediate $AdS_3$ geometry.}
Examples of such terms are $A+6C$ and $B$ in \refb{eexactchor}.
We shall describe our analysis in the context of the 
five dimensional black hole, but it will be clear that
the result we derive is also valid in four dimensions, the only
difference being the absence of any reference to the $SU(2)_L$
symmetry and the associated anomaly coefficient $k_L$ in four
dimensions.

We begin by recollecting some relevant 
results from \S\ref{smacro}. Recall that 
$c_L^{bulk}$ is computed in \S\ref{smacro} via the relation
\be \label{ed2}
c_L^{bulk}=c_{grav}^{bulk} + 6 k_R^{bulk}\, ,
\ee
where $k_R^{bulk}$ and $c_{grav}^{bulk}$
are the coefficeints of the $SU(2)_R$ and Lorentz Chern-Simons terms
in the intermediate $AdS_3$ geomery. 
On the other hand $k_L^{bulk}$ was given by the coefficient of the
$SU(2)_L$ Chern-Simons term in the $AdS_3$ geometry.
Part of the contribution to these Chern-Simons terms came from
integrating ten dimensional
Chern-Simons terms on $K3\times S^3$, 
but this left open the possibility of constant one loop
corrections to these coefficients
which arise after compactification on $S^3$.
Now imagine that instead of doing this reduction on the
$K3\times S^3$ that arises in the intermediate $AdS_3$ region,
we do this in the asymptotic region where the geometry is
locally $K3\times R^6$.\footnote{Recall that we have taken the
asymptotic radius of $S^1$ to infinity so that we have a (5+1)
dimensional asymptotic space-time.}
Let us take a thick spherical shell of large radius
around the origin, 
bounded by the hypersurfaces $r=r_1$ and
$r=r_2$ for large $r_1$, $r_2$, and regard
this space as locally 
$R^3 \times K3\times S^3$, with
$S^3$ labelling the angular coordinates and $R^3$ containing the
time coordinate, the radial coordinate $r$ and the coordinate along $S^1$.
We can now formally dimensionally reduce the ten dimensional action
on $K3\times S^3$ to calculate the coefficients of the Lorentz
and $SU(2)_R\times SU(2)_L$ Chern-Simons terms on $R^3$. 
The calculation is identical to the one described in
appendix \ref{sd} for the 
intermediate $AdS_3$ geometry,
except that
this time we do not
expect any additional one loop correction due to compactification
on $S^3$ since we are really
doing the computation in $K3\times R^6$ rather than on
$K3\times S^3\times AdS_3$. 
Thus the result for these coefficients will be identical to
$c_{grav}^{bulk}$, $k_R^{bulk}=c_R^{bulk}/6$ 
and $k_L^{bulk}$ computed in
\S\ref{smacro} and appendix \ref{sd}
except
for the constant one loop shifts. We shall denote these coefficients
by $c_{grav}^{asymp}$, $k_R^{asymp}$ and $k_L^{asymp}$ respectively. 
For completeness we shall list below the values of 
$c^{asymp}_{grav}$,
$k^{asymp}_R$ and $k_L^{asymp}$ 
for each of the systems analyzed in
\S\ref{smacro}: 
\begin{enumerate}
\item D1-D5-p system in type IIB on $K3\times S^1$
in the type IIB Cardy limit:
\be \label{exp1}
c^{asymp}_{grav} = 0, \qquad k^{asymp}_R = Q_1 Q_5,
\qquad k^{asymp}_L = Q_1 Q_5\,.
\ee
\item D1-D5-p system  in type IIB on $K3\times S^1$
in the type IIA Cardy limit:
\be \label{exp2}
c^{asymp}_{grav} = 12Q_5, \qquad 
k^{asymp}_R = Q_5(n+1), \qquad k^{asymp}_L = Q_5(n-1)\,.
\ee
\item Four dimensional black hole in M-theory on $K3\times T^2
\times S^1$:
\be \label{exp3}
c^{asymp}_{grav} = 12K, \qquad 
k^{asymp}_R = K(Q_1Q_5+2)\,.
\ee
\item D1-D5-p system in type IIB on $T^4\times S^1$
in the type IIB Cardy limit:
\be \label{exp4}
c^{asymp}_{grav} = 0, \qquad k^{asymp}_R = Q_1 Q_5,
\qquad k^{asymp}_L = Q_1 Q_5\,.
\ee
\item D1-D5-p system in type IIB on $T^4\times S^1$
in the type IIA Cardy limit:
\be \label{exp5}
c^{asymp}_{grav} = 0, \qquad k^{asymp}_R = Q_1 Q_5, 
\qquad k^{asymp}_R = Q_1 Q_5\,.
\ee
\item Four dimensional black hole in M-theory on $T^6\times S^1$:
\be \label{exp6}
c^{asymp}_{grav} = 0, \qquad 
k^{asymp}_R = KQ_1Q_5\,.
\ee
\end{enumerate}

We shall now try to express the difference between the
Chern-Simons coefficients
calculated in the asymptotic geometry and the intermediate
$AdS_3$
geometry in terms of some known quantities and in the process
gain knowledge about the constant terms in the expression for the
Chern-Simons coefficients in the intermediate $AdS_3$ region.
For this we note that the coefficients of the Chern-Simons terms
can also  be
interpreted as certain anomaly coefficients. For example $k_R^{bulk}$
and $k_L^{bulk}$
reflect the change in the effective action in the bulk theory
by certain boundary terms in the intermediate
$AdS_3$ geometry
under $SU(2)_R$ and $SU(2)_L$ gauge transformations, and 
$c_{grav}^{bulk}$
reflects a similar change under local Lorentz transformations. 
$k_R^{asymp}$, $k_L^{asymp}$ and $c_{grav}^{asymp}$ reflect similar
anomalies under local $SU(2)_R$, $SU(2)_L$ and 
Lorentz transformations
in the asymptotic region.
Thus the difference between $k_R^{asymp}$ and $k_R^{bulk}$
 must be accounted for by the contribution to the $SU(2)_R$
anomaly due to the exterior
degrees of freedom sitting between the asymptotic
observer and the $AdS_3$. We shall denote this by 
$k_R^{exterior}$.
An identical argument holds for $k_L$ and $c_{grav}$.
Thus we have
\be \label{ed3}
k_R^{asymp} = k_R^{bulk}+k_R^{exterior}, \quad
k_L^{asymp} = k_L^{bulk}+k_L^{exterior}, \quad
c_{grav}^{asymp}  =  c_{grav}^{bulk} 
+c_{grav}^{exterior} \, .
\ee
Using \refb{edeft9}, \refb{ed2} and \refb{ed3} we get
\ben \label{ed4}
&& c_{L,eff}^{macro}  = c_{grav}^{asymp}  - c_{grav}^{exterior}
+ 6 (k_R^{asymp}-k_R^{exterior}) 
+ c_{L,eff}^{exterior} =  c_{grav}^{asymp} + 6 k_R^{asymp} 
+\Delta\, ,\nonumber \\
&& k_{L,eff}^{macro} = k_L^{asymp} + \delta\, ,
\een
where
\be \label{edefDelta}
\Delta \equiv - 6 k_R^{exterior}
- c_{grav}^{exterior} + c_{L,eff}^{exterior}= 
- 6 k_R^{exterior}
- (c_{L}^{exterior}-c_R^{exterior}) + c_{L,eff}^{exterior}\, ,
\ee
\be \label{edefdelta}
\delta = k_{L,eff}^{exterior} - k_L^{exterior}\, .
\ee
Now we have already argued that the results for $c_{grav}^{asymp}$,
$k_R^{asymp}$ and $k_L^{asymp}$ are identical to those of
$c_{grav}^{bulk}$,
$k_R^{bulk}$ and $k_L^{bulk}$ in \S\ref{smacro} except for the
constant one loop shifts. This if we can show that $\Delta$ and
$\delta$ vanish, then we would prove that the effect of the exterior
contributions 
is to precisely cancel these constant shifts in the $AdS_3$
central charges.

We shall now show that 
$\Delta$ and $\delta $ vanish.  For this we shall need to make
some assumptions on the structure of the exterior modes.
We make the following assumptions:
\begin{enumerate}
\item The exterior modes consist of free massless 
scalars and fermions
belonging to singlet and/or spinors representations of
$SU(2)_L$ and $SU(2)_R$.
\item The scalar modes which transform in the vector (2,2)
representation of the transverse rotation group $SO(4)=
SU(2)_L\times SU(2)_R$ are non-chiral. Physically this assumption
stems from the fact that such modes arise from the  oscillations of the
center of mass mode of the black string which is non-chiral.
Due to this assumption the contribution to the $SU(2)_R$ and
$SU(2)_L$ anomalies from any scalar in the $(2_L,2_R)$ 
representation
of $SU(2)_L\times SU(2)_R$ always vanishes. Taking advantage
of this fact we can assign the contribution to 
$(k_L,k_R)$  from a
left-moving $(2_L,2_R)$ scalar to be $(a,b)$ and a right-moving
$(2_L,2_R)$ scalar to be $(-a,-b)$ for any arbitrary pair of
numbers
$(a,b)$.  We shall choose $(a,b)=(-1,-1)$ for convenience.
\end{enumerate}
To this we shall add the information that 
the (1+1) dimensional conformal
field theory of exterior modes is
invariant under (0,4) supersymmetry. This follows from the
supersymmetry of the solution outside the $AdS_3$ region.
{\it We shall not make the assumption that the $SU(2)$ R-symmetry
current of this superconformal algebra has any relation to the
spatial rotation group $SU(2)_R$. Thus we shall not have any
relation between $c_R^{exterior}$ and $k_R^{exterior}$.}

We shall now separately 
evaluate the contribution to
$\Delta$ and $\delta$ 
from each type of field that could appear as part
of the exterior degrees of freedom. 
For this we need to calculate $k_L$, $k_R$, $c_L-c_R$,
$c_{L,eff}$ and $k_{L,eff}$ from each field. 
This is done with the help of the following 
observations:
\begin{enumerate}
\item
The
calculation of $(k_R, k_L, c_L-c_R)$ is straighforward since
these are given by the contribution to $SU(2)_L$, $SU(2)_R$
and gravitational anomalies. 
\item
The calculation of
$c_{L,eff}$ and $k_{L,eff}$ involves computing the contribution
from these fields to the index $\wt C_{2k}^{exterior}
\equiv Tr(-1)^{2J_R} (2J_R)^2
e^{2\pi i p\tau + 4\pi i J_L z}$. To this end 
we note that the factor of $(2J_R)^2$ is needed to soak up the
$SU(2)_R$ doublet
fermion zero modes. Thus after taking the trace over the
fermion zero modes we are left with
$Tr(-1)^{2J_R}e^{2\pi i p\tau + 4\pi i J_L z}$ 
from the oscillator modes.
Due to supersymmetry
this receives contribution {\it only from the left-moving modes}.
\item
Since $(-1)^{2J_R}=1$ for the $SU(2)_R$ singlet fields,
the $SU(2)_R$ singlet left-moving 
fields contribute in the
same way to the index and the degeneracy. Thus
for them $c_{L,eff}=c_L$, 
and $k_{L,eff}=k_L$.
\item 
 $SU(2)_R$ doublet left-moving
fields have the property
that the contribution to $\wt C_{2k}^{exterior}$
from a left-moving scalar oscillator,
given by $\left(1 - e^{2\pi i p_{osc}
\tau + 4\pi i J_{L,osc} z}\right)^{-1}$,
can be regarded as the
inverse of the contribution to the partition function
from a left-moving fermionic oscillator, and
the contribution to 
$\wt C_{2k}^{exterior}$ from
a left-moving fermionic oscillator, given by
$\left(1 - e^{2\pi i p_{osc}\tau + 4\pi i J_{L,osc} z}\right)$,
can be regarded as the
inverse of the contribution to the partition function from
a left-moving bosonic oscillator.
Thus their contribution to $c_{L,eff}$ and $k_{L,eff}$ can
be computed by replacing the fermions by bosons and vice
versa, and including an extra $-$ sign in front of the corresponding
values of $c_L$ and $k_L$.
\end{enumerate} 
This gives the following contribution to $\Delta$ and $\delta$
from various
fields:
\ben \label{elist}
&& \hbox{left-moving $(1_L,1_R)$ scalar:}
\nonumber \\
&& \qquad 
k_R = 0, \quad k_L=0, \quad c_R=0, \quad c_L = 1,
 \quad c_{L,eff} = 1, \quad k_{L,eff}= 0, \quad \Delta = 0,
 \quad \delta = 0\, ,
\nonumber \\
&& \hbox{left-moving $(2_L,2_R)$ scalar:}
\nonumber \\
&& \qquad 
k_R = -1, \quad k_L=-1,
\quad c_R=0, \quad c_L = 4,
 \quad c_{L,eff} = -2, \quad k_{L,eff}= -1, \quad \Delta = 0, \quad
 \delta = 0\, ,
\nonumber \\
&& \hbox{left-moving $(2_L,1_R)$ fermion:}
\nonumber \\
&& \qquad
k_R = 0, \quad k_L={1\over 2}, 
\quad c_R=0, \quad c_L = {1},
 \quad c_{L,eff} = {1}, \quad k_{L,eff}={1\over 2},
 \quad \Delta = 0, \quad \delta = 0\, ,
\nonumber \\
&& \hbox{left-moving $(1_L,2_R)$ fermion:}
\nonumber \\
&& \qquad 
k_R = -{1\over 2}, \quad k_L=0, \quad 
\quad c_R=0, \quad c_L = 1,
 \quad c_{L,eff} = -2, \quad k_{L,eff}= 0, \quad \Delta = 0,
 \quad \delta =0\, ,
\nonumber \\
&& \hbox{right-moving $(1_L,1_R)$ scalar:}
\nonumber \\
&& \qquad 
k_R = 0, \quad k_L=0, \quad c_R=1, \quad c_L = 0,
 \quad c_{L,eff} = 0, \quad k_{L,eff}=0, \quad \Delta = 1,
 \quad \delta=0\, ,
\nonumber \\
&& \hbox{right-moving $(2_L,2_R)$ scalar:}
\nonumber \\
&& \qquad 
k_R = {1}, \quad k_L = 1, \quad 
\quad c_R=4, \quad c_L = 0,
 \quad c_{L,eff} = 0, \quad k_{L,eff} = 0, \quad \Delta = -2, \quad
 \delta = -1\, ,
\nonumber \\
&& \hbox{right-moving $(2_L,1_R)$ fermion:}
\nonumber \\
&& \qquad 
k_R = 0, \quad k_L = -{1\over 2}, \quad c_R={1}, 
\quad c_L = {0},
 \quad c_{L,eff} = 0, \quad k_{L,eff} = 0,
 \quad \Delta = {1}, \quad \delta ={1\over 2}\, ,
\nonumber \\
&& \hbox{right-moving $(1_L,2_R)$ fermion:}
\nonumber \\
&& \qquad 
k_R = {1\over 2}, \quad k_L=0, \quad 
\quad c_R=1, \quad c_L = 0,
 \quad c_{L,eff} = 0, \quad k_{L,eff}=0, \quad \Delta = -2,
 \quad \delta =0\, .
 \nonumber \\
\een
Note that in evaluating the contribution to $k_L$ and $k_R$
from the $(2_L,2_R)$ scalars we have exploited the freedom of
choice mentioned earlier.
{}From this table we see that the left-moving exterior modes do not
contribute to $\Delta$ or $\delta$. 
On the other hand since we have supersymmetry
acting on the right-movers, and since the supersymmetry generators
are doublets of $SU(2)_R$, a right-moving $SU(2)_R$ doublet
scalar must be accompanied by a pair  of $SU(2)_R$ singlet
fermions and a right-moving $SU(2)_R$ doublet fermion must be
accompanied by a pair of $SU(2)_R$ single 
scalars.\footnote{We emphasize that that this does not
imply that $SU(2)_R$ is the zero mode part of the right-moving
R-symmetry current. As already remarked, the latter acts
trivially on all the left-moving fields while the former
has non-trivial action on some left-movers.}
{}From \refb{elist} we see that the net contribution to $\Delta$
and $\delta$ still vanishes for such fields. 

Using $\Delta=0$ and $\delta=0$ we get from \refb{ed4} that
\be \label{ed6}
c_{L,eff}^{macro}
= c_{grav}^{asymp} + 6 k_R^{asymp}, \qquad k_{L,eff}^{macro}
= k_L^{asymp}\, .
\ee
As already argued before, $c_{grav}^{asymp} + 6 k_R^{asymp}$
and $k_L^{asymp}$ are  given respectively by the same
computation as $c_L^{bulk}$ and $k_L^{bulk}$
of \S\ref{smacro} except that the
constant shifts are absent. This proves that the effect of the inclusion
of the exterior contribution is to remove the constant term in the
central charges due to one loop corrections.
Note also that in 
\refb{elist}
the values of $c_L$ and $c_{L,eff}$ differ for several of the modes.
Thus if we had focussed on the absolute degeneracy rather than the
index then its growth will not be controlled solely by the anomaly
coefficients since for the contribution due to the exterior modes
$c_{L,eff}$ will now be replaced by $c_L$.

\section{Microscopic Results \label{smicro}}

In this section we shall examine the computation of the
microscopic indices of
certain black holes in four and five dimensions, and show that 
these agree with the results of explicit macroscopic
calculations given in \S\ref{smacro} and \S\ref{shair}.

\subsection{D1-D5-p System in type IIB on $K3\times S^1$}
\label{sd1d5p}

In this section we shall examine in detail the
microscopic formul\ae\ for the index of the D1-D5-p
system in type IIB string theory compactified
on $K3\times S^1$ in various limits.
We consider
a  system of 1 D5-brane wrapped on
$K3\times S^1$ and $Q_1+1$
D1-branes wrapped on $S^1$, carrying $n$ units of left-moving
momentum
along
$S^1$ and $SU(2)_L$ angular momentum $J_L=J/2$.
Since a D5-brane wrapped on K3 carries $-1$ unit of D1-brane
charge, $Q_1$ represents the physical D1-brane charge carried
by this system.
We consider the index:
\be \label{edefind}
d_{micro}(n,Q_1,J)\equiv C_2(n,Q_1,J) =
-{1\over 2!}\,
Tr\left[ (-1)^{2 J_R} \, (2 J_R)^2
\right]\, ,
\ee
where the trace is taken over all states carrying fixed $Q_1$, $n$
and $J_L=J/2$ but
different values of $J_R$.
The partition function
$Z_{5D}(\rho,\sigma,v)$, defined through the relation
\be \label{epart}
Z_{5D}(\rho,\sigma,v) \equiv \sum_{Q_1,n,J}
e^{2\pi i (\rho n  + \sigma Q_1 + v J)} \, (-1)^J \, d_{micro}(n,Q_1,J)\, ,
\ee
is
given by \cite{9608096,9903163}
\ben\label{ep2}
Z_{5D}(\rho,\sigma,v) &=&
e^{-2\pi i  \sigma
 }\prod_{k,l,j\in \zzz
\atop k\ge 1, l\ge 0}
\left( 1 - e^{2\pi i ( \sigma k   +  \rho l + v j)}
\right)^{-c(4lk - j^2)} \nonumber \\
&& \times \left\{\prod_{l\ge 1} (1-e^{2\pi i(l\rho+v)})^{-2} \,
(1-e^{2\pi i(l\rho-v)})^{-2} \, (1-e^{2\pi il\rho})^4 \right\}\,
(-1)\, (e^{\pi i v}-
e^{-\pi i v})^2 \nonumber \\
\een
where $c(u)$ are defined via the relations:
\be\label{e1.5copy}
F(\tau,z) =
\sum_{j,n\in \zzz} c(4n -j^2)
e^{2\pi i n\tau + 2\pi i jz}\, .
\ee
\be \label{efdef}
F(\tau, z) = 8\left[ {\vartheta_2(\tau,z)^2
\over \vartheta_2(\tau,0)^2} +
{\vartheta_3(\tau,z)^2\over \vartheta_3(\tau,0)^2}
+ {\vartheta_4(\tau,z)^2\over \vartheta_4(\tau,0)^2}\right]\, .
\ee
The first line of \refb{ep2} is the contribution from
the relative motion between the D1 and D5 branes \cite{9608096}
and
the second line represents the contribution from
the center of mass modes \cite{0901.0359}.
Strictly speaking we should subtract from this the contribution
from the half-BPS states carrying zero momentum, but
as long as we use this formula to extract the index of states
carrying non-zero momentum along $S^1$, we shall not make
any error. The $-(2J_R)^2/2!$ factor in the trace has
been absorbed by the four fermion zero modes associated with
the center of mass motion carrying $(J_L, J_R)=(0,\pm{1\over 2})$,
and  the factor of $-(e^{\pi i v} - e^{-\pi i v})^2$
comes from the contribution from the four
fermion zero modes on the D1-D5 world-volume
carrying $(J_L, J_R)=(\pm{1\over 2}, 0)$.

Eq.\refb{ep2} may be rewritten as
\be \label{ehor}
Z_{5D}(\rho,\sigma,v)
= -\left(e^{\pi i v}-
e^{-\pi i v}\right)^4\, {\eta(\rho)^{24} 
\over \Phi_{10}(\rho,\sigma, v)}
\, , \ee
where
\be \label{ephi10}
\Phi_{10}(\rho,\sigma, v)
= e^{2\pi i \sigma + 2\pi i \rho + 2\pi i v
 }\prod_{k,l,j\in \zzz
\atop k,l\ge 0,  j<0 \, \hbox{{\small for}}\, k=l=0}
\left( 1 - e^{2\pi i ( \sigma k   +  \rho l + v j)}
\right)^{c(4lk - j^2)}\, ,
\ee
is the Igusa cusp form.
In going from \refb{ep2} to \refb{ehor} we have used
$c(0)=20$, $c(-1)=2$.
{}From \refb{epart}, \refb{ehor} we get
\be \label{ehh2}
d_{micro}(n,Q_1, J)
= (-1)^{J+1}\,
\int_0^1 d\rho \int_0^1 d\sigma \int_0^1 dv\, \left(e^{\pi i v}-
e^{-\pi i v}\right)^4\, 
e^{-2\pi i (\rho n + \sigma Q_1 + J v)}\,
{\eta(\rho)^{24} \over \Phi_{10}(\rho,\sigma, v)}\, .
\ee

We shall be interested in studying the behavior of
$d_{micro}(n,Q_1,J)$
in two different limits:
\begin{enumerate}
\item Type IIB Cardy limit: $n$ large at fixed $Q_1$ and
$Q_1 - {J^2 \over 4n}> K_1$ for some fixed positive number
$K_1$.
\item Type IIA Cardy limit \cite{0807.0237}: $Q_1$ large at fixed
$n$ and $ n - {J^2\over 4 Q_1}>K_2$ for some fixed
positive number $K_2$.
\end{enumerate}
Estimates for $K_1$, $K_2$ can be found in appendix
\ref{sa}.
In both these limits the combination $\Delta\equiv (4 Q_1 n - J^2)$
becomes large. In this case the asymptotic expansion of
$d_{micro}(n,Q_1, J)$ is governed by the residue of the integrand
in \refb{ehh2} on the
subspace \cite{9607026,0412287,0605210,0708.1270}
\be \label{epole}
\rho\sigma - v^2 + v =0\, ,
\ee
where the integrand has a pole. Since the analysis in
 \cite{9607026,0412287,0605210,0708.1270}
were carried out in a different limit where $n$, $Q_1$ and $J$
were all large and of same order, we have given a careful analysis
in
appendix \ref{sa} showing that even in the two limits we are
considering the dominant contribution comes from this pole.
Near this pole
\be \label{eres}
{1\over \Phi_{10}(\rho,\sigma, v)} = -(4\pi^2)^{-1} \,
\rho^{10} \,  \check v^{-2}
\eta(\check\rho)^{-24}  \, \eta(\check\sigma)^{-24} + \hbox{non-singular}\, ,
\ee
where
\be \label{edefwc}
\check\rho = {\rho\sigma - v^2\over \rho}, \quad
\check\sigma = {\rho\sigma - (v-1)^2 \over \rho}, \quad
\check v = {\rho\sigma - v^2 + v\over \rho}\, .
\ee
Picking up the residue at the pole at \refb{epole}
restricts the three dimensional
integral to a two dimensional subspace.
This is best done by changing the variables of integration to
$(\check\rho,\check\sigma,\check v)$, and using
\be \label{echvar}
d\rho\wedge d\sigma\wedge dv = -(2\check v-
\check\rho-\check\sigma)^{-3}\,
d\check\rho\wedge d\check\sigma\wedge d\check v\, .
\ee
In these variables the
residue at the pole at $\check v=0$
can be calculated easily using standard
procedure.
Introducing the variables $(\tau_1,\tau_2)$ via
\be \label{edeftau}
\check\rho = \tau_1+i\tau_2, \qquad \check\sigma=-\tau_1+i\tau_2\, ,
\ee
we have near the $\check v=0$ subspace:
\be \label{etaudef}
\rho={i\over 2\tau_2}+ {1\over 2\tau_2^2}\,
\check v + \OO(\check v^2), \quad 
\sigma=i{\tau_1^2 + \tau_2^2\over 2\tau_2}
+ {\tau_1^2+\tau_2^2 \over 2\tau_2^2}\,
\check v + \OO(\check v^2),
\quad v = {1\over 2} - i{\tau_1\over 2\tau_2}
- {\tau_1\over 2\tau_2^2}\,
\check v + \OO(\check v^2)\, .
\ee
Then the contribution to the integral from the residue at
$\check v=0$ is
given by \cite{0412287,0605210,0708.1270}\footnote{In
 \cite{0412287,0605210,0708.1270} the analysis was carried out for
the four dimensional black hole for which the integrand in
\refb{ehh2} involves $1/\Phi_{10}$ instead of $\eta(\rho)^{24}
/\Phi_{10}$. Eqs.\refb{efcont}, \refb{edefF} are obtained by
multiplying the integrand of  \cite{0412287,0605210,0708.1270}
by a factor of $\eta(\rho)^{24}$, and then picking up the residue at
$\check v=0$.
This procedure is similar to the ones followed in
 \cite{0807.0237,0807.1314},
except that we have included in our analysis the contribution
from the center of mass degrees of freedom of the D1-D5-brane
system and removed the contribution due to the
fermion zero modes associated with the hair.}
\be \label{efcont}
d_{micro}(n,Q_1, J) \simeq
\int{d^2 \tau\over \tau_2^2} \, e^{-F(\tau_1, \tau_2)}
\, ,
\ee
where
\ben \label{edefF}
F(\tau_1, \tau_2)
&=&  -{\pi\over \tau_2} \left[ n + Q_1(\tau_1^2 + \tau_2^2)
- \tau_1 J \right] + 24 \ln \eta(\tau_1+i\tau_2)
+ 24 \, \ln\, \eta(-\tau_1 +i\tau_2) 
\nonumber \\
&& + 12 \ln (2\tau_2) - 24 \, \ln \eta\left({i\over 2\tau_2}\right)
- 4\ln \left\{2\cosh\left({\pi\tau_1\over 2\tau_2}\right)
\right\}\nonumber \\ &&
- \ln \left[{1\over 4\pi} \left\{ 26 + {2\pi\over \tau_2}
\left( n + Q_1(\tau_1^2 + \tau_2^2)
- \tau_1 J\right) + i{24\over \tau_2}\, {\eta'(i/2\tau_2)\over
\eta(i/2\tau_2)} + 4\pi {\tau_1\over \tau_2}
\tanh{\pi\tau_1\over 2\tau_2}\right\}\right]\, .
\nonumber \\
\een
$\simeq$ in \refb{efcont} implies equality up to exponentially
suppressed contributions.
Although we have not been careful
to keep track of the sign, this can
be done by carefully following each step as in  \cite{0708.1270}.
The result is that the
$\tau_1$, $\tau_2$ integrations run
along the imaginary $\tau_1$, $\tau_2$ directions through the saddle
points of $F(\tau_1,\tau_2)$ and the integration measure $d^2\tau$
represents $d(Im\tau_1) d (Im\tau_2)$. Thus the leading
contribution to $d_{micro}(n,Q_1,J)$ is positive.

The integration over $\tau_1$, $\tau_2$ can be evaluated using the
method of steepest descent.
First of all note that if we ignore all terms except the one inside the
first square bracket on the right hand side of \refb{edefF}, the extremum
of $F(\tau_1, \tau_2)$ lies at
\be \label{eextr1}
\tau_1 = {J\over 2 Q_1}, \qquad \tau_2 = \sqrt{4 n Q_1 - J^2
\over 4 Q_1^2}\, .
\ee
If $Q_1$, $n$ and $J$ become large at the same rate then
$(\tau_1,\tau_2)$ are of order unity and the first term in the square
bracket in \refb{edefF}
dominates over the other term. However since
we want to take
different limits we need to keep track of the contribution from the
rest of the terms.

\begin{enumerate}

\item In the type IIB Cardy limit we have
$n\to \infty$ at fixed values
of $Q_1$, and $Q_1- {J^2\over 4n}> K_1$. In this case
we get from \refb{eextr1}
$\tau_2\sim \sqrt n$ and $\tau_1\lsim \sqrt n$. Since
$\tau_2$ is large, we have
\be \label{elim2}
24 \ln \eta(\tau_1 + i\tau_2) \simeq 2\pi i(\tau_1+i \tau_2), \quad
24 \ln \eta(-\tau_1 + i\tau_2) \simeq 2\pi i(-\tau_1+i \tau_2), \quad
24 \ln \eta({i\over 2\tau_2}) \simeq -4\pi\tau_2\, .
\ee
Substituting this into \refb{edefF} we see that in the rest of
the terms other than those contained in the first square bracket
the terms linear in $\tau_1$ and
$\tau_2$ cancel, and at \refb{eextr1} the net contribution
from these terms is small
compared to the first term in the square bracket. Thus the leading
contribution to $-\ln d_{micro}$ will be obtained by evaluating
the first term in the square bracket at the saddle point
\refb{eextr1}. This gives
\be \label{eiibcardy}
\ln d_{micro}(n,Q_1, J) \simeq \pi \sqrt{4n Q_1-J^2}\, .
\ee
In this equation $\simeq$ denotes equality up to power
suppressed corrections. In the rest of this section $\simeq$
in the expression for $d_{micro}$
will denote corrections suppressed by powers of $n$ ($Q_1$)
in the type IIB Cardy (type IIA Cardy) limit.
In principle we can compute these power suppressed corrections
by systematically carrying out the integration over $(\tau_1,\tau_2)$
about this saddle point.

If we have $Q_5$ D5-branes instead of one D5-brane
with $\gcd(Q_1,Q_5)=1$ then by
duality invariance the result for the index depends on the
combination $Q_1Q_5$. Thus the result for general $Q_5$ is
obtained by replacing $Q_1$ by $Q_1Q_5$ in \refb{eiibcardy}:
\be \label{eiibgen}
\ln d_{micro}(n,Q_1, Q_5,J) \simeq \pi \sqrt{4n Q_1 Q_5-J^2}\, .
\ee
The result is valid for large $n$ with
$Q_1Q_5 - {J^2 \over 4n}> K_1$. This result is in perfect agreement
with the result of the direct macroscopic calculation given in
\refb{dhor1}.

It is worth comparing the result for the index
with the result for the
degeneracy. For simplicity we shall
sum over all the $J$ values keeping the other charges fixed.
In this case the index grows as $\exp[\pi \sqrt{4n Q_1 Q_5}]$.
For computing the degeneracy we shall apply the Cardy formula.
Since the relative motion of the D1-D5 system
is described by a super-conformal field theory
whose target space
is the symmetric product of $(Q_1Q_5+1)$ copies of $K3$, we get
a central charge of $6(Q_1Q_5+1)$ from the dynamics of these
modes. The center of mass motion in the transverse directions will
give a superconformal field theory with target space $R^4$, and gives
a central charge 6. Thus the total central charge of this system
is $c^{micro}=6(Q_1Q_5+2)$, both for the left and the right-moving modes.
Since the black hole microstates are identified as the left-moving
excitations in this CFT, we get the expected growth of degeneracy
to be  $\exp[2\pi \sqrt{c^{micro}n/6}]\sim
\exp[2\pi\sqrt{(Q_1Q_5+2)n}]$. This is different from the
rate of growth $\exp[2\pi \sqrt{n Q_1 Q_5}]$ of the index.

\item
In the type IIA Cardy limit we have
$Q_1\to \infty$ at fixed values
of $n$, and $n-{J^2\over 4Q_1}>K_2$. Thus \refb{eextr1} gives
$\tau_2\sim 1/\sqrt{Q_1}$ and $\tau_1\lsim 1/\sqrt{Q_1}$.
Since $(\tau_1 + i\tau_2)$ is small, it is natural to define
\be \label{enewvar}
\pm\sigma_1+ i \sigma_2 = - {1\over \pm\tau_1+ i\tau_2}\, .
\ee
At \refb{eextr1},
$\sigma_2=\sqrt{4nQ_1-J^2}/2n$. This is
large in the limit we are considering, and hence
we have
\ben \label{elim3}
&& 24 \ln \eta(\tau_1 + i\tau_2) \simeq 2\pi i(\sigma_1+i\sigma_2)
, \quad
24 \ln \eta(-\tau_1 + i\tau_2) \simeq 2\pi i(-\sigma_1+i\sigma_2),
\nonumber \\
&& 24 \ln \eta({i\over 2\tau_2}) \simeq -{\pi(\sigma_1^2 +\sigma_2^2)
\over \sigma_2}\, .
\een
Each of these terms is of order $\sqrt{Q_1}$
at the saddle point and they do not cancel.
Since in the limit of large $Q_1$,  the terms inside
the first square bracket of \refb{edefF}
and the contribution from the rest of the terms
are both of order  $\sqrt{Q_1}$,  it is no longer appropriate to
neglect the rest of the terms. Instead we must evaluate the saddle
point by taking into account the
contribution from all the terms.
We shall proceed with the ansatz
that at the saddle point $\sigma_2$  is of
order $\sqrt{Q_1}$;
this will be verified at the end
to check the self-consistency of our approximation.
With this assumption
we can approximate the $\eta$ functions by \refb{elim3} and get the
leading terms in $F(\tau_1,\tau_2)$ to be:
\be \label{fappra}
-{\pi\over \sigma_2} \left[ Q_1 + n(\sigma_1^2 + \sigma_2^2)
+ \sigma_1 J \right]  - 4\pi \sigma_2 +{\pi(\sigma_1^2 + \sigma_2^2)
\over \sigma_2}
\, .
\ee
This has an extremum at
\be \label{esigext}
\sigma_1 = -{J\over 2(n-1)}, \qquad \sigma_2 =
\sqrt{\left(Q_1 -{J^2\over 4(n-1)}\right) / (n+3)}\, ,
\ee
 and at
this extremum
\be \label{efnewext}
F=-2\pi\sqrt{(n+3) \left(Q_1 - {J^2 \over 4(n-1)}\right)}\, .
\ee
This gives
\be \label{ehetcardy}
\ln d_{micro}(n, Q_1, J) \simeq 2\pi
\sqrt{(n+3) \left(Q_1 - {J^2 \over 4(n-1)}\right)}\, ,
\ee
up to power suppressed corrections.
Furthermore from \refb{esigext} we see that
$\sigma_2\sim \sqrt{Q_1}$
in agreement with our ansatz.

We can
write down the result for $Q_5$ number of D5-branes 
with $\gcd\{Q_1,Q_5\}=1$ by
replacing $Q_1$ by $Q_1 Q_5$ in \refb{ehetcardy}:
\be \label{ehetgen}
\ln d_{micro}(n, Q_1, Q_5,J) \simeq 2\pi\sqrt{(n+3) \left(Q_1Q_5
- {J^2 \over 4(n-1)}\right)}\, .
\ee
This result is valid when $Q_1Q_5$ is large, and
$ n - {J^2\over 4 Q_1Q_5}>K_2$. This is
again in perfect agreement with the
result of the macroscopic calculation given in
\refb{efinent}.

To first subleading order
in an expansion in powers of $1/n$ and $J^2$
this agreement was found in
 \cite{0807.0237}.

\end{enumerate}

\subsection{D1-D5-p-KK monopole system in type IIB on
$K3\times T^2$} \label{sfourmicro}

We consider now the same D1-D5-p system analyzed in \S\ref{sd1d5p}
and place it at the center of a Taub-NUT space. This gives a four
dimensional black hole, with the asymptotic circle $\wt S^1$ of
the Taub-NUT space identified as a new compact direction.
Since the black hole breaks 12 of the 16 supersymmetries of the theory,
the relevant index is $B_6$.
The Taub-NUT background has three effects on the index computation:
it first of all converts the angular momentum $2J_L=J$ to
momentum along $\wt S^1$ \cite{0503217}, it shifts the momentum
along $S^1$ by $-1$ units, and it gives
additional contribution to the `partition function' for the
index \cite{0605210}.
We shall denote by $d_{micro}(n, Q_1, J)$ the negative of
the sixth helicity trace index for these dyons.
Then \cite{9607026,0412287,0505094,0605210}
\be \label{e4dpart}
d_{micro}(n, Q_1, J)
= (-1)^{J+1}\,
\int_0^1 d\rho \int_0^1 d\sigma \int_0^1 dv\,
e^{-2\pi i (\rho n + \sigma Q_1 + J v)}\,
{1 \over \Phi_{10}(\rho,\sigma, v)}\, .
\ee
We shall be interested in the
behavior of this quantity in the limit of large $n$ at
fixed values of $Q_1$, and $J=0$. The analysis proceeds as
in \S\ref{sd1d5p} and we arrive at the
result \cite{0412287,0605210,0708.1270}:
\be \label{efour1}
d_{micro}(n,Q_1, J=0)
\simeq
\int{d^2 \tau\over \tau_2^2} \, e^{-F(\tau_1, \tau_2)}
\, ,
\ee
where
\ben \label{efour2}
F(\tau_1, \tau_2)
&=&  -{\pi\over \tau_2} \left[ n + Q_1(\tau_1^2 + \tau_2^2)
\right] + 24 \ln \eta(\tau_1+i\tau_2)
+ 24 \, \ln\, \eta(-\tau_1 +i\tau_2) \nonumber \\
&& + 12 \ln (2\tau_2)
- \ln \left[{1\over 4\pi} \left\{ 26 + {2\pi\over \tau_2}
\left( n + Q_1(\tau_1^2 + \tau_2^2)
\right)\right\}\right]\, .\nonumber \\
\een
Using $\tau_1\to -\tau_1$ symmetry we can set $\tau_1=0$ at the
saddle point. To extract the behavior of this integral for large
$n$ we shall proceed with the ansatz that $\tau_2$ is large,
of order $\sqrt{n}$ at the saddle point. In this case we can
approximate $F(\tau_1=0, \tau_2)$ by
\be \label{eapp23}
F(\tau_1=0, \tau_2)
=  -{\pi\over \tau_2} \left[ n + Q_1\tau_2^2
\right] - 4\pi \tau_2\, .
\ee
This has an extremum at
\be \label{ext3}
\tau_2 = \sqrt{n/(Q_1+4)}\, .
\ee
Thus at the extremum $\tau_2\sim\sqrt n$, satisfying our ansatz.
Evaluating $F(0,\tau_2)$ at the extremum we get
\be \label{e4dent}
\ln \left(d_{micro}(n, Q_1, J=0)\right) \simeq -F(0, \tau_2)|_{extremum}
= 2\pi \sqrt{(Q_1+4) n}\, .
\ee
We can in fact find the full asymptotic expansion by
replacing the
$-12\ln(2\tau_2)$\break
$+ \ln \left[{1\over 4\pi} \left\{ 26 + {2\pi\over \tau_2}
\left( n + Q_1(\tau_1^2 + \tau_2^2)
\right)\right\}\right]$ factor in the exponent by a
multiplicative factor of\break 
$(2\tau_2)^{-12}
\left[{1\over 4\pi} \left\{ 26 + {2\pi\over \tau_2}
\left( n + Q_1(\tau_1^2 + \tau_2^2)
\right)\right\}\right]$ 
in the integrand
and approximating $\eta(\tau)$ by $e^{2\pi i\tau/24}$
as in \refb{eapp23}. The $\tau_1$ integral
then becomes a gaussian integral which can be
evaluated, and the $\tau_2$ integral gives sum of
Bessel functions. Using appropriate identities among 
Bessel functions we can bring the integral to the
form
\be \label{efuffform}
d_{micro} = C_0 \left({n\over Q_1+4}\right)^{-23/4}
I_{23/2} (2\pi \sqrt{n(Q_1+4)})\, ,
\ee
where $C_0$ is a constant independent of
$n$ and $I_\nu$ denotes the
standard Bessel function with imaginary argument.
This is precisely the leading term in the Rademacher
expansion\cite{appear}. 

The final answer \eqref{efuffform} can be readily determined 
directly using standard facts about the Rademacher expansion 
of modular forms and Jacobi forms as follows.  Doing the $\sigma$ 
integral first, we pick up the $Q_{1}$-th Fourier coffecient of  the 
partition function.  Since $1/{\Phi_{10}}$ is a Siegel modular form 
of weight $-10$, this Fourier coefficient $\psi(\tau, z)$ is a weak 
Jacobi form in two variables of weight $-10$ and index $Q_{1}$.  
 Furthermore, $\psi$ is known to be  the  partition function of a 
 $(0, 4)$ SCFT of central charge $C = 6Q_{1} +24$. For a Jacobi 
 form of  weight $-k$, the index of the Bessel function   and the 
 power of the prefactor in the Rademacher expansion\footnote{The 
 usual Rademacher expansion of weak Jacobi forms assumes 
 that the Jacobi form is holomorphic. In our case,  turns out to be 
 meromorphic because of the poles in partition function and the 
 Rademacher expansion is modified but by terms that exponentially 
 subleading \cite{appear}.} is controlled by $(k+3/2)$  which in our 
 case is $23/2$. The argument of the Bessel function and the 
 prefactor are, on the other hand, given by  $2\pi \sqrt{C n/6}$ 
 which in our case gives  $2\pi \sqrt{n (Q_{1}+4)}$.

If we take a system with $Q_5$ D5-branes instead of a single
D5-brane with $\gcd(Q_1,Q_5)=1$
then the $B_6$ index must depend on $Q_1$ and $Q_5$
through the duality invariant combination $Q_1 Q_5$. This gives
\be \label{e4dgen}
 \ln \left(d_{micro}(n, Q_1,Q_5, J=0)\right) \simeq
 2\pi \sqrt{(Q_1Q_5+4) n}\, .
\ee
What if we have $K$ KK-monopoles instead of a single KK monopole
associated with $\wt S^1$? As long as $\gcd(Q_1, Q_5)=1$ and
$\gcd(n, K)=1$, we can find a duality transformation that maps this
charge vector to the one considered above with $n$ replaced by
$n\, K$ \cite{0712.0043,0801.0149}. Thus we have
\be \label{e4dgengen}
 \ln d_{micro}(n, Q_1,Q_5, K, J=0) \simeq
  2\pi \sqrt{(Q_1Q_5+4) n\, K}\, .
\ee
This is in perfect agreement with the macroscopic result
\refb{efbh}, computed by describing the system as a black hole
in M-theory on $K3\times T^3$, carrying M5-brane charges and
momentum along a circle.

When the above arithmetic condition on $(n,K,Q_1,Q_5)$ fails
to hold there is no duality transformation that maps this
charge vector to the one for which we carried out the analysis.
Nevertheless the answer for $B_6$ for these more general
charge vectors is
known \cite{0802.0544,0802.1556,0803.2692} and, in
the limit of large $n$, differs from \refb{e4dgengen} by
exponentially suppressed terms. Thus we can continue to use
\refb{e4dgengen} for the general dyon.

\subsection{Black holes in toroidally compactified type II string theory}
\label{etorusmicro}

In this section we shall generalize the analysis of the previous
sections to toroidally compactified type IIB string theory.
Since the D1-D5-p system on $T^4\times S^1$ describes a
1/8 BPS state in a theory with 32 unbroken supercharges, the
relevant index is $C_{6}$ defined in \refb{edefi2k}.
This index was computed in  \cite{9903163}.
For simplicity we shall set $Q_5=1$ and denote the corresponding
index $C_6(n,Q_1,J)$ by $d_{micro}(n,Q_1,J)$; at the end we can
recover the result
for general $Q_5$ satisfying $\gcd(Q_1,Q_5)=1$ 
by replacing $Q_1$
by $Q_1Q_5$. The result of  \cite{9903163} for the
index may be
expressed as
\be \label{ei6full}
\sum_J (-1)^J\, d_{micro}(n,Q_1,J) \, e^{2\pi i J v} 
=  \left(e^{\i\pi v} - e^{-i\pi v}\right)^4
\sum_{j\in \zzz}\sum_{s|n,Q_1,j}
s \, \wh c\left({4 Q_1 n - j^2\over s^2}\right) e^{2\pi i v j}\, ,
\ee
where $\wh c(\Delta)$ is defined through the relation:
\be \label{ek6.5}
-\vt_1(z|\tau)^2 \, \eta(\tau)^{-6} \equiv \sum_{k,l} \wh c(4k-l^2)\,
e^{2\pi i (k\tau+l z)}\, .
\ee
$\vt_1(z|\tau)$ and $\eta(\tau)$ are respectively the odd Jacobi
theta function and the Dedekind eta function.
The $(-1)^J$ factor in \refb{ei6full} appears from the inclusion
of an extra $(-1)^J$ factor in the definition of the index
in \cite{9903163}.
In the limit when $Q_1n$ is large only the $s=1$ term is important
and we get
\be \label{etx5}
d_{micro}(n,Q_1,J) \simeq (-1)^{J+1}\,
\int_0^1 d\tau \, \int_0^1
\, dv\, e^{-2\pi i Q_1 n \tau-2\pi i J v}
\, (e^{\pi i v} - e^{-\pi i v})^{4}\,
{\vt_1(v|\tau)^2\over \eta(\tau)^6}\, ,
\ee
up to exponentially suppressed corrections.
We shall evaluate the integral over $\tau$ and $v$ using the saddle
point method. We proceed with the ansatz that at the saddle point
$\tau$ is small and $v\sim 1$, and  verify this at the end.
In this case we can express the integrand in \refb{etx5} as
\be \label{eappr}
(-1)^J\, e^{-2\pi i Q_1 n \tau-2\pi i J v}
\, (e^{\pi i v} - e^{-\pi i v})^{4}\,  e^{-2\pi i v^2/\tau}\,
e^{2\pi i v/\tau}\,
(1 - e^{-2i\pi v/\tau})^2
\, (-i\tau)^2\, .
\ee
Extremizing the integrand with respect to $v$ and $\tau$ we find the
approximate saddle point in the rangle
$0\le Re(v)<1$ at
\be \label{eap2}
v = {1\over 2} -{J\over 2} \tau +\cdots, \qquad \tau = i/\sqrt{4nQ_1
-{J^2
}}+\cdots \, ,
\ee
where $\cdots$ denote subleading terms. The value of the integrand
at this saddle point is
\be \label{eap3}
\exp[\pi \sqrt{4nQ_1 - J^2} +\cdots]\, .
\ee
This gives the leading contribution to $d_{micro}(n,Q_1,J)$.
We can recover the results for $Q_5\ne 1$ with $\gcd(Q_1,Q_5)=1$
by replacing $Q_1$ by $Q_1Q_5$ in \refb{eap3}. This gives
\be \label{eap4}
\ln d_{micro}(n,Q_1, Q_5, J) \simeq \pi\sqrt{4nQ_1Q_5 - J^2}\, .
\ee
This is
in perfect
agreement with the
macroscopic result given in \refb{etx1} and \refb{etx2}.
Note that in the microscopic analysis there is no distinction
between type IIB Cardy limit ($n\to\infty$) and type IIA Cardy
limit ($Q_1\to\infty$) since the result depends on the combination
$Q_1n$.

If instead of using the index we had computed the absolute degeneracy
then the results would change as follows. The motion of $Q_1$ D1-branes
inside a single D5-brane gives us $4Q_1$ bosonic degrees of freedom
and their $4Q_1$ fermionic partners. Besides this we have
four extra bosonic modes associated with the D1-D5 center of mass motion
and four more bosonic modes associated with the Wilson lines on the
D5-brane along $T^4$. Thus we have eight extra bosonic modes and
their fermionic superpartners. This would give a total contribution of
$6(Q_1+2)$ to the left-handed central charge, and the logarithm
of the degeneracy computed from this would grow as
$\pi\sqrt{4n(Q_1+2)}$ for $J=0$. This is clearly different from
\refb{eap3} for $J=0$.

Finally consider the four dimensional system containing
$Q_5$ D5-branes along $T^4\times S^1$,
$Q_1$ D1-branes along $S^1$ and K Kaluza-Klein monopoles
associated with $\wt S^1$, carrying $n$ units of momentum along
$S^1$. This is U-dual to the M5-brane configuration discussed in
\S\ref{storusmacro}.
We shall restrict our analysis to the case
$\gcd\{Kn, Q_1Q_5, KQ_1, KQ_5, nQ_1, nQ_5\}=1$.
The exact $B_{14}$ index of these states is known, and up to
exponentially suppressed corrections, the index
is given by \cite{0506151,0803.1014,0804.0651}
\be \label{et72}
-B_{14} \simeq  -\wh c(4Q_1 Q_5 K n)\, ,
\ee
with $\wh c(\Delta)$ defined as in \refb{ei6full}.
For
large $\Delta$ we have \cite{0911.1563}
\be \label{easymp}
\wh c(\Delta) \sim (-1)^{\Delta +1}\,
\Delta^{-2}\, \exp(\pi\sqrt{\Delta})\, .
\ee
Eq.\refb{et72} now shows that the logarithm of the index
$-B_{14}$ grows as
$2\pi \sqrt{Q_1Q_5K n}$. This gives 
the microscopic prediction for the
logarithm of the index of the four dimensional black hole:
\be \label{emicpred}
\ln d_{micro}(n,Q_1,Q_5,K) \simeq 2\pi \sqrt{Q_1 Q_5 K n}\, .
\ee
This is
in perfect
agreement with the
macroscopic result given in \refb{etx3}.

\section{MSW Analysis for M5-branes on
$K3\times T^3$ and $T^7$}
\label{smsw}

In \S\ref{sfour} we described a black hole whose microscopic description
contains M5-branes wrapped on a 5-cycle of $K3\times T^3$
or $T^7$.
However while computing the microscopic index of this
system in \S\ref{sfourmicro} we used an indirect method by mapping it
to a D1-D5-p-KK monopole system in type IIB string
theory. In this section we shall
directly compute the microscopic index of the M5-brane
system following  \cite{9711053}, and show that the results agree
with those obtained in \S\ref{sfourmicro}.

\subsection{M5-brane on $K3\times T^3$} \label{scommsw}

We begin by recalling the system of M5-branes described in
\S\ref{sfour}.
We consider
M-theory on $K3\times
S^1\times \wh S^1\times S^1_M$, and take a brane configuration
consisting of $Q_1$ M5-branes along $C_2\times S^1 \times
\wh S^1\times S^1_M$,
$Q_5$ M5-branes wrapped along $\wt C_2\times S^1 \times \wh S^1
\times S^1_M$,
and $K$ M5-branes wrapped along $K3\times S^1$,
carrying $n$ units of
momentum along $S^1$.
The $B_6$ index of this configuration
can be calculated following the procedure described in
 \cite{9711053,9711067}.
In order to follow the notation of  \cite{9711053}, we introduce
some new notation for the charges, denoting the electric charges by
$(q_0,q_a)$ and magnetic charges by
$(p^0,p^a)$. The charge $q_0$ corresponds to momentum along the
circle $S^1$ while $q_a$ corresponds to exciting the self-dual
antisymmetric tensor field on the 5-brane, carrying
charges corresponding to wrapping
M2-branes
on various 2-cycles of $K3\times \wh S^1\times S^1_M$.
The magnetic charge
$p^0$ corresponds to a Kaluza-Klein monopole
associated with the circle $S^1$.
The other magnetic charges are associated with an
M5-brane wrapping $P\times S^1$
with $P$ a four cycle of $K3\times \wh S^1\times S^1_M$.
For the configuration
we are considering, $p^0$ and $q_a$ for $a\ne 0$ vanish,
the charges $p^a$ can be identified with the triplet
$(Q_1,Q_5,K)$ and the charge $q_0$ can be identified with
$n$.
Using the isomorphism between 4-cycles and 2-forms we can associate
with $P$ a 2-form on $M$ which we shall also denote by $P$.
In this case
we can write the magnetic
charge vector in cohomology language, i.e, $P=p^a\Sigma_a$ with
$\Sigma_a\in H^2(M,\mathbb{Z})$,
$M\equiv K3\times \wh S^1\times S^1_M$.

If we take the limit in which the circle $S^1$ has a size much
larger than the size of $K3\times \wh S^1\times S^1_M$, then
the low energy limit of the effective theory
describing the dynamics of
the 5-brane on $P\times S^1$ is a two dimensional $(0,4)$ CFT.
The BPS states in this theory involve left-moving excitations and
the growth of degeneracy
of these states for large momentum is determined in terms
of the left-moving central charge $c^{micro}_L$ via the Cardy formula.
$c^{micro}_L$ in turn is given by $N^B_L +{1\over 2} N^F_L$
where $N^B_L$ and $N^F_L$ are the numbers of left-handed bosons
and fermions respectively.
If instead of the degeneracy we consider the helicity trace index
$B_6$, then the computation proceeds as follows. The requirement of
unbroken supersymmetry forces the right-movers into their ground
state. The $(2h)^6$ factor in the trace is soaked up by the
12 fermion zero modes associated with the broken supersymmetry
generators. Thus we are left with the trace over the
left-handed bosonic and fermionic non-zero mode oscillators, weighted by
$(-1)^F$ where $F$ denotes fermion number. The growth of this
trace for large momentum along $S^1$ is controlled by a Cardy like
formula, but with an effective central charge
\be \label{effcena}
c^{micro}_{L,eff}= N^B_L-N^F_L
\, .
\ee
This follows from the fact that the insertion of $(-1)^F$ into the trace
does not affect the contribution to the partition function due to a
bosonic oscillator, but the contribution to the partition function
due to a fermion is now given by the inverse of the contribution from
a boson.
Note that if $N^F_L=0$ then $c^{micro}_{L,eff}=c^{micro}_L$,
but otherwise they are
different.

Now
the numbers of left and right-moving
bosons are given by \cite{9711053}
\begin{eqnarray} \label{nbos}
 N^B_L&=&d_p(P)+b_2^{-}(P)+3, \nonumber\\
 N^B_R&=&d_p(P)+b_2^{+}(P)+3.
\end{eqnarray}
Here $d_p$ is the dimension the moduli space of deformations
of $P$ inside $M$, 3 accounts for the center of mass
translations and $b_2^{-},b_2^+$, denoting the number of
anti-self-dual and self-dual two forms of $P$, count the scalar
fields arising from the
reduction of the 2-form field living
on the 5-brane. For fermions we 
have \cite{9711053,0712.3166}
\begin{eqnarray} \label{nfer}
 N^F_L&=&4 h_{1,0}(P), \nonumber \\
 N^F_R&=&4\, h_{2,0}(P)+4\, .
\end{eqnarray}
Under the assumption that
the Calabi-Yau 3-fold $M$ does not have 1-cycle and that the
4-cycle
$P$ is ample,
the authors of  \cite{9711053} gave a formula for $d_p(P)$ and
used it to compute the number of left- and right-moving fermions
and bosons.
We however have a Calabi-Yau manifold with two 1-cycles $\wh S^1$
and $S^1_M$,
and hence the formul\ae\ of  \cite{9711053} are not directly
applicable.
Thus we need to proceed a little differently following
 \cite{0712.3166}.
On a compact K\"ahler manifold we have the relations:
\be \label{eckah}
b_2\equiv b_2^+ + b_2^- = 2 h_{2,0} + h_{1,1},
\qquad b_2^- = h_{1,1}-1\, .
\ee
Substituting this into \refb{nbos} and \refb{nfer} we get
\be \label{ediff}
N^B_R - N^F_R = d_p(P) - 2 h_{2,0}(P)\, .
\ee
Now since supersymmetry acts on the right-movers, the number of
right-moving bosons and fermions must be equal. This gives
\be \label{edp}
d_p(P) = 2 h_{2,0}(P)\, .
\ee
This agrees with the result given in  \cite{9711067}.
Substituting this into \refb{nbos} and \refb{nfer} we get \cite{9711067}
\be \label{enleft}
N^B_L = 2 h_{2,0}(P) + h_{1,1}(P) + 2=
b_{even}(P), \quad N^F_L = 4h_{1,0}(P)=
b_{odd}(P)\, ,
\ee
where $b_{even}(P)$ and $b_{odd}(P)$ are the dimensions of the
even and odd cohomologies of $P$. Thus $c^{micro}_{L,eff}$ given in
\refb{effcena} is just the Euler character of $P$. This in turn has
a simple expression in terms of the 2-form $P$ representing the
4-cycle $P$ \cite{9711053}:
\be \label{echi}
c^{micro}_{L,eff} = 
\chi(P) = \int_M \, (P\wedge P\wedge P + P\wedge c_2(M))\, .
\ee
Evaluating this for the particular brane configuration we have, we get
\be \label{eval}
c^{micro}_{L,eff} = 6\, K\, (Q_1 Q_5 + 4)\, .
\ee
This is in perfect agreement with the formula for the
index of the D1-D5-p-KK system given in \refb{e4dgengen},
which in turn is in agreement with the macroscopic
result given in \refb{emiccentot}. If instead we had calculated the
central charge that controls the
growth of absolute degeneracy, then
we would get the result \cite{0712.3166}
\be \label{ecenm}
c^{micro}_L = N^B_L + {1\over 2} N^F_L = c^{micro}_{L,eff}
+ {3\over 2} N^F_L = 6\, (K Q_1 Q_5 + 4K + 1)\, ,
\ee
since $N^F_L=4h_{1,0}(P) = 4h_{1,0}(M) = 4$. As
noted in  \cite{9906094,0712.3166}, \refb{ecenm} fails to agree
with the macroscopic result \refb{emiccentot}.
Thus we see that the apparent puzzle in  \cite{9906094,0712.3166} arose
from comparing the microscopic degeneracy with the macroscopic index,
and there is no disagreement as long as we compare the index on both
sides.

\subsection{M5-brane on $T^7$} \label{smswt7}

We shall now repeat the analysis of  \S\ref{scommsw}
with K3
replaced by $T^4$, \i.e.\ directly compute the microscopic index
of the system of M5-branes wrapped on $T^7$ without
mapping it to the D1-D5-p-KK monopole system.
Let us label the $T^7$
by coordinates 1-7. In this theory we
consider a configuration with $Q_1$
M5-branes wrapped along 12345 directions,
$Q_5$
M5-branes wrapped along 12367 directions and
$K$
M5-branes wrapped along 14567 directions, carrying momentum $n$
along the 1-direction.
This configuration breaks 28 out of 32 supersymmetries
of the theory and hence
the relevant helicity trace index is $B_{14}$.  Following the analysis
of \S\ref{scommsw} we arrive at the same result
\refb{echi} for the effective central
charge $c^{micro}_{L,eff}$.
However since $c_2$ vanishes on $T^6$, we get
\be \label{et71}
c^{micro}_{L,eff}= 6\, Q_1\, Q_5\, K\,,
\ee
and hence
\be \label{edhorpred}
\ln d_{micro}(n,Q_1,Q_5,K) = 2\pi \sqrt{Q_1Q_5K n}\, .
\ee
This
agrees with the result \refb{emicpred} computed from the D1-D5-p-KK
monopole system, in agreement with the duality symmetry.
More importantly for us, it agrees with the macroscopic prediction
\refb{etx3}. If instead of using the effective central charge we had
used the actual central charge computed in the limit of free theory,
we would get $c^{micro}_L
=6 (Q_1 Q_5 K +3)$ since we now have $h_{1,0}(P)
= h_{1,0}(M)=3$. This would not agree with the macroscopic result.


\section{Why do the Microscopic and Macroscopic Results Agree?}
\label{sdis}

So far we have computed the index of various systems in the
macroscopic and the microscopic sides and shown that they agree.
However given that on the macroscopic side the index is expressed
in terms of the coefficients of the Chern-Simons terms in the action,
one might hope that this agreement can be proved in general
without having to explicitly compute the index in each case. We shall
now show that this is indeed the case. This argument is closely 
related to the one given in  \cite{0506176}, but takes into account
the additional subtlety that arises due to the failure of the identification
R-symmetry group of the brane world-volume theory with the spatial
rotation group.
For definiteness we shall
present the argument for five dimensional black holes; the only
change in four dimensions will be that we need to drop all references
to the $SU(2)_L$ part of the spatial rotation group and interprete
$SU(2)_R$ as the full rotation group.

The argument goes as follows. For black holes of the type
considered here, the low energy
dynamics of the system of branes underlying the 
microscopic description of the black hole 
is described by a (0,4) superconformal field
theory. We shall divide the system into two parts. One part
which we shall call the regular part has the property that
the right-moving SU(2) R-symmetry current,
associated with the (0,4) superconformal symmetry on the
world-sheet of the branes, can be identified with the 
$SU(2)_R$ subgroup of the spatial
rotation group. Furthermore the action of
the $SU(2)_L$ subgroup of the
spatial rotation group on the regular part 
must correspond to the group
generated by the zero modes of a left-moving SU(2) current algebra
on the brane world-sheet theory.
The second part  does
not satisfy this property, and will be called the irregular part.
This in particular will contain the center of mass degrees of
freedom for which the non-chiral scalars are charged under
both $SU(2)_L$ and $SU(2)_R$. 
Clearly this decomposition is not unique since
we can include part of the regular modes into the
irregular part, and we can utilise this freedom to choose the
irregular part to our convenience.
We can now express the total contribution
to the index as a combination of the contribution from the two
parts as in \S\ref{sdegind}, treating the regular part in the same
way as the
modes associated with the bulk of $AdS_3$ and the irregular part
in the same way
as the exterior modes. In particular if we denote by 
$c_{L,eff}^{micro}$
and $k_{L,eff}^{micro}$ the quantities which control the growth of the
microscopic index, we have the relation analogous to 
\refb{edeft9}:
\be \label{ekx1}
c_{L,eff}^{micro} \equiv c_L^{reg} 
+ c_{L,eff}^{irreg},
\qquad k_{L,eff}^{micro} \equiv k_L^{reg} + k_{L,eff}^{irreg}
\, .
\ee
As in \S\ref{sdegind}, we shall denote by $k_L$, $k_R$ and
$c_{grav}$ the contribution to $SU(2)_R$, $SU(2)_L$ and
gravitational anomaly from various fields on the brane world-volume.
In \refb{ekx1} we have used the fact that for
the regular part the identification of the R-symmetry
group with the spatial rotation group allows us to
conclude, as in the case of the bulk modes, that the
quantities which control the growth of the index are the same as
the ones which control the growth of degeneracy, that is
the central charge $c_L^{reg}$ of the left-moving Virasoro
algebra and the anomaly $k_L^{reg}$ of 
$SU(2)_L$.\footnote{An indirect evidence for the presence
of the irregular part follows from the observations of
\S\ref{smicro}, \S\ref{smsw} that in the microscopic theory the
index and degeneracies do not always agree. Since for the
regular part the index and the degeneracy grow in the same
manner, the difference can be attributed to the presence of the
irregular part. Later we shall explicitly see examples of
irregular parts of the microscopic system.}
Furthermore we also have the relations:
\be \label{ekx2}
c_{grav}^{reg} = c_L^{reg}-c_R^{reg}, \qquad c_R^{reg} = 6 k_R^{reg}
\, .
\ee
Let us denote by $k_L^{micro}$, $k_R^{micro}$ and $c_{grav}^{micro}$
the total contribution to the $SU(2)_L$, $SU(2)_R$ and the
gravitational anomaly from all the microscopic degrees of freedom.
Then we have the relations:
\be \label{ekx3}
k_L^{micro} = k_L^{reg}+ k_L^{irreg}, \quad
k_R^{micro} = k_R^{reg}+ k_R^{irreg}, \quad
c_{grav}^{micro} = c_{grav}^{reg}+ c_{grav}^{irreg}\, .
\ee
Using \refb{ekx1}-\refb{ekx3} we get
\be \label{ekx4}
c_{L,eff}^{micro} =c_{grav}^{micro} + 6 k_R^{micro} + \Delta_{micro},
\quad k_{L,eff}^{micro} = k_L^{micro} +\delta_{micro}\, ,
\ee
where
\be \label{ekx5}
\Delta_{micro} \equiv - 6 k_R^{irreg}
- c_{grav}^{irreg} + c_{L,eff}^{irreg}= 
- 6 k_R^{irreg}
- (c_{L}^{irreg}-c_R^{irreg}) + c_{L,eff}^{irreg}\, ,
\ee
\be \label{ekx6}
\delta_{micro} = k_{L,eff}^{irreg} - k_L^{irreg}\, .
\ee
These are the analogs of eqs.\refb{edefDelta} and 
\refb{edefdelta} in the macroscopic theory. We can now proceed
in the same way as in \S\ref{shair} to show that $\Delta_{micro}$ and
$\delta_{micro}$ vanish. For this we need to make the same assumptions
on the structure of the irregular modes as we had to do on the
structure of the exterior modes in \S\ref{shair}. Thus we get
\be \label{ekx7}
c_{L,eff}^{micro} =c_{grav}^{micro} + 6 k_R^{micro} ,
\quad k_{L,eff}^{micro} = k_L^{micro} \, .
\ee
Finally we make use of
the observation that the coefficients of the gauge and Lorentz
Chern-Simons terms in the bulk theory are related to the gauge and
gravitational anomalies on this brane
configuration  \cite{9808060,0506176}. 
This allows us to
conclude that $c_{grav}^{micro}$,  $k_R^{micro}$ and
$k_L^{micro}$ 
must be equal to
$c_{grav}^{asymp}$, $k_R^{asymp}$ and
$k_L^{asymp}$ 
-- the coefficients of the Lorentz, $SU(2)_R$ and  $SU(2)_L$
Chern-Simons
term in the effective action. Thus from \refb{ed6} we get
\be \label{ekx8}
 c_{L,eff}^{micro} = c_{L,eff}^{macro}, \quad
 k_{L,eff}^{micro} = k_{L,eff}^{macro}\, .
 \ee
This establishes the equivalence of the macroscopic and the
microscopic index.

We shall now explicitly compute the coefficients 
$c_{grav}^{micro}$,  $k_R^{micro}$ and
$k_L^{micro}$ in some examples by computing the
anomalies due to the world-volume fields and show that the
results agree with the explicit microscopic results for the
index given in \S\ref{smicro} and \S\ref{smsw}. During this
analysis we shall also identify the irregular modes in various
systems.
We begin with the D1-D5-p system on $K3\times S^1$ in the
type IIB Cardy limit,
For simplicity we shall take $Q_5=1$. 
Since a D5-brane wrapped on K3 carries $-1$ unit of D1-brane charge,
we need $(Q_1+1)$ D1-branes to produce $Q_1$ units of D1-brane
charge.
In this case the 
world-volume bosonic degrees of freedom consist of $4(Q_1+1)$
scalars describing D1-brane motion along $K3$ and 4 scalars
describing the overall motion of the D1-D5-brane system in the
transverse direction.
The former are all
neutral under the $SU(2)_L\times SU(2)_R$ rotation group
in the space transverse to the D1-D5-brane world-volume, while the
latter are in the $(2_L,2_R)$ representation of $SU(2)_L\times SU(2)_R$.
Since these scalars are non-chiral they do not contribute to
$SU(2)_L\times SU(2)_R$ anomaly. 
In order to determine the $SU(2)_L\times SU(2)_R$ quantum numbers
of the fermions we can use the (4,4) supersymmetry of the
world-volume theory.
Since the left/right
moving modes are paired by supercharges which are doublets of
$SU(2)_L / SU(2)_R$, the fermionic partners of the
$4(Q_1+1)$ neutral
scalars consist of a total of $4(Q_1+1)$ left-moving fermions
in the representation $(2_L,1_R)$ and $4(Q_1+1)$ right-moving fermions
in the representation $(1_L,2_R)$. 
On the other hand the fermionic partners
of the $(2_L,2_R)$ 
scalars representing the transverse motion will consist of
4 left-moving fermions
in the representation $(1_L,2_R)$ and 4  right-moving fermions
in the representation $(2_L,1_R)$. Thus as far as the $SU(2)_L$ group
is concerned, we have altogether $4(Q_1+1)$ left-moving
  fermions
and 4 right-moving  fermions belonging to the doublet representation
of $SU(2)_L$.
This gives a total contribution of $Q_1+1-1=Q_1$ to
the $SU(2)_L$ anomaly coefficient $k^{micro}_L$.
A similar counting gives $k^{micro}_R=Q_1$. 
On the other hand since the spectrum on the brane is left-right
symmetric, the gravitational anomaly $c_{grav}^{micro}$
vanishes. Eq.\refb{ekx7} now gives $c_{L,eff}^{micro} = 6 Q_1$
and $k_{L,eff}^{micro}=Q_1$.
This is
in agreement with the microscopic result \refb{eiibcardy}.

This analysis also
throws some light on the origin of the discrepancy between
$c_{L,eff}^{micro}=6Q_1$ -- the quantity that controls the
growth of the index on the microscopic side,
and
$c_L^{micro}=(Q_1+2)$ -- 
the quantity that controls the growth of the
microscopic degeneracy at weak coupling.
As argued before, for regular part $c_L=c_{L,eff}$; so the
difference must be due to the irregular part.
In this case the irregular part comes from the
$(2_L,2_R)$ 
scalars representing the transverse motion of the brane
and their fermionic partners.
As argued above these include 4 left-moving fermions
in the representation $(1_L,2_R)$ and 4  right-moving fermions
in the representation $(2_L,1_R)$. Now the SU(2) R-symmetry
current on the brane world-volume, associated with the
(0,4) superconformal algebra, is right-moving. Hence all the
left-moving fermions and bosons
must be neutral under it. In contrast we see that the left-moving
components of the 
$(2_L,2_R)$ scalars and the left-moving $(1_L,2_R)$ fermions
are in the doublet representation of the $SU(2)_R$ spatial rotation.
Thus on these fields the SU(2) R-symmetry action cannot
be identified as the action of the $SU(2)_R$ spatial rotation,
and they must be considered as part of the irregular modes.
Indeed by carefully examining the computation of $c_{L,eff}^{micro}$
given above one can easily see that it is due to the presence of
these irregular modes that $c_{L,eff}^{micro}$ and $c_L^{micro}$
differ. Similarly for regular modes we also require that the spatial
$SU(2)_L$ rotation acts as the zero mode of a left-moving 
$SU(2)$ current algebra. Thus all the right-moving regular
modes must
be neutral under $SU(2)_L$. This  fails for the right-moving
$(2_L,2_R)$ scalars and $(2_L,1_R)$ fermions, showing
that they must
also be part of the irregular modes.

The explicit computation of $c_L^{micro}$ and $k_L^{micro}$
for the D1-D5-p system on $T^4\times S^1$ in the type IIB
Cardy limit is almost
identical. In this case the D5-brane on $T^4$ does not carry
any D1-brane charge and we  have $4Q_1$ bosons associated with the
motion of the D1-brane inside the D5-brane and 4 extra bosons
associated with Wilson line on the D5-brane along $T^4$. All
of these are neutral under $SU(2)_L\times SU(2)_R$. We also
have four transverse bosons in the $(2_L,2_R)$ representation of the
$SU(2)_L\times SU(2)_R$. Thus the total spectrum of bosons is
identical to that in the case of D1-D5-p system on $K3\times S^1$,
and due to supersymmetry the fermionic spectrum is also
identical. Thus we still have $k_R^{micro}=Q_1$, $k_L^{micro}=Q_1$,
$c_{grav}^{micro}=0$, and eq.\refb{ekx7} leads to
$c_{L,eff}^{macro}=6Q_1$, in agreement with the microscopic
result for the index given in \refb{eap3}.

For the D1-D5-p system in the type IIA Cardy limit the underlying
microscopic system is the system of $Q_5$ NS5-branes and
$Q_1$ fundamental strings. The dynamics of this system is not well
understood and hence we do not have an independent calculation
of $c_{grav}^{micro}$, $k_R^{micro}$ and $k_L^{micro}$ from
the computation of anomalies in the microscopic theory.
Nevertheless the macroscopic results for these quantities, as well
as the exact results for the microscopic index derived in the dual
type IIB frame, tells us what these anomaly coefficients should be.

A similar analysis can be carried out for the MSW
string \cite{0712.3166} analyzed in \S\ref{smsw}.
We consider
M-theory on $M\times S^1$ where $M$ can be either $K3\times T^2$
or $T^6$ and take an M5-brane wrapped on a four cycle $P$ in
$M$ times $S^1$.
According to Eqs.\refb{nbos}-\refb{enleft} the
number of left- and right-moving bosons and fermions are
given by:
\ben \label{enbosfer}
N^B_L = 2 h_{2,0}(P) + h_{1,1}(P) +2, \quad &&
\quad N^F_L = 4 h_{1,0}(P), \nonumber \\
 N^B_R = 4 h_{2,0}(P)+4, \quad && \quad N^F_R = 4h_{2,0}(P)
+ 4\, .
\een
This gives the gravitational anomaly coefficient in the microscopic
theory to be
\be \label{egrav}
c_{grav}^{micro} 
= N^B_L + {1\over 2} N^F_L -N^B_R - {1\over 2} N^F_R
= h_{1,1}(P) -4 h_{2,0}(P) +2 h_{1,0}(P) - 4\, .
\ee
Next we turn to the computation of $k^{micro}_R$,
-- the anomaly in the spatial rotation symmetry.\footnote{Note
that in this case there is no $SU(2)_L$ symmetry since we are
considering a black hole in 3+1 dimensions.}
The chiral bosons
associated with the component of the 2-form field along the
M5-brane world-volume are neutral under $SU(2)$ and hence
cannot contribute to the $SU(2)$ anomaly. The non-chiral bosons
of course also do not contribute to the $SU(2)$ anomaly. The
$N^F_R$ right-moving fermions are doublets of $SU(2)$ and
give a contribution of $N^F_R/4$ to $k^{micro}_R$ whereas the $N^F_L$
left-moving fermions are also doublets of $SU(2)$ and give a
contribution of $-N^F_L/4$. Thus the net contribution
to $k^{micro}_R$ is given by
\be \label{egauge}
k^{micro}_R = {1\over 4}
(N^F_R - N^F_L) =  h_{2,0}(P) -  h_{1,0}(P)+1\, .
\ee
Using \refb{ekx7}, \refb{egrav} and \refb{egauge} we get
\be \label{ecombab}
c_{L,eff}^{micro} = c_{grav}^{micro} + 6 k^{micro}_R =
h_{1,1}(P) + 2 h_{2,0}(P) - 4 h_{1,0}(P) +2
= \chi(P)
\, .
\ee
This agrees with the  microscopic
result for $c^{micro}_{L,eff}$ given in \refb{echi}.

Note that \refb{ecombab}
does not agree with the microscopic central charge
\be \label{emiccen2}
c_L^{micro}=N^B_L + {1\over 2} N^F_L
= h_{1,1}(P) + 2 h_{2,0}(P) + 2 h_{1,0}(P) +2\, .
\ee
Again the difference can be traced to the contribution from
the irregular modes. For example there are $4h_{1,0}(P)$ 
left-moving fermions
which transform as doublets of the spatial SU(2) rotation group.
Since the left-moving fermions must be neutral under the
right-moving R-symmetry current, on these fermions the R-symmetry
and spatial rotation act differently. Thus they must be considered as
part of the irregular modes.

\subsection*{Acknowledgments}

It is a pleasure to thank  Nabamita Banerjee, Justin David, Bernard de Wit,
Rajesh Gopakumar,
Per Kraus,
Shiraz Minwalla and Jan Troost
for valuable discussions. The work of A.~D. was supported in 
part by the Excellence Chair of the Agence Nationale de la Recherche (ANR).
The work of J.~G was supported in part by Fundac\~{a}o para 
Ci\^{e}ncia e Tecnologia (FCT). The work of S.~M.
is supported in part by the European Commission 
Marie Curie Fellowship under the
contract PIIF- GA-2008-220899. The work of A.~S. was supported in
part by the J. C.  Bose fellowship of the Department of
Science and Technology, India and by
the Chaires Internationales de Recherche Blaise Pascal, France.

\appendix

\section{Chern-Simons Contribution from Higher Derivative 
Terms}
\label{sd}

In this section we describe, following
 \cite{0606230}, how to compute the gauge and Lorentz
Chern-Simons terms in $AdS_3$ by starting with a six dimensional
action and dimensionally reducing it on $AdS_3\times S^3$. The six
dimensional theory will be assumed to have metric and a 2-form field
$B$ as the fundamental fields, but inclusion of other fields in the
discussion is straightforward. We shall denote by 
$H=dB$ the 3-form field strength. First consider a theory with 
manifestly gauge and general coordinate invariant
Lagrangian density given as a function of $H$, $g_{\mu\nu}$,
the Riemann tensor and covariant derivatives of these fields.
Dimensional reduction of the metric on $S^3$ produces $SO(4)$
gauge fields. When all the fluctuating fields around the
$AdS_3\times S^3$ background, including these $SO(4)$ gauge
fields, are set to zero then the
background 3-form field on $AdS_3\times S^3$ takes the form:
\be \label{exz1}
H_3 = {a\over 4}\, \epsilon_3+b\ast\epsilon_3\, ,
\ee
where $\epsilon_3$ is the 
unit 3-sphere volume form, normalized so that
$\int_{S^3}\eps_3=16\pi^2$, $*$ denotes Hodge dual
in six dimensions and $a$ and $b$ are two
constants. 
We shall normalize the 2-form field so that $\int H_3$
is quantized in integer units.
The quantized electric and magnetic charges
$Q$ and $P$ associated with this background are now defined
through the equations:
\begin{equation} \label{exz2}
 \int_{S^3} H_3=4\pi^2 P,
\end{equation}
and\footnote{While regarding $\delta S_0/\delta H_3$
as a 3-form, we need to lower the indices using
the $\varepsilon$ 
tensor as $(\delta S_0/\delta H_3)_{\mu\nu\rho}
= (\delta S_0/\delta (H_3)_{\alpha\beta\gamma})
\varepsilon_{\alpha\beta\gamma\mu\nu\rho}$.}
\begin{equation} \label{exz3}
 \int_{S^3} \left({\delta S_0\over \delta H_3}\right) ={Q \over 2\pi},
\end{equation}
where $S_0$ is the action obtained by integrating the gauge
and diffeomorphism invariant lagrangian density over
$AdS_3\times S^3$.
Eq.\refb{exz1}
gives
\be \label{exz3.5}
a = P\, .
\ee
$b$ is related to $Q$ but this relation depends on the form of the
action $S$.

Let us now consider the effect of switching on the fields
describing fluctuations around the $AdS_3\times S^3$ background.
Dimensional reduction of the metric on $S^3$ produces a set of
$SO(4)=SU(2)_L\times SU(2)_R$ gauge fields
$A_L, A_R$ on $AdS_3$. When these
gauge fields are non-zero we need to 
replace \refb{exz1} by \cite{0606230}
\begin{equation} \label{exz4}
 H_3=4\pi^2 a\left(e_3(A)-\chi_3(A)\right)+b\ast \epsilon_3\, .
\end{equation}
Here $e_3(A)$ is 3-form on $AdS_3\times S^3$ 
defined in  \cite{0606230} and has the property that 
$\int_{S^3} e_3=1$ and that when
the $SO(4)$ gauge fields are set to zero $e_3$ reduces to
$\eps_3/16\pi^2$. $\chi_3$ is the Chern-Simons term for the
$SO(4)$ gauge fields:
\be \label{exz5}
\chi_3 = {1\over 8\pi^2} \left(\omega(A_R) - \omega(A_L)
\right)\, ,
\ee
\be \label{edefomega}
\omega(A) \equiv Tr\left( A \wedge d A+{2\over 3}
A\wedge A\wedge A\right)
\, .
\ee
The trace is taken over the fundamental representation of
$SU(2)$.
Note that since $\int_{S^3} e_3=1$ and $\chi_3$ is directed along
the $AdS_3$ component, the background \refb{exz4}
continues to carry magnetic charge $P=a$ defined via
\refb{exz2}.
Now one can show that $e_3(A)$ is invariant under $SO(4)$
gauge transformation \cite{0606230}, but due to the 
presence of $\chi_3$
in \refb{exz4},
$H_3$ is no longer gauge invariant. Under an $SO(4)$ 
gauge transformation denoted by $\delta$, we have
\begin{equation} \label{exz6}
 \delta 
 H_3=-4\pi^2 a\, d\chi_2= -4\pi^2 P\, d\chi_2,
\end{equation}
where $\chi_2$ is defined via the equation:
\be \label{exz7}
\delta \chi_3 = d\chi_2\, .
\ee
The variation of the action under this gauge transformaion
is then given by
\begin{equation}
 \delta S_0 = 4\pi^2 
 P\int d\chi_2\wedge \left(\frac{\delta S_0}{\delta H_3}
 \right)
\, .
 \end{equation}
 Now since $d\chi_2$ has components only along $AdS_3$,
 we must pick the component of 
 $\left(\frac{\delta S_0}{\delta H_3}
 \right)$ along $S^3$. Using \refb{exz3} we now 
 get\footnote{We are using the sign convention that
 $\int_{AdS_3\times S^3} B_{AdS_3}\wedge A_{S^3} 
 = (\int_{S^3} A_{S^3}) (\int_{AdS_3} B_{AdS_3})$ for 3-forms
 $A$ and $B$ on $S^3$ and $AdS_3$ respectively.}
 \be \label{exz8}
 \delta S_0= 2\pi PQ\int_{AdS_3} d\chi_2,
\end{equation}
which is the gauge variation of a three dimensional Chern-Simons terms
\begin{equation} \label{exz9}
 2\pi\, PQ\, \int_{AdS_3}\chi_3
 = 
 {PQ\over 4\pi}\,  \int_{AdS_3}\left[-\omega(A_L)+\omega(A_R)
 \right]\, .
\end{equation}
Using the
standard relation between the coefficients of the
Chern-Simons terms and the level 
$(k_R^{bulk}, k_L^{bulk})$ 
of the current algebra in the boundary theory 
\cite{0506176,0508218,0609074} we get
from \refb{exz9}
\be \label{erefbulk}
k_R^{bulk}=k_L^{bulk} = PQ\, .
\ee 
For the case of D1-D5 system in type IIB Cardy limit it
follows from \refb{epnorm}, \refb{erotate},
\refb{exz2} and \refb{exz3} that we have
$P=Q_5$, $Q=Q_1$ and hence $PQ=Q_1Q_5$. 
In the type IIA Cardy limit the system
is an NS5-brane fundamental string system and we have
$P=Q_5$, $Q=n$ and hence $PQ= Q_5n$.

So far we have assumed that the six dimensional Lagrangian
density is gauge and diffeomorphism invariant.
Let us now discuss the effect of the Chern-Simons term in the
six dimensional action of the form
\be \label{ecsx1}
S_{CS} = -{\beta\over 32\pi^3}\, \int_{AdS_3\times S^3} 
H_3\wedge \omega_v(\Gamma)
= {\beta\over 32\pi^3}\, \int_{AdS_3\times S^3}
\omega_v(\Gamma)\wedge H_3\, ,
\ee
where $\Gamma$ is the six-dimensional spin connection, and
$\omega_v(\Gamma)$ is the Lorentz Chern-Simons term
\be \label{ecsx2}
\omega_v(\Gamma) = Tr_v\left( \Gamma\wedge d\Gamma +
{2\over 3}\Gamma\wedge \Gamma\wedge \Gamma\right)\, ,
\ee
the trace being taken over the vector representation
of $SO(6)$.
For field configurations of the type we are considering we have
\begin{equation}
\omega_v(\Gamma)= \omega_v(\Gamma_{AdS_3})
 +\omega_v(A),
\end{equation}
where $\Gamma_{AdS_3}$ denotes the spin connection
in $AdS_3$ and $A$ denotes the $SO(4)$ gauge fields associated
with the compactification on $S^3$.
After integrating over $S^3$ the 
Chern-Simons term \refb{ecsx1} reduces to
\be \label{ecsx5}
{\beta\over 8\pi} \, P \, \int_{AdS_3}
\left[ \omega_v(\Gamma_{AdS_3})
 +\omega_v(A)\right]\, .
\ee
Now the gauge field $A$ can be decomposed into $SU(2)_L$
and $SU(2)_R$ parts $A_L$ and $A_R$, and the trace over
the vector representation of $SO(4)$ will give twice the trace
over the fundamental representation of $SU(2)_L$ and 
$SU(2)_R$. This enables us to write \refb{ecsx5} as
\begin{equation}  \label{ech2}
 \int_{AdS_3}\left[
 \frac{\beta P}{8\pi}\omega_v(\Gamma_{AdS_3})
 +\frac{\beta P}{4\pi}\omega(A_R) + \frac{\beta P}{4\pi}
 \omega(A_L) \right]\, ,
\end{equation}
where in computing $\omega(A_{R,L})
= \text{Tr$_f$}\left(A_{L,R}\wedge dA_{L,R}
+\frac{2}{3}A_{L,R}\wedge A_{L,R} \wedge A_{L,R}\right)$ we
compute the trace in the fundamental representation.
Using the standard relation between the Chern-Simons
coefficients and the central charges \cite{0506176,0508218,0609074} 
we now get the
following one loop corrections to the various central charges:
\be \label{eonelooppre}
\Delta  c_{grav}^{bulk} = 12 \beta P, \qquad \Delta 
 k_R^{bulk} = \beta P,
\qquad \Delta  k_L^{bulk} = - \beta P\, .
\ee

Finally we shall briefly discuss possible effect of Chern-Simons terms
on the definition of the charges. For this we note first that the
correct definition of the electric and magnetic charges is via
eq.\refb{exz2} and \refb{exz3}, but with the $S^3$ located at
infinity instead of in the intermediate $AdS_3$ region. Thus the
question is whether the value of the integrals change as we move
the integration surface from the intermediate $AdS_3$ region to 
asymptotic infinity. Since $H_3=dB$, the integral \refb{exz2} does
not change. On the other hand due to the presence of the
Chern-Simons term in the action we have from the equation of
motion of $B$,
\be \label{echar1}
d \left({\delta S_0\over \delta H_3}\right) 
\propto Tr(R\wedge R)\, ,
\ee
where $R$ is the six dimensional Riemann tensor. Since the
topology of the region bounded by asymptotic infinity and
the intermediate $AdS_3$ geometry has the form of
$\RRR\times S^3$, integral of $Tr(R\wedge R)$ over this region
vanishes. Thus we see that the presence of the Chern-Simons
term does not change the definition of the electric charge
either.\footnote{Note that if instead we place the system
at the center of Taub-NUT space to get a four dimensional
black hole\cite{0503217}, then the near horizon geometry and hence the
entropy remains the same, but the charge of the system
receives an additional contribution from the 
Chern-Simons term\cite{0807.0237}.
This can be seen in two ways; by integrating $Tr(R\wedge R)$
between the horizon and the asymptotic space, or by dimensionally
reducing the action on a circle so that the Chern-Simons term
takes a covariant form and the contribution of this term to the
charge can be calculated using the
entropy function formalism.}

\section{Asymptotic Expansion} \label{sa}

In this appendix we shall analyze carefully the behavior of the
index associated with the D1-D5-p system in various limits
and check that possible corrections to the results derived in
\S\ref{smicro} are indeed subleading.
Our starting point is the integral representation for the index
\be \label{ehhx}
d_{micro}(n,Q_1, J)
= (-1)^{J+1}\,
\int_0^1 d\rho_1 \int_0^1 d\sigma_1 \int_0^1 dv_1\,
e^{-2\pi i (\rho n + \sigma Q_1 + J v)}\, f(\rho,\sigma, v)\, ,
\ee
where $(\rho,\sigma,v)\equiv (\rho_1+i\rho_2, \sigma_1+i\sigma_2,
v_1+iv_2)$ are three complex parameters and 
$f(\rho,\sigma, v)= (e^{\pi i v} - e^{-\pi i v})^4
{\eta(\rho)^{24} / \Phi_{10}(\rho,\sigma, v)}$ for five dimensional
black holes and $1/\Phi_{10}(\rho,\sigma, v)$ for four dimensional
black holes. While carrying out this integral we fix
$(\rho_2,\sigma_2,v_2)$   at
\be \label{estr2}
\rho_2 = \Lambda {Q_1 \over \sqrt{4 n Q_1 - J^2}}, \quad
\sigma_2 = \Lambda {n \over \sqrt{4 n Q_1 - J^2}}, \quad
v_2 =- \Lambda {J\over 2\sqrt{4 n Q_1 - J^2}}
\, ,
\ee
where $\Lambda$ is a large positive number. For four dimensional
black holes this choice gives the degeneracy of single centered
black holes \cite{0706.2363}.

We now consider a family of contours
\be \label{estr3}
\rho_2 = \lambda {Q_1 \over \sqrt{4 n Q_1 - J^2}}, \quad
\sigma_2 = \lambda {n \over \sqrt{4 n Q_1 - J^2}}, \quad
v_2 =- \lambda {J\over 2\sqrt{4 n Q_1 - J^2}}
\, ,
\ee
where $\lambda$ is a real number. At $\lambda=\Lambda$ we recover
the original contour.
But we now deform the contour by reducing $\lambda$.
As long as the contour does not cross any pole of the integrand
the value of the integral remains unchanged.
Now the poles of the integrand are given by the divisors of the function
$\Phi_{10}(\rho,\sigma, v)$ which are the surfaces
\be
\label{divdefa}
n_2 ( \rho\sigma- v^2) + j v + n_1 \sigma  - m_1 \rho + m_2 = 0\ ,
\ee
where $j$ is any odd integer
and  the 5 integers $(m_1,m_2,n_1, n_2,j)$
are constrained to satisfy
\be
\label{deldefa}
 j^2 + 4 (m_1 n_1 + m_2 n_2)-1 = 0\ .
\ee
$n_{2} $ can be chosen to be non-negative.
The intersection of the codimension 3 subspace given in
\refb{estr3} and the codimension 2 subspace given in \refb{divdefa}
describes a one dimensional curve in the six dimensional space
spanned by $(\rho,\sigma,v)$. For fixed $(\rho_2,\sigma_2,v_2)$
it is an easy exercise to find this
curve in the $(\rho_1,\sigma_1,v_1)$ space
and we arrive at the result:
\ben \label{curve}
&& \rho_1 = -{n_1\over n_2}-
{1\over \sigma_2}\, \left\{ \rho_2 \left(\sigma_1
- {m_1\over n_2}\right)- 2 v_2 \left(v_1 - {j\over 2n_2}\right)\right\}
\nonumber \\ &&
{\rho_2\over \sigma_2} \left(\sigma_1
- {m_1\over n_2}\right)^2 + \left(v_1 - {j\over 2n_2}\right)^2
- 2{v_2\over \sigma_2} \left(\sigma_1
- {m_1\over n_2}\right) \left(v_1 - {j\over 2n_2}\right)
= {1\over 4 n_2^2} - (\rho_2\sigma_2-v_2^2)\, . \nonumber \\
\een
The last equation describes an ellipse in the
$(\sigma_1,v_1)$ plane for $(\rho_2\sigma_2 - v_2^2)
< (4n_2^2)^{-2}$ and has no solution otherwise. Using
\refb{estr3} the condition for the absence of a solution to
\refb{curve} reduces to
\be \label{eabsence}
\lambda > {1\over n_2}\, .
\ee
This shows that as long as $\lambda$ is larger then 1, none of the
poles of the integrand intersect the contour and hence the integral will
have the same value for all $\lambda>1$. We shall however deform
the contour to $\lambda = {1\over 2}+\eps$ where $\eps$ is a
small positive number. During the deformation of $\lambda$ from
$\Lambda$ to ${1\over 2}+\eps$ the contour crosses the $n_2=1$
poles. The contribution from the residue at this
pole was analyzed in \S\ref{smicro}.\footnote{For $n_2=1$ we can
use the three shift symmetries $\rho\to\rho+1$, $\sigma\to
\sigma+1$ and $v\to v+1$ to set $n_1=m_1=m_2=0$ and
$j=1$ \cite{9607026}.}
Our goal will be to
analyze the contribution from the contour at $\lambda={1\over 2}
+\eps$ and argue that this integral is subdominant compared to the
residue at the $n_2=1$ pole.

Our strategy will be to estimate each term appearing in the integrand
separately and then multiply the results to estimate the integrand.
First consider the exponential factor in
\refb{ehhx}. For the choice of $(\rho_2,\sigma_2,v_2)$ given in
\refb{estr3} with $\lambda={1\over 2}+\eps$, this factor is given by
\be \label{expbound}
\exp\left[\left({1\over 2}+\eps\right)\, \pi\, \sqrt{4 n Q_1 - J^2}\right]\, ,
\ee
up to a phase.

Next consider the $(e^{\pi i v} - e^{-\pi i v})^4
\eta(\rho)^{24}$ factor that is present in the
five dimensional index. Since for
\refb{estr3} $|e^{\pi i v} - e^{-\pi i v}|^4\sim 1$
and $|\eta(\rho)|<1$, we can drop
 this while estimating an
 upper bound for the integrand. This will allow us to study the
 corrections to the
 four and the five dimensional degeneracies together since they
 differ only due to the presence of the  $\eta(\rho)^{24}$ factor.
 This will also
 have the advantage that for the five dimensional black holes
 once we estimate the correction term in the type IIB Cardy limit, we
 can get the result for the type IIA Cardy limit by exchanging $n$
 and $Q_1$ since the only term in the integral that breaks this
 symmetry is the $\eta(\rho)^{24}$ factor.

Finally we turn to an estimate of $1/\Phi_{10}$. On the subspace
\refb{estr3} $\rho_2\sigma_2-v_2^2$ is finite, but in the two
limits we are interested in, either $\rho_2$ or $\sigma_2$ becomes
small. We do not have a way to find a direct estimate of $\Phi_{10}$
in this region; so we shall use an intuitive reasoning. First of all note that
if $\lambda=1/n_2$ then the equations \refb{estr3},
\refb{curve} have a unique
solution:
\be \label{etaudefnewa}
\rho = {i\over 2 n_{2} \tau_2} - {n_{1} \over n_{2}}, \quad \sigma=i{
\tau_1^2 + \tau_2^2\over 2 n_{2} \tau_2} + {m_1 \over n_{2}},
\quad v = {j\over 2n_{2}} - i{\tau_1\over 2 n_{2} \tau_2}\, .
\ee
where
\be \label{etaudefa}
\tau_1 = {J\over 2 Q_1}, \qquad \tau_2 =
\sqrt{4nQ_1-J^2\over 4 Q_1^2}\, .
\ee
This represents the unique point on the surface \refb{estr3} with
$\lambda=1/n_2$ which  also
lies on the divisor \refb{divdefa}.
Thus $1/\Phi_{10}$ diverges
there. For $\lambda={1\over n_2}+\eps$ the surface \refb{estr3} does not
intersect the divisor \refb{divdefa},
but for sufficiently small $\eps$ the two subspaces come close
near a point near \refb{etaudefnewa}.
Since
$1/\Phi_{10}$ has a double pole near the divisor \refb{divdefa}
we expect that as we move along \refb{estr3}, $1/|\Phi_{10}|$
reaches a local
maximum near the point of closest approach to the
divisor  \refb{divdefa}, which in turn is close to \refb{etaudefnewa}.
Assuming that the dominant contribution to the integral comes
from near this local maximum, we can estimate
$1/\Phi_{10}$ by its behavior
near this divisor.
This was analyzed in  \cite{0810.3472,0904.4253}.
We shall here follow the
notation of  \cite{0904.4253} where the analysis was carried out for
general value of $n_2$.
The analysis uses the fact that all the divisors lie
in an orbit of $Sp(2,\IZ)$ under which the $\Phi_{10}$
is a Siegel modular form of weight $10$.
At the diagonal divisor $v=0$,
\be
\label{resv0}
{1 \over \Phi_{10}(\rho,\sigma,v)} = -{1\over 4\pi^2}\, 
\frac{1}{v^2\, \eta^{24}(\rho) \, \eta^{24}(\sigma)} + \cO(v^0)
\  .
\ee
One then finds the explicit $Sp(2,\IZ)$ transformation
which maps the divisor $v=0$ to the generic divisor
(\ref{divdefa}), and then uses the modular property
of the function $\Phi_{10}$ to find the residue at the
generic pole. Thus near such a generic pole we shall have
\be \label{egen1}
 {1 \over \left|\Phi_{10}(\rho,\sigma,v)\right|} \sim \frac{1}
 {\left|v_0^2\, \eta^{24}(\rho_0) \, \eta^{24}(\sigma_0)\right|}
\sim \exp\left[-2\ln |v_0| - 24 \ln |\eta(\rho_0)\eta(\sigma_0)|
\right] \, ,
 \ee
 where $(\rho_0, \sigma_0, v_0)$ are related to $(\rho,\sigma, v)$
 by this specific $Sp(2,\IZ)$ transformation. In writing
 \refb{egen1} we have ignored some additional factors related
 to the modular weight of $\Phi_{10}$, but they do not affect
 the estimate to leading order.
 The dominant contribution to the exponent comes from the
 $- 24 \ln |\eta(\rho_0)\eta(\sigma_0)|$ terms. Thus our goal will be to
 estimate this term. For sufficiently small $\eps$ we can
 estimate this
 by evaluating $\rho_0$ and $\sigma_0$ at the
 point \refb{etaudefnewa}. This in turn requires knowing the
 $Sp(2,\ZZZ)$ transformation that relates $(\rho,\sigma,v)$ to
 $(\rho_0,\sigma_0,v_0)$.

Before we proceed we need to define some number
theoretic quantities.
First, define $r \equiv {\rm gcd}(n_1,n_2)$, so we can
write $r=k_{2}n_{1} - k_{1}n_{2}$ for some $k_1,k_2\in\ZZZ$.
Since (\ref{deldefa}) is satisfied,
$r$ must  divide $(j^2-1)/4$.  We can then uniquely
decompose  $r=r_1 r_2$ into a product
of relatively prime factors, where $r_1$ divides
$(j+1)/2$ and $r_2$ divides $(j-1)/2$.
In this convention the result of
 \cite{0904.4253}
for $(\rho_0,\sigma_0)$ are
\be \label{rho0sig0a}
\rho_{0} =   \delta_1
+  {r_{2}^{2} \over n_{2}} (- \tau_{1} + i \tau_{2}) , \quad
\sigma_{0}  = \delta_2 + {r_{1}^{2} \over n_{2}}
( \tau_{1} + i \tau_{2}) \ ,
\ee
where $\delta_1$ and $\delta_2$ are constants determined
in terms of $m_i,n_i,j$.
In the type IIB Cardy limit we get from \refb{etaudefa} that
$\tau_2$ is large. In this limit we get
\be \label{est1}
|\eta^{-24}(\rho_0) \eta^{-24}(\sigma_0)|
\sim \exp\left[{2 \pi \over n_{2}} (r_{1}^{2}
+r_{2}^{2}) \tau_{2}\right] \sim
\exp\left[ {2 \pi \over n_{2}} (r_{1}^{2}
+r_{2}^{2}) \sqrt{4nQ_1-J^2\over 4 Q_1^2}\right]\, .
\ee
Let us now focus on the case $n_2=2$ since our goal is to estimate
the integrand on the contour $\lambda={1\over 2}+\eps$.
Since $r_1r_2$ is a divisor of $n_2$, for $n_2=2$ we have
$r_1^2+r_2^2\le 5$. Thus
\refb{est1} gives
\be \label{est2}
|\eta^{-24}(\rho_0) \eta^{-24}(\sigma_0)|\lsim
\exp\left[ 5\pi   \sqrt{4nQ_1-J^2\over 4 Q_1^2}\right]\, .
\ee
The result for the type IIA Cardy limit may be obtained by exchanging
$Q_1$ and $n$ in \refb{est2}:
\be \label{est3}
|\eta^{-24}(\rho_0) \eta^{-24}(\sigma_0)|\lsim
\exp\left[ 5\pi   \sqrt{4nQ_1-J^2\over 4 n^2}\right]\, .
\ee
Combining \refb{expbound} with \refb{est2}, \refb{est3}
we arrive at the following estimates for the correction
$\delta d_{micro}$ to the
index $d_{micro}$ at $\lambda = {1\over 2}+\eps$.
In the type IIB Cardy limit we have
\be \label{est5}
\delta d_{micro}\lsim
\exp\left[\left({1\over 2}+\eps\right)\, \pi\,
\sqrt{4 n Q_1 - J^2} + 5 \pi
\sqrt{4nQ_1-J^2\over 4 Q_1^2}\right]
\ee
and in the type IIA Cardy limit
\be \label{est6}
\delta d_{micro}\lsim
\exp\left[\left({1\over 2}+\eps\right)\, \pi\,
\sqrt{4 n Q_1 - J^2} + 5\pi    \sqrt{4nQ_1-J^2\over 4 n^2}
\right]\, .
\ee
Comparing \refb{est5} with the result given in \refb{eiibcardy}
we see that
the correction terms are smaller than \refb{eiibcardy} if
\be \label{ecomp1}
\sqrt{Q_1 -{J^2\over 4n}} > {1\over 2} \, \sqrt{Q_1 -{J^2\over 4n}}
+ {5 \over 2 Q_1} \sqrt{Q_1 -{J^2\over 4n}}\, .
\ee
This holds for $Q_1> 5$.
Similarly comparing \refb{est6}  with the result given in
\refb{ehetcardy} we see that the correction terms are subdominant
in the region:
\be \label{ecomp2}
\sqrt{(n+3) \left(1 - {J^2 \over 4(n-1)Q_1}\right)}
> {1\over 2} \, \sqrt{n -{J^2\over 4Q_1}}
+ {5 \over 2 n} \sqrt{n -{J^2\over 4Q_1}}\, .
\ee
This can be easily satisfied for example by requiring
\be \label{ecomp3}
n - {J^2\over 4Q_1} \ge 7\, .
\ee
Neither of these are the best bounds possible, particularly since
we have dropped the $(\eta(\rho))^{24}$ factor from the integrand in
estimating the correction term.
However this analysis shows the existence of the constants $K_1$,
$K_2$ appearing in the definition of the type IIB and type IIA Cardy
limits beyond which our result for the asymptotic
behavior of the microscopic index holds.
Finally the leading
contribution to the
four dimensional index in the $n\to\infty$ limit, given in
\refb{e4dent}, is always larger than the five dimensional
index \refb{eiibcardy} in the type IIB Cardy limit, and hence will
dominate over the correction given in \refb{est5} when
$Q_1>  5$.

\bibliographystyle{utphys}

\end{document}